\documentclass[11pt,tightenlines,eqsecnum,floats,aps,amsmath,amssymb,amsfonts,showpacs,nofootinbib,superscriptaddress,prd]{revtex4-2}

\usepackage{LQC-symbols}
\usepackage{makecell}

\newcommand{\blA}{\mathbf{A}}
\newcommand{\blB}{\mathbf{B}}
\newcommand{\blM}{\mathbf{M}}
\renewcommand{\id}{\openone}

\usepackage{enumerate}
\usepackage{graphicx}
\usepackage{tikz}
\usepackage{pgfplots}
\usetikzlibrary{calc,fadings,decorations.markings,pgfplots.fillbetween}
\usepackage{xparse}
\usetikzlibrary{shapes}
\usepackage{epsfig}
\usepackage{color}
\usetikzlibrary{quotes,angles}
\usepackage{tkz-euclide}
\usepackage{euscript,amssymb}
\usepackage{mathrsfs}
\usepackage{amsbsy}
\usepackage{epsfig}
\usepackage{color}

\usepackage{amsfonts}
\usepackage{enumitem}
\usepackage{amsmath}
\usepackage{amssymb}

\usepackage{fancyhdr}

\usepackage{esint}
\usepackage[unicode=true, pdfusetitle,
bookmarks=true,bookmarksnumbered=false,bookmarksopen=false,
breaklinks=false,pdfborder={0 0 1},backref=false,colorlinks=false]
{hyperref}

\usepackage{colordvi}

\begin{document}

%%%%%%%%%%%%%%%%%%%%%%%%%%%%%%%%%%%%%%%%%%%%%%%%%%%%%%%%%%%%%%%%%%%%
\title{Rainbow Black Hole From Quantum Gravitational Collapse}

\author{Aliasghar Parvizi}
\email{a.parvizi@ut.ac.ir}
\affiliation{Department of Physics, University of Tehran, 14395-547 Tehran, Iran}
\affiliation{School of Particles and Accelerators, Institute for Research in Fundamental Sciences (IPM) \\ 
	P.O. Box 19395-5531, Tehran, Iran}

\author{Tomasz \surname{Paw{\l}owski}}
\email{tomasz.pawlowski@uwr.edu.pl}
\affiliation{Institute for Theoretical Physics, Faculty of Physics and Astronomy, University of Wroc{\l}aw, pl. M. Borna 9, 50-204  Wroc{\l}aw, Poland}

\author{Yaser Tavakoli}
\email{yaser.tavakoli@guilan.ac.ir}
\affiliation{Department of Physics, University of Guilan, 41335-1914 Rasht, Iran}
%\affiliation{School of Astronomy, Institute for Research in Fundamental Sciences (IPM),	 19395-5531, Tehran, Iran}

\author{Jerzy Lewandowski}
\email{jerzy.lewandowski@fuw.edu.pl}
\affiliation{Faculty of Physics, University of Warsaw, Pasteura 5, 02-093 Warsaw, Poland}

%%%%%%%%%%%%%%%%%%%%%%%%%%%%%%%%%%%%%%%%%%%%
\begin{abstract}

Quantum evolution of a scalar field's modes propagating on quantum spacetime of a collapsing homogeneous dust ball is written effectively, as an evolution of the same quantum modes on a (semiclassical) dressed geometry. When the backreaction of the field is discarded, the classical spacetime singularity is resolved due to quantum gravity effects and is replaced by a quantum bounce on the dressed collapse background. In the presence of backreaction, the emergent (interior) dressed geometry becomes mode dependent and the energy density associated with the backreaction of each mode scales as a radiation fluid. Semiclassical dynamics of this so-called {\em rainbow} dressed background is analyzed. It turns out that the backreaction effects speed up the occurrence of the bounce in comparison to the case where only a dust fluid is present.
By matching the interior and exterior regions  at the boundary of dust, a mode-dependent  black hole geometry emerges as the exterior spacetime. Properties of such a rainbow black hole are discussed. That mode dependence causes, in particular, a chromatic aberration in the gravitational lensing process of which maximal magnitude is estimated via calculation of the so-called Einstein angle.

\end{abstract}

\date{\today}

%%%%%%%%%%%%%%%%%%%%%%%%%%%%%%%%%%%%%%%%%%%%%%%%%%%%%%%%
%\keywords{loop quantum gravity, quantum field theory}
\pacs{04.60.-m, 04.60.Pp, 98.80.Qc, 04.60.Bc}

\maketitle

\section{Introduction}

In contrast to classical  general relativity (GR), quantum gravity models based on discrete spacetime structure often predict that the local Lorentz invariance may be modified or broken at sufficiently high energy \cite{Gambini:1998it,Alfaro:1999wd}. This leads, in particular, to the deformation of dispersion relations for the propagation of particles \cite{Gambini:1998it,Alfaro:1999wd,Magueijo:2002xx,Mattingly:2005re,Lafrance:1994in}. Consequently, it has some  phenomenological implications that  can  provide an empirical  ground to test  quantum  gravity theories \cite{AmelinoCamelia:1997gz, Mattingly:2005re} (for a review on such an issue, see \cite{Addazi:2021xuf}). The effects of possible Lorentz invariance violation are expected, in particular, to be present within frameworks implementing the quantum theory of fields propagating on a  quantized background spacetime.

An example of such a framework, where  significant progress has been achieved recently, is a quantum field theory (QFT) on a spherically symmetric quantum spacetime described by loop quantum gravity (LQG) \cite{Ashtekar:2009mb,Dapor:2013pka,Lewandowski:2017cvz}. It was shown that the evolution of quantum fields in a quantum spacetime  leads to emerging  an effective (semiclassical) dressed background metric. The components of this dressed metric depend on the fluctuations of the background quantum geometry. In the presence of the backreaction of the fields, the emergent dressed metric's  components  depend further on the energy of the field modes \cite{Dapor:2012jg,Lewandowski:2017cvz}, which is called ``rainbow metric" in the literature \cite{Lafrance:1994in,Magueijo:2002xx}. Propagation of electromagnetic signals (or massless scalar perturbations)  on this rainbow background is superluminal, which violates the local Lorentz symmetry \cite{Lewandowski:2017cvz}. [See other  scenarios for violation of the Lorentz symmetry, e.g., due to an emerging rainbow metric from a {\em massive} quantum   field on a loop quantum cosmology (LQC) spacetime \cite{Assaniousssi:2014ota}, or due to polymer quantization of the field \cite{Garcia-Chung:2020zyq}.]

The highest observable energies in the Universe are provided by cosmic gamma rays and cosmic rays. Considering the fact that all long-duration gamma ray bursts (GRBs) are physically connected with the core-collapse supernovae (SNe) \cite{AmelinoCamelia:1997gz,Bernardini:2014oya, Yi:2005ht,Meszaros:2006rc}, it is not unreasonable to expect that when observing  GRBs, we directly observe the gravitational collapse of a massive and compact star core. Therefore, it is of pertinence to  search for Lorentz violation signals in the gravitational collapse of a massive star. Gravitational collapse of a fluid with a variety of  matter fields has been well studied  within the framework  of LQG \cite{Tavakoli:2013tpa,Tavakoli:2013rna,Goswami:2005fu,Bojowald:2005qw}. It turns out that, when considering a  background quantum spacetime for gravitational collapse of a spherically symmetric star, a rainbow interior region would emerge which effects can be carried out to the exterior region  through appropriate junction conditions at the boundary of the star. This provides a fruitful scenario for the formation of a rainbow black hole as the final state of  gravitational collapse in quantum gravity. If such black holes exist in nature, high-energy astrophysical observations from them can raise the possibility of tests of the Lorentz symmetry and quantum gravity theories.

The present work is concerned with the phenomenological  issues of quantum gravity in the context of gravitational collapse. It is organized as follows. In Sec. \ref{Classic-Dust} we present the dynamics of the gravitational collapse of a spherically symmetric dust cloud. In Sec. \ref{QFT0} and \ref{QFT2} we study the quantum theory of a massless scalar field on the background spacetime of a collapsing dust ball. Then, we quantize the background due to LQG and show that the theory of quantum field on this quantum background corresponds to a quantum theory of the same field on an effective, dressed background. By considering the backreaction of the field, we will show that the components of the dressed background metric depend on the energy of the field. Next, we expand the dynamics of the emerging dressed background spacetime  by means of the higher-order quantum corrections provided by fluctuations due to moments of the quantum spacetime state through a semiclassical regime. In Sec. \ref{exterior}, we match the interior spacetime to a convenient exterior geometry. We will show that the quantum gravity effects in the interior region are carried out to the collapse of radiation exterior spacetime by matching and a rainbow black hole will emerge. We will discuss some optical properties of the emerging rainbow black hole in Sec. \ref{GL}. Finally, in the Sec. \ref{conclusion}, we will present the conclusion of our work.

%%%%%%%%%%%%%%%%%%%%%%%%%%%%%%%%%%%%%%%%%%%%%%%%
\section{Gravitational collapse of a dust field}
\label{Classic-Dust}

Our purpose in this section is to construct a classical  model  of gravitational collapse with an interior region filled with an {\em irrotational dust}  field $T$, such that  in the late time stages of the collapse (cf. next sections), when the fluid enters the Planck regime, the quantum gravity effects could alter the nature of singularity or/and development of trapped surfaces in spacetime. Thus, we consider  a homogeneous, isotropic barotropic fluid for the matter content of the interior collapse background which can have a Friedmann-Lema\^{i}tre-Robertson-Walker (FLRW) metric, equipped with the coordinates $(x_0, \mathbf{x})$ and a scale factor $a(x_0)$. We assume that $x_0\in \mathbb{R}$ is a generic time coordinate and $\mathbf{x}\in \mathbb{T}^3$ is the spatial coordinates [$\mathbb{T}^3$ is the three-torus with coordinates $x^j\in (0, \ell)$].

For the background matter source, being the  irrotational dust  field $T$, the Lagrangian density is given by \cite{Husain:2011tk}
\begin{equation}
\mathcal{L}_{T}\, =\,  - \tfrac{1}{2}\sqrt{-g}\rho_T\big(g^{\mu \nu}\partial_{\mu}T\partial_{\nu}T +1\big),
\end{equation}
where $\rho_T$ is a multiplier enforcing the gradient of the dust field to be timelike (e.g., see also \cite{Brown:1994py,Bojowald:2010qpa}).
We further consider a (inhomogeneous) {\em massless scalar field   perturbation} $\phi(t, \mathbf{x})$,  with the Lagrangian
\begin{equation}
\mathcal{L}_{\phi}\, =\,  - \tfrac{1}{2}\sqrt{-g}\, g^{\mu \nu}\partial_{\mu}\phi \partial_{\nu}\phi,
\end{equation}
which propagates on the background spacetime of the collapsing cloud. The corresponding action for the background geometry, coupled to dust  and the scalar perturbation,  reads
\begin{eqnarray}
S =  \int d^4x \sqrt{-g} \Big[\frac{\mathcal{R}}{16\pi G} +   \mathcal{L}_{T}\Big] + S_{\phi}\, .
\end{eqnarray}
Then, on the full phase space, total Hamiltonian density  is written as 
\begin{eqnarray}
{\cal H}  &=& {\cal H}_{\rm grav} + {\cal H}_{T} + {\cal H}_{\phi},
\label{Hamiltonian-tot}
\end{eqnarray}
where, ${\cal H}_{\rm grav}$, ${\cal H}_{T}$ and ${\cal H}_{\phi}$ are respectively, the Hamiltonian densities of the gravitational sector, dust  and the scalar field. 

Using a general background metric in Arnowitt-Deser-Misner decomposition,
\begin{equation}
\rd s^2  = -N^2 \rd x_0^2 + q_{ab}(N^a\rd x_0 + \rd x^a)(N^b\rd x_0 +\rd x^b),
\end{equation} 
the Hamiltonian densities of the scalar perturbation, ${\cal H}_{\phi}$, and the background dust field, ${\cal H}_T$, are written respectively, as
\begin{eqnarray}
{\cal H}_\phi = \frac{N}{2}\left[\frac{P_\phi^2}{\sqrt{q}} + \sqrt{q}q^{ab}(\nabla_a\phi)(\nabla_b\phi) \right],\quad \quad 
\label{Hamil-scalar}
\end{eqnarray}
and
\begin{eqnarray}
{\cal H}_T = \frac{1}{2}\left[\frac{p_T^2}{\rho_T\sqrt{q}} + \frac{\rho_T\sqrt{q}}{p_T^2}\left(p_T^2 + q^{ab}C^D_aC^D_b\right)\right]. \quad \quad 
\end{eqnarray} 
In the above equations, $N$ and $N^a$ denote, respectively, the lapse function and the shift vector, and $q_{ab}$ is the (spatial) three-metric with the conjugate momentum $\pi^{ab}$. Moreover, $C^D_a=-p_T\partial_aT$, where $p_T$ is the momentum  conjugate to $T$,  given by
\begin{eqnarray}
p_T = (\sqrt{q}\rho_T/N)(\partial_0T + N^a\partial_aT).
\end{eqnarray}

The stress-energy tensor for the dust field  can then be obtained as
\begin{eqnarray}
{\rm T}_{\mu \nu}^{\rm (D)} = \frac{2}{\sqrt{-g}}\frac{\delta S_T}{\delta g^{\mu \nu}} = \rho_T u_{\mu} u_{\nu}\, ,
\label{tensor-Dust}
\end{eqnarray}
where $u_{\mu}=\partial_{\mu}T$ is the four-velocity field in the dust coordinate frame which satisfies the condition $g^{\mu\nu}u_{\mu}u_{\nu}=-1$. 
From the equation of motion for $\rho_T$, we obtain  \cite{Husain:2011tk}
\begin{eqnarray}
\rho_T = \frac{1}{\sqrt{q}}\frac{p_T^2}{\sqrt{p_T^2+q^{ab}C^D_aC^D_b}}\, . 
\label{rho_T}
\end{eqnarray}
Equation~(\ref{tensor-Dust}) represents the stress-energy tensor of a perfect fluid with the energy density $\rho_T$ and a vanishing pressure, so $\rho_T$ is regarded as the  dust energy density.
Now, by substituting Eq.~\eqref{rho_T} in ${\cal H}_T$ and fixing time gauge $x_0=T$, the total Hamiltonian density (\ref{Hamiltonian-tot})  becomes
\begin{eqnarray}
{\cal H}  = {\cal H}_{\rm grav} + p_{T} + {\cal H}_{\phi}.
\label{Hamiltonian-tot2}
\end{eqnarray}
The dynamics of the gravity-matter system is then generated by the regulated integral 
\begin{eqnarray}
NH = \int_{\cal{V}} N \left( {\cal H}_{\rm grav} + p_{T} + {\cal H}_{\phi} \right),
\label{Hamiltonian-tot3}
\end{eqnarray}
where, for simplicity, we have considered $\cal{V}$ to be a cell of unit volume.

We will consider the internal spacetime model  in the classical regime
given by a marginally bound ($k = 0$) case, 
\begin{equation}
ds^2 = -dx_0^2 + a^2(x_0)\, d\mathbf{x}^2,
\end{equation}
so that the dust fluid begins to collapse from a very
large physical radius.
Then, the physical trajectories lie on the surface of the Hamiltonian constraint
\begin{eqnarray}
- p_{T} =  H_{\rm grav} +  H_{\phi},
\label{Hamiltonian-constraint}
\end{eqnarray}
where $H_{\rm grav}$ is given by
\begin{eqnarray}
H_{\rm grav} = - \frac{3\pi G}{2 \alpha_o} b^2 |v|.
\end{eqnarray}
In LQC, the gravitational part of the phase space is conveniently coordinatized by a canonically conjugate pair $\{b, v\}=2$, where $v=a^3/\alpha_o$ is the oriented volume  and $b$ is the Hubble parameter $b=\gamma(\dot{a}/a)$. As usual, a ``dot"  refers to a derivative with respect to the proper time $x_0$. 
Moreover, $\alpha_o=2\pi \gamma \sqrt{\Delta}  \;\ell_{\rm Pl}^2$, in which $\Delta \equiv 4 \sqrt{3}\pi \gamma \ell_{\rm Pl}^2$ is the LQC {\em area gap} \cite{Ashtekar:2006rx}.

By solving the constraint equation (\ref{Hamiltonian-constraint}), we find the evolution equation for the collapse as $(b/\gamma)^2=(\dot{a}/a)^2=(8\pi G/3) \rho$, being the standard Friedmann equation, where $\rho$ is  the total density of the system including the energy densities of dust and scalar perturbation. Moreover, notice that for the collapsing process here, $\dot{a}<0$.
%%%
For the homogeneous spatial slices  here we get $C^D_a=0$, so $p_T=a^3\rho_T(\partial_0 T)$. Therefore, the energy density $\rho_T$ in Eq.~(\ref{rho_T}) reduces to $\rho_T=p_T/a^3$. From the Hamilton equation of motion $\dot{p}_T=\{p_T,  H\}=0$, it turns out  that $p_T$ is a constant of motion, so as expected, the energy density of the dust becomes $\rho_T=p_T\, a^{-3}$.
	
In order to fix sufficient  initial conditions for the collapse, we assume that $\rho_0$ and $a_0$ are, respectively, the total energy density and the scale factor of the collapsing cloud at  the initial time, $x_0=0$. On large scales in  the classical region, the energy density of the scalar perturbation is negligible so that,  at the initial configuration of the collapse,  the energy density of the dust field dominates; we thus assume that $\rho\approx\rho_T$. Nevertheless, as the collapse enters the quantum regime, the backreaction of the scalar field, $\phi$, on the background quantum geometry will become important. Our aim in the next sections, therefore, will be to investigate the effects of this backreaction on the evolution of trapped surfaces and the emergence of mode-dependent exterior spacetime.
In fact, the scenario that we will consider is to discuss, by taking a full quantum system, the effects of the backreaction of scalar perturbation on the evolution of the background quantum geometry, and to explore a suitable (semiclassical) geometry for the exterior region. In particular, we shall assume that the homogenous classical sector of the scalar field, $\hat{\phi}$, has no effect on the background;  by definition $\hat{\phi} = \langle \hat{\phi} \rangle \id + \delta \hat{\phi}$, this means that we assume $\langle \hat{\phi} \rangle = 0$ for the vacuum expectation value of $\hat{\phi}$, while $\delta \hat{\phi}$ describes the inhomogeneous part of $\hat{\phi}$. So, there is no backreaction on the background geometry caused by the homogenous part and we will consider only the excitations and particle states of the massless field.  At the semiclassical level, these excitations give only an effective description (sum over all modes) for spacetime.

In the classical region, we  write the energy density of the collapsing dust cloud in the form  
\begin{eqnarray}
\rho = {\rho_T}_0(a_0/a)^3,
\label{dustdensity}
\end{eqnarray}
where  $p_T={\rho_T}_0a_0^3$. As the collapse evolves, the energy density of the dust grows and ultimately diverges at $a=0$. 
Therefore, a singularity will form at the end state of the collapse.  This singularity will be covered by a Schwarzschild  horizon during the dynamical evolution of the collapse which is extracted through a suitable matching at the boundary of the dust cloud. At a given time $x_0$ and for a fixed shell with the radius $r$, the  mass of the  dust cloud reads  $M=(4\pi/3)\rho R^3$, where $R=ra$ is the physical radius of the collapsing shell. It turns out that, for the dust energy density (\ref{dustdensity}), the mass $M$ is constant and equals the initial mass, $M_0=(4\pi/3)\rho_0 r^3a^3_0$, at $x_0=0$. This is the mass of the exterior Schwarzschild black hole with the horizon radius $R_{\rm S}=2GM$.

%%%%%%%%%%%%%%%%%%%%%%%%%%%%%%%%%%%%%%%%%%%%%%%%%%%%%%%%%%%%%%%%%%
\section{QFT in quantized interior background}
\label{QFT0}

In this section, we will first show  that the quantum theory of the scalar field perturbation, $\phi$, propagating on the interior quantum background  of the dust cloud, corresponds to an emerging quantum theory for the field on an effective, dressed background spacetime. 

\subsection{Perturbed background}
\label{QFT1}

In canonical quantum gravity coupled with dust field \cite{Husain:2011tk}, in the case when the effect of a scalar perturbation (denoted by $\hat{\mathcal{H}}_{\phi}$) in Eq.~(\ref{Hamiltonian-tot}) is negligible, the total Hamiltonian  operator, $\hat{H}_{\rm geo}=\hat{H}_{\rm grav}+\hat{H}_T$, of the  system  is well defined on $\mathscr{H}_{\rm kin}^o=\mathscr{H}_{\rm grav}\otimes\mathscr{H}_T$, where $\mathscr{H}_{\rm grav}$ is a suitable Hilbert space for the gravity sector and $\mathscr{H}_T$ is the dust sector of the kinematical Hilbert space, which is quantized according to the Schr\"odinger picture with the Hilbert space $L^2(\mathbb{R}, dT)$. The physical states $\Psi_o(v, T)\in\mathscr{H}_{\rm kin}^o$ are those lying on the kernel of $\hat{H}_{\rm geo}$. Therefore, the states $\Psi_o(v, T)$ are solutions to the self-adjoint Hamiltonian constraint $\hat{H}_{\rm geo}\Psi_o=0$, so that
\begin{align}\label{eq:Psi0-Sch}
i\hbar\partial_T\Psi_o(v, T) = \hat{H}_{\rm grav}\Psi_o(v, T).
\end{align}
Here, $\hat{H}_{\rm grav}$ is a well-defined, self-adjoint operator acting on 
$\mathscr{H}_{\rm grav}$. 
In quantum theory, the polymer representation of the Poisson algebra of $v$ and $b$ is characterized by the Hilbert space $\mathscr{H}_{\rm grav} = L^2(\bar{\re},\rd\mu_{\Bohr})$, where $\bar{\re}$ is the Bohr compactification of the real line and $d\mu_{\rm Bohr}$ is the Haar measure on it \cite{Ashtekar:2003hd}. Thereby, the gravitational Hamiltonian operator is expressed as \cite{Husain:2011tm}
\begin{equation} \label{Ham_grav}
\hat{H}_{\rm grav} = \frac{3\pi G}{8\alpha_o} \sqrt{\hat{v}} \big(\hat{N}^2 - \hat{N}^{-2}\big)^2 \sqrt{\hat{v}} \, ,
\end{equation}
where, $\hat{v}|v\rangle=v|v\rangle$, and the operator $\hat{N}\equiv\widehat{\exp(ib/2)}$ acts on the  basis $\{|v\rangle\}$, i.e. the eigenstates of $\hat{v}$, as  $\hat{N}|v\rangle = |v +1\rangle$, so that,  $[\hat{b}, \hat{v}]=2i\hbar$.

When the contribution of the scalar field $\phi$  in the  Hamiltonian constraint  (\ref{Hamiltonian-tot})  is significant,  the kinematical Hilbert space for the full, quantized gravity-matter system (dust  plus scalar perturbation) becomes $\mathscr{H}_{\rm kin}=\mathscr{H}_{\rm grav}\otimes\mathscr{H}_T\otimes\mathscr{H}_\phi$, where the perturbation sector is quantized due to the Schr\"odinger picture with $\mathscr{H}_\phi=L^2(\mathbb{R}, d\phi)$. Now, the  total states $\Psi\in \mathscr{H}_{\rm kin}$ of the system are different from  the pure geometrical states $\Psi_o$
and are solutions to a new evolution equation 
\begin{eqnarray}
i\hbar\partial_T\Psi(v, \phi, T) = \big(\hat{H}_{\rm grav} + \hat{H}_{\phi}\big)\Psi(v, \phi, T).
\label{Schroedinger1}
\end{eqnarray}
On the quantized background here, the gravitational sectors of the quantized Hamiltonian (\ref{Hamil-scalar}) of the perturbation turn out to be operators on $\mathscr{H}_{\rm kin}$; thus, the quantum Hamiltonian of the massless scalar perturbation becomes
\begin{eqnarray}
\hat{H}_\phi = \tfrac{1}{2}\Big[\hat{V}^{-1}\otimes\hat{P}_\phi^2 + \hat{V}^{1/3}\otimes(\nabla_i\hat{\phi})^2 
%+ m^2\widehat{a^3}\otimes \hat{\phi}^2
\Big] ,
\label{Hamil-scalar2}
\end{eqnarray}
where  $V$, defined as $V=\ell^3a^3=\alpha_o v$, is the physical volume of the collapsing cloud. For convenience, we  set $\ell=1$ throughout this section and will bring it back again into our formalism in Sec. \ref{exterior}.

By using the  Fourier expansion, we can rewrite $\hat{H}_\phi$ in Eq.~(\ref{Hamil-scalar2}) as  an assembly of the Hamiltonians of decoupled harmonic oscillators,  each represented by a pair of canonically conjugate variables $(Q_{\mathbf{k}}, P_{\mathbf{k}})$ \cite{Ashtekar:2009mb}. It reads
\begin{equation}
\hat{H}_\phi := \sum_{\mathbf{k}\in{\cal L}} \hat{H}_\mathbf{k} = \frac{1}{2}\sum_{\mathbf{k}\in{\cal L}} \Big[\hat{V}^{-1}\otimes\hat{P}_\mathbf{k}^2 + k^2\, {V}^{1/3}\otimes \hat{Q}_{\mathbf{k}}^2
\Big],
\label{Hamil-scalar2b}
\end{equation}
where, the wave vectors $\mathbf{k}~(\in2\pi\mathbb{Z})$ span a three-dimensional lattice ${\cal L}$. We will focus on a linear response theory and will  study the quantum theory of a single mode $\mathbf{k}$ of the scalar field on the background quantum spacetime. Thereby,  Eqs.~(\ref{Schroedinger1}) and (\ref{Hamil-scalar2b}) for a given mode $\mathbf{k}$ yield
\begin{equation}
i\hbar\partial_T\Psi_{\mathbf{k}}(v, Q_\mathbf{k}, T) = \big(\hat{H}_{\rm grav} + \hat{H}_{\mathbf{k}}\big)\Psi_{\mathbf{k}}(v, Q_\mathbf{k}, T).\quad \quad
\label{Schroedinger12}
\end{equation}

In order to find the quantum theory of the test field $Q_\mathbf{k}$ on an emergent background spacetime, we should simplify the Schr\"odinger equation (\ref{Schroedinger12}) and represent it as an effective equation for the state $\psi_\mathbf{k}(Q_\mathbf{k})\in\mathscr{H}_\mathbf{k}$ only ($\mathscr{H}_\mathbf{k}$ represents the Hilbert space of each mode $\mathbf{k}$). To do so, we employ the following algorithm:
%%%%%%%%%%%%%%%%%%%%%%%%%%%%%%%%%%%%%%%
\begin{enumerate}[label=\roman*)]
	\item To decompose the heavy (gravity) and light (scalar perturbation) degrees of freedom in the wave function, as $\Psi_\mathbf{k}(v, Q_\mathbf{k}, T) = \Psi(v, T) \otimes \psi_\mathbf{k}(Q_\mathbf{k}, T)$, we will employ the Born-Oppenheimer (BO) approximation. This approximation enables us to take into account the backreaction between the field and the geometry.
	
	\item To make our resulting evolution
	comparable to that of a quantum field on a classical {\em dynamical} background, instead of working in the ``interaction picture" used in Ref.~\cite{Ashtekar:2009mb}, we will trace out the heavy and light degrees of freedom in Eq.~(\ref{Schroedinger12}) to drive the evolution equation for the scalar perturbation  and the background states, respectively, as  
	\begin{align}
	\qquad i\hbar\partial_T\psi_{\mathbf{k}}(Q_\mathbf{k}, T) &=  \hat{H}_{\mathbf{k}}\, \psi_{\mathbf{k}}(Q_\mathbf{k}, T), \label{Schroedinger-field}
	\\
	\quad\quad   i\hbar\partial_T\Psi(v, T) &= \big(\hat{H}_{\rm grav} + \langle\hat{H}_{\mathbf{k}}(\hat{v})\rangle\big)\Psi(v, T), \label{Schroedinger-geo}
	\end{align}
	where, {\small $\langle \hat{H}_{\mathbf{k}}(\hat{v}) \rangle  = \langle\psi_\mathbf{k}(T)|\hat{H}_{\mathbf{k}}(\hat{v})| \psi_\mathbf{k}(T) \rangle$}. In this pattern,  quantum geometry and field are described using the Schr\"odinger picture, in which expectation values evolve over the time parameter and give results of Ref.~\cite{Ashtekar:2009mb} as the mean (test) field  approximation. 
	
	\item We plan to solve Eqs.~\eqref{Schroedinger-field} and \eqref{Schroedinger-geo} step by step perturbatively. At the first step, we will consider the mean field limit, where $\langle\hat{H}_{\mathbf{k}}(\hat{v})\rangle = 0$, and find the unperturbed geometry state $\Psi_o(v, T)$. This state is then used to evaluate the expectation values of the geometry component operators present as components of $\hat{H}_k$, which allows us to evaluate the solutions to Eq.~\eqref{Schroedinger-field} to find the scalar field's eigenfunctions, {\small $\chi^{n}_{\mathbf{k}}(Q_{\mathbf{k}};g^{(1)},g^{(2)})$}, parametrized by said geometry expectation values, $g^{(1)}$ and $g^{(2)}$, given by
 \begin{align}
    g^{(1)} \equiv (1/2)\langle \hat{V}^{-1} \rangle_o  \quad \text{and} \quad 
    g^{(2)} \equiv (1/2)\langle \hat{V}^{1/3} \rangle_o . \label{eq:g12-def}
  \end{align}
  A procedure analogous to this step has already been performed in Ref.~\cite{Ashtekar:2009mb}. In the next step, in order to construct the backreacted state $\Psi_1(v, T)$, we will use the obtained eigenfunctions {\small $\chi^{n}_{\mathbf{k}}(Q_{\mathbf{k}};g^{(1)},g^{(2)})$} to evaluate $\langle \hat{H}_{\mathbf{k}}(\hat{v})\rangle$. The result is then put into Eq.~\eqref{Schroedinger-geo}; however, here the variables $g^{(1)}$ and $g^{(2)}$ are again promoted to geometry operators (powers of $\hat{V}$) according to the form specified in Eq.~\eqref{eq:g12-def}. This, in turn, allows us to find  the modified eigenfunctions $\xi^\mu_\mathbf{k}(v)$ of the geometry. This step is the second-order modification to the so-called {\em test field approximation}, presented in Ref.~\cite{Ashtekar:2009mb}, wherein the  backreaction effects were discarded.
\end{enumerate} 
%%%%%%%%%%%%%%%
In BO approximation, the total wave function consists of the products of two sets of eigenstates: The first one is  
the (discrete) field mode's eigenstate $\chi^{n}_{\mathbf{k}}$, being  the solution to the stationary state equation
\begin{eqnarray}
\hat{\tilde{H}}_\mathbf{k}\, \chi^{n}_{\mathbf{k}}\big(Q_{\mathbf{k}}; g^{(1)},g^{(2)}\big)  = \epsilon^n_{\mathbf{k}}\big(g^{(1)},g^{(2)}\big)\, \chi^{n}_{\mathbf{k}}\big(Q_{\mathbf{k}}; g^{(1)},g^{(2)}\big), \label{Eigenvalue-field1}
\end{eqnarray}
where $\hat{\tilde{H}}_{\mathbf{k}} \in \mathscr{H}_{\mathbf{k}}$ is the Hamiltonian operator of the scalar field propagating on the specified (fixed)  quantum geometry. The second one is a chosen (usually semiclassical) state of the background wave function $\Psi(v, T)$, being a solution to Eq.~\eqref{Schroedinger-geo}.
More precisely, Eq.~\eqref{Eigenvalue-field1} is constructed by a partial tracing over the geometry degrees of freedom, Eq.~\eqref{Schroedinger12} [defined on the full Hilbert space with the product state $\Psi(v, T)\otimes\psi_\mathbf{k}(Q_\mathbf{k}, T)$].
 In this approximation, $\hat{\epsilon}^n_{ \mathbf{k}}$, the energy eigenvalue of the test field $Q_\mathbf{k}$, is still an operator on the gravitational Hilbert space, $\mathscr{H}_{\rm grav}$.

Partial tracing of Eq.~\eqref{Schroedinger12} over the geometry degrees of freedom (d.o.f) yields
\begin{subequations}\label{eigenvalue-total}
\begin{align}
\hat{\tilde{H}}_\mathbf{k}   &=  g^{(1)}\, \hat{P}_{\mathbf{k}}^2 + k^2 g^{(2)}\, \hat{Q}_{\mathbf{k}}^2 \, , \label{Hamiltonian-semi} \\
\epsilon^n_{\mathbf{k}} (g^{(1)},g^{(2)})  &=  \langle\Psi_o(T)|\, \epsilon^n_{\mathbf{k}}\big(\hat{g}^{(1)},\hat{g}^{(2)}\big) |\Psi_o(T) \, \rangle , 
\label{Eigenvalue-semi}
\end{align}
\end{subequations}
where $\hat{g}^{(1)}$ and $\hat{g}^{(2)}$ are composite operators expressed via $\hat{v}$.
Thus, the scalar perturbation behaves as that of a (geometry-dependent) quantum harmonic oscillator. The resulting eigenvalue problem, thus, takes the form
\begin{eqnarray}
	g^{(1)}\, \frac{d^2 |\chi^n\rangle}{dQ_{\mathbf{k}}^2} + k^2 g^{(2)}\, Q_{\mathbf{k}}^2\, |\chi^n\rangle  
	&=& \epsilon^n_{\mathbf{k}} \, |\chi^n\rangle.
	\label{Hamilt-SF}
\end{eqnarray}
The solutions to differential equation (\ref{Hamilt-SF}) are well known, given by
\begin{subequations}
\begin{align}
\epsilon^n_{\mathbf{k}}  &= \sqrt{g^{(1)} g^{(2)}} \left(n+\tfrac{1}{2}\right)\hbar k , \label{epsilon} \\
|\chi^n\rangle  &=  a_n k^{1/4} B^{1/8} \exp \left(-x^2/2\right) H_n(x), \label{field-sol1}
\end{align}
\end{subequations}
where, we have defined
\begin{eqnarray}
a_n &\equiv& \left(\frac{1}{\pi 2^{2n}(n!)^2}\right)^{\frac{1}{4}},  \nonumber \\
B &\equiv& \frac{g^{(2)}}{g^{(1)}\hbar^2} \quad {\rm and} \quad
x \equiv k^{1/2} B^{1/4}Q_{\mathbf{k}}.
\end{eqnarray}

In further treatment, we would like to follow the procedure introduced in Ref.~\cite{Giesel:2009at}. 
There, an essential step was treating the emergent description of the matter field (in our case, the scalar field) as parametrized by a single geometric variable, the volume $V^{\prime} = \langle\Psi_o(T)| \hat{V} |\Psi_o(T)\rangle$.
In subsequent steps of the considered procedure that parameter was promoted back to quantum operator. Here, however, the scalar field description involves two expectation values: $\langle \hat{V}^{-1} \rangle$ and $\langle \hat{V}^{1/3}\rangle$. In order to introduce the parametrization analogous to that in Ref.~\cite{Giesel:2009at}, we note that the functions $g^{(i)}$ can be expanded in terms of the central Hamburger moments corresponding to the volume \cite{Bojowald:2005cw}, namely, 
\begin{equation}
	\langle \hat{V}^{\alpha} \rangle = \langle \hat{V} \rangle^{\alpha} + \sum_{i=1}^\infty    \left({\begin{array}{c}
		\alpha \\
		i  \\
		\end{array} } \right) \langle \hat{V} \rangle^{\alpha-i} G^{i00} \, ,
\end{equation}
where $G^{i00}=\langle (\delta \hat{V})^i \rangle$. Since the background state $\Psi_o$ is \emph{chosen}, both $V^{\prime}$ and $G^{i00}$ are determined as functions of $T$. Had $V^{\prime}(T)$ been invertible (which happens, for example, in geometrodynamics if we restrict ourselves to the post-big-bang epoch), we would be able to define 
\begin{equation}
	G^{i00}(V^{\prime}) = G^{i00}(T(V^{\prime})) .
\end{equation}
In LQC, however, there are two reasons preventing us from doing so:
\begin{enumerate}
	\item The dynamics of the background state features a bounce (see, for example, \cite{Husain:2011tm}), thus, the function $V^{\prime}(T)$ is not globally invertible. One could, in principle, choose a state symmetric with respect to the bounce, that is, such that
		\begin{subequations}\begin{align}
			V^{\prime}(T_B+\delta T) &= V^{\prime}(T_B-\delta T) , \\ G^{i00}(T_B+\delta T) &= G^{i00}(T_B-\delta T) , 
		\end{align}\end{subequations}
		where, $T_B$ is the time of the bounce, however, this choice is a fine-tuning and there is no physical reason distinguishing it.
	\item The expectation value of $\hat{V}$ never drops below $V^{\prime}_B \propto -\langle\hat{H}_{\rm grav}\rangle$; thus $T(V^{\prime})$ is not defined for 
		$V^{\prime}<V^{\prime}_B$. 
\end{enumerate}

As a consequence, in order to introduce the desired parametrization, we have to neglect at this step all the second- and higher-order quantum corrections 
(encoded in $G^{i00}$) of the background state, leaving only the quantum imprint on the trajectory. Then, $g^{(i)}$'s in Eq.~(\ref{eq:g12-def}) reduce to
\begin{equation}
	g^{(1)} = (1/2)\langle\hat{V}\rangle^{-1}_o , \quad g^{(2)} = (1/2) \langle\hat{V}\rangle^{1/3}_o,
\end{equation}
and in consequence
\begin{subequations}\begin{align}
\hat{\epsilon}^n_{\mathbf{k}}  &= \left(b_\mathbf{k}^{\dagger n} b_\mathbf{k}^n+\frac{1}{2}\right)\langle\hat{V}\rangle^{-1/3}_o \, \hbar k \, , \label{epsilon-2} \\
|\chi^n\rangle &=     
a_n \left(\frac{k}{\hbar}\right)^{\frac{1}{4}} \langle\hat{V}\rangle^{1/12}_o 
 \exp \left(-\frac{k\langle\hat{V}\rangle^{1/3}_o}{2\hbar}Q_{\mathbf{k}}^2\right)\, H_n\left(\sqrt{\frac{k}{\hbar}} \langle\hat{V}\rangle^{1/6}_o Q_{\mathbf{k}}\right).  
\label{field-sol2}
\end{align}\end{subequations}
%%%%%%%%%%%%%%%%%%%%%%%%%%%%%%%%%%%%%%%
Having defined $|\chi^n_{\mathbf{k}}\rangle$ as the eigenfunctions of $\hat{\tilde{{H}}}_\mathbf{k}$ with eigenvalues $\epsilon^n_{\mathbf{k}}$, we can now turn back to the geometry eigenfunction components $\xi^\mu_{\mathbf{k}}(g^{(1)},g^{(2)})$ of the perturbed quantum geometry state \eqref{Schroedinger-geo}. After tracing out the scalar field degrees of freedom, the evolution equation for eigenfunctions of geometry is obtained as
\begin{equation}
\big[\hat{H}_{\rm gr} + \epsilon^n_{\mathbf{k}}(\hat{v}) \big] \xi^\mu_{\mathbf{k}}(v) =  E^\mu_{\mathbf{k}} \; \xi^\mu_{\mathbf{k}}(v),
\label{Eigenvalue-grav1}
\end{equation}
where 
\begin{align}\label{eq:beta}
\epsilon^n_{\mathbf{k}}(\hat{v})  &= \left(n+\dfrac{1}{2}\right) \hbar k \hat{V}^{-1/3} \nonumber \\ & =: N_k \hbar k\, \hat{V}^{-1/3},
\end{align}
represents the energy of the mode of test field state, in which $\hat{N}_k$, in the adiabatic regime, reduces to the number operator of a harmonic oscillator \cite{Berger:1975ag}.

The eigenvalue problem \eqref{Eigenvalue-grav1} differs from the one of the background state (studied in Ref.~\cite{Husain:2011tm})
only by the bounded potential 
quickly decaying to zero as $v$ increases. As a consequence, the operator on its left-hand side will share the spectral property of the background one: Its spectrum is nondegenerate and continuous and consists of the entire real line. The eigenvectors $\xi^\mu_\mathbf{k}(v)$ can be found by numerical means via methods used in Refs.~\cite{Ashtekar:2006wn,MenaMarugan:2011me}.
The properties of these eigenvectors are quite similar to those of the background Hamiltonian, 
$\hat{H}_{\rm grav}$.
Each has a form of the reflected wave further featuring the region of 
exponential suppression around $v$ of the size depending on $\mu$ and 
$\mathbf{k}$. In the next subsection we will find eigenfunctions $\xi^\mu_\mathbf{k}(v)$ of Eq.~\eqref{Eigenvalue-grav1} and show that, for large $v$, they feature the following asymptotic behavior:
\begin{align}
\xi^\mu_{\mathbf{k}}(v) &= \frac{C}{v^{1/4}} (1+f(\mu,\mathbf{k}))\cos\left(\mu v^{1/2} + \varphi(\mu,\mathbf{k})\right) 
+ \mathcal{O}(v^{-9/4}) , \label{bases}
\end{align}
where $\varphi(\mu,\mathbf{k})$ is a phase shift and $C$ is a normalization factor. This asymptotic actually provides for us a precise definition of the label $\mu$ in the choice of which we have a freedom due to the continuity of the spectrum of the studied operator.
The form of the asymptotic implies that $\xi^\mu_{\mathbf{k}}$ are Dirac-delta normalizable. We can thus form out of them an orthonormal basis (for each value of ${\mathbf{k}}$ independently), setting
\begin{equation}
(\xi^\mu_{\mathbf{k}}|\xi^{\mu'}_{\mathbf{k}}) = \delta(\mu-\mu').
\end{equation}

%%%%%%%%%%%%%%%%%%%%%%%%%%%%%%%%%%%%%%%%%%%%%%%%%%%%
\subsection{Rate of convergence of bases}\label{rateOFcon}

To solve Eq.~\eqref{Eigenvalue-grav1} numerically, we need to explicitly show that the convergence rate \eqref{bases} exists. This specific rate has applications in numerical calculations of LQC \cite{MenaMarugan:2011me}, results of which will be presented in a separate paper \cite{ModifiedDR:2022}. To study the rate of convergence, we compare  $\xi^\mu_\mathbf{k}(v)$ with eigenfunctions $\ub{e}^{\mu}_\mathbf{k}(v)$ of the Wheeler-DeWitt (WDW) analog of Eq.~\eqref{Schroedinger1} at the asymptotic region. 

The quantum Hamiltonian constraint in WDW theory can be expressed as a differential analog of LQC evolution operator $\Theta$, where an action of the operator $\Theta$ equals
\begin{equation}\label{eq-evolution}
\begin{split}
[\Theta\psi](v) &= f_-(v)\Psi(v-4) - f_o(v)\Psi(v)  + f_+(v)\Psi(v+4) ,
\end{split}
\end{equation}
with
\begin{align}
	f_\pm(v) &= (3\pi G/8\alpha_o) (v\pm 4)^{1/2}  v^{1/2} , \nonumber \\  
	f_o(v) &= (3\pi G/4\alpha_o) v - N_k \hbar k \alpha_o^{-1/3}v^{-1/3} .
\label{f0pm}
\end{align}
To arrive to the WDW equation, we select the factor ordering
consistent with the one of Eq.~\eqref{Ham_grav}, so we get
\begin{align}
i \partial_{T}\ub{\Psi}(v,\phi) &=  \ub{\Theta}\,\ub{\Psi}(v,\phi) \nonumber \\
&:= \frac{6\pi G}{\alpha_o}\, |v|^{1/2} \partial_v \partial_v |v|^{1/2} \, \ub{\Psi}(v,\phi).
\label{wdw}
\end{align}
One can build a basis in $\ub{\mathscr{H}}_{\grav}$ (Hilbert space of WDW theory) out of the eigenfunctions $\ub{e}_{\mu}(v)$ corresponding to non-negative eigenvalues
\begin{equation}\label{WDWeignvalue}
[\ub{\Theta} \; \ub{e}_{\mu}](v) = - \omega^2(\mu) \ub{e}_{\mu}(v),
\end{equation}
where $\omega(\mu) = \sqrt{3\pi G /2\alpha_o}\mu$ with $\mu>0$. A general solution to Eq.~\eqref{WDWeignvalue} is
\begin{equation}
\ub{e}_{\mu}(v) = c_1 J_1\left(2 \sqrt{v} \omega \right)+2 c_2 Y_1\left(2 \sqrt{v} \omega \right),
\end{equation}
where $J_1\left(2 \sqrt{v} \omega \right)$ and $Y_1\left(2 \sqrt{v} \omega \right)$ are Bessel functions of the first and second kind, respectively. These solutions asymptotically tend to the orthonormal basis:
\begin{equation} \label{wdw-eigenf}
\ub{e}^{\pm}_{\mu}(v) \approx \dfrac{|v|^{-1/4}}{\sqrt{4\pi}} \; e^{\pm i \mu |v|^{1/2}}.
\end{equation}

To verify asymptotes of $\xi^\mu_\mathbf{k}(v)$, we start 
 with  rewriting Eq.~\eqref{eq-evolution},
the second-order difference equation, in a first-order
form, introducing the vector notation
\begin{equation}\label{eig-vector}
	\vec{\xi}^\mu_\mathbf{k}(v):= \left( \begin{array}{l} \xi^\mu_\mathbf{k}(v) \\ \xi^\mu_\mathbf{k}(v-4) \end{array} \right).
\end{equation}
Using it, Eq.~\eqref{eq-evolution} turns to
\begin{equation}\label{firstorder}
	\vec{\xi}^\mu_\mathbf{k}(v+4) = \blA(v)\,\vec{\xi}^\mu_\mathbf{k}(v) ,
\end{equation}
where the matrix $\blA$ can be expressed as
\begin{equation}\label{A-matrix}
	\blA(v) = \left(\begin{array}{cc} 
		\frac{f_o(v)-\omega^2(\mu)}{f_+(v)} & -\frac{f_-(v)}{f_+(v)} \\
		1 & 0
	\end{array}\right) .
\end{equation}
To relate $\xi^\mu_\mathbf{k}(v)$ with $\ub{e}^{\pm}_{\mu}$, we note that the value of $\xi^\mu_\mathbf{k}(v)$ at each pair of consecutive points $v$ and $v+4$ can be encoded as a linear combination of the WDW components of $\ub{e}^{\pm}_{\mu}$, that is,
\begin{equation}\label{eig-coeff}
	\vec{\xi}^{\mu}_{\mathbf{k}}(v+4) = \blB^{\mu}_{\mathbf{k}}(v) \vec{\chi}^{\mu}_{\mathbf{k}}(v+4) ,
\end{equation}
where the transformation matrix $\blB^{\mu}_{\mathbf{k}}$ is defined as follows:
\begin{equation}
	\blB^{\mu}_{\mathbf{k}}(v) := \left( \begin{array}{ll} 
		\ub{e}^{+}_{\mu}(v+4) & \ub{e}^{-}_{\mu}(v+4) \\
		\ub{e}^{+}_{\mu}(v) & \ub{e}^{-}_{\mu}(v)
	\end{array} \right) .
\end{equation}
Using the objects defined above, we can rewrite Eq.~\eqref{firstorder} as the iterative equation for the vectors of coefficients $\vec{\chi}^\mu_\mathbf{k}$:
%\begin{equation}
\begin{align}
\label{M-matrix}
\vec{\chi}^{\mu}_{\mathbf{k}}(v+4) &= \blB^{\mu^{-1}}_{\mathbf{k}}(v) \blA(v) \blB^{\mu}_{\mathbf{k}}(v-4) \vec{\chi}^{\mu}_{\mathbf{k}}(v) \nonumber \\
&=: \blM^{\mu}_{\mathbf{k}}(v) \vec{\chi}^{\mu}_{\mathbf{k}}(v).
\end{align}
%\end{equation}
The exact elements of the matrix $\blM^{\mu}_{\mathbf{k}}(v)$ can be calculated explicitly for the coefficients of the evolution operator \eqref{eq-evolution}. By straightforward calculations, one can find that it has the following asymptotic behavior:
\begin{equation}
	\blM^{\mu}_{\mathbf{k}}(v) = \id + \mathcal{O}(v^{-1/3}).
\end{equation}
This result does not grant the rate of convergence needed for Eq.~\eqref{bases}. We can improve the level of convergence by replacing the components $\ub{e}^{\pm}_{\mu}(v)$ in Eq.~\eqref{wdw-eigenf} with functions,
\begin{align}
\ub{e}^{\pm}_{\mu,\mathbf{k}}(v) &= \frac{|v|^{-1/4}}{\sqrt{4 \pi}} \left[ 1 + \sum_{n=1}^{5} a_n\;|v^{-n/3}| \right] \exp\left[\pm i \mu |v|^{1/2} \left(1 + \sum_{n=1}^{7} b_n |v^{-n/3}|	\right)\right], 
\label{def:eigenfunction-mod}
\end{align}
where coefficients $a_n$ and $b_n$ are presented in Appendix \ref{subleadingT}. Direct inspection of the asymptotics of $\blM^{\mu}_{\mathbf{k}}(v)$ shows that
\begin{equation}
\blM^{\mu}_{\mathbf{k}}(v) = \id + \mathcal{O}(v^{-3}) , 
\end{equation}
which now admits the level of convergence needed for Eq.~\eqref{bases}. Modified eigenfunctions \eqref{def:eigenfunction-mod} will be used in a subsequent paper for the normalization procedure in LQC numerical calculations \cite{ModifiedDR:2022}.

%%%%%%%%%%%%%%%%%%%%%%%%%%%%%%%%%%%%%%%%%%%%%%%%%%%%%%
\subsection{Emerging mode-dependent dressed cosmological background}
\label{QFT1b}

At this point, we have at our disposal the matter field basis eigenfunctions $|\chi^{n}_\mathbf{k}\rangle$, 
corresponding to the discrete eigenvalues $\epsilon^{n}_{\mathbf{k}}$, parametrized by $n$ and a family of bases of the gravitational Hilbert space formed of eigenfunctions $|\xi^\mu_{\mathbf{k}})$ which can be determined numerically and normalized using Eq.~\eqref{def:eigenfunction-mod}. To construct the complete wave function $|\Psi_1 \rangle$, we need to determine the spectral profiles $c_{\mathbf{k}}(\mu)$ which may differ from the profile of the original background state. In order to determine $\Psi_1 (v, T)$ and describe the eigenfunctions $|\xi^\mu_{\mathbf{k}})$ in a convenient manner, which is more suitable for our perturbation treatment, we expand $|\xi^\mu_{\mathbf{k}})$ as follows, distinguishing the hierarchy of corrections by 
\begin{eqnarray}
|\xi^\mu_{\mathbf{k}}) \, =:\,  \underbar{N} \Big[|\xi^\mu_o) +|\delta\xi^\mu_{\mathbf{k}})\Big],
\label{eigenstate-pert1}
\end{eqnarray}
where $\underbar{N}$ is the overall normalization 
factor determined from orthonormality of the bases. A first-order solution to Eq.~(\ref{Schroedinger-geo}) can be constructed using profile $c_{\mathbf{k}}(\mu)$ and eigenfunctions $\xi^\mu_{\mathbf{k}}(v)$ as \cite{MenaMarugan:2011me, Ashtekar:2006rx}
\begin{eqnarray}
\Psi_1(v, T) &=&  \int_{\mu \in \mathbb{R}} d\mu\,  c_{\mathbf{k}}(\mu)\,  \xi^\mu_{\mathbf{k}}(v) \, e^{i \omega(k) T}. \qquad 
\label{BO1}
\end{eqnarray}
Here, we are interested in considering only backreaction effects of the field on geometry and ignoring any correlation effects between these two. Within this approximation, the perturbed wave function can be expressed [by substituting Eq.~(\ref{eigenstate-pert1}) in the wave function (\ref{BO1})] as
\begin{align}
\Psi_1(v, T)\, &= \,  \int_{\mu \in \mathbb{R}} d\mu\,   c(\mu)\, \xi^\mu_o(v) \, e^{i \omega(k) T}  +  
\int_{\mu \in \mathbb{R}} d\mu\,   c(\mu)\, \delta\xi^\mu_{\mathbf{k}}(v) \, e^{i \omega(k) T} \nonumber \\
%%%%%%%%%%%%%%%%%%%%%%%%
&=:\, \Psi_o(v, T) + \delta\Psi_{\mathbf{k}}(v, T) ,
\label{BO1-Perturbation}
\end{align}
where, we used the same profile for the perturbed and unperturbed states. The first term on the right-hand side above denotes the unperturbed wave function, while the second term represents the corrections in the geometry quantum state induced by backreaction of each mode of the field on the geometry.
To build first the order total wave function, following BO approximation, we focus on the situation where the geometry and field components of the above backreaction term are uncorrelated (separable), that is,
\begin{equation}
\Psi^1_{\mathbf{k}}(v, Q_{\mathbf{k}}, T)\, =\,  \Psi_1(v, T) \otimes  \psi_\mathbf{k}(Q_\mathbf{k}, T), \quad
\label{BO1-Perturbation-2}
\end{equation}
where, $\psi_\mathbf{k}(Q_\mathbf{k}, T)$ is constructed using eigenfunctions $\chi^{n}_{\mathbf{k}}(Q_{\mathbf{k}}; v)$. This indicates that, the wave function $\Psi_1(v, T)$ of the background, perturbed by field's backreaction, depends on the energy of the mode $\mathbf{k}$. In other words, due to backreaction, each mode  of the field induces different changes in the geometry state and probes a specific background geometry which depends on $k=|\mathbf{k}|$.

In zeroth order, in mean field limit $\langle \hat{H}_{\mathbf{k}}(\hat{v})\rangle = 0$, Eq.~\eqref{Schroedinger-geo} leads us to an eigenvalue equation for the geometry part, which is 
\begin{equation}
\hat{H}_{\rm grav}|\xi^\mu_o) \, = \, E^\mu_o|\xi^\mu_o).
\label{ground-energy}
\end{equation}
Tracing out the geometrical d.o.f. in Eq.~\eqref{Schroedinger-field},  using the state $\Psi_o(v, T)$ which is constructed from eigenfunctions \eqref{ground-energy}, yields the unperturbed Schrodinger-like equation
\begin{eqnarray}
i\hbar\partial_T\psi_\mathbf{k} = \frac{1}{2}\Big[\langle\hat{V}^{-1}\rangle_o\, \hat{P}_{\mathbf{k}}^2 + k^2\, \langle\hat{V}^{\frac{1}{3}}\rangle_o\, \hat{Q}_{\mathbf{k}}^2\Big]\psi_{\mathbf{k}}. \qquad \quad
\label{Schroedinger2}
\end{eqnarray}
Having found the perturbed eigenfunctions $|\xi^\mu_{\mathbf{k}})$ of the geometry eigenvalue equation \eqref{Eigenvalue-grav1} and constructing the perturbed state $\Psi_1(v, T)$, we get the following Schrodinger-like equation for each mode of the scalar field:
\begin{eqnarray}
i\hbar\partial_T\psi_\mathbf{k} = \tfrac{1}{2}\Big[\langle\hat{V}^{-1}\rangle \, \hat{P}_{\mathbf{k}}^2 + k^2\, \langle\hat{V}^{\frac{1}{3}}\rangle\, \hat{Q}_{\mathbf{k}}^2\Big]\psi_{\mathbf{k}},  \qquad\quad 
\label{Schroedinger3}
\end{eqnarray}
in which we have defined the expectation values $\langle \cdot \rangle$ with respect to the total perturbed state $\Psi_1(v, T)$, i.e.,
\begin{subequations}
\begin{equation}
\langle\hat{V}^{-1}\rangle := \langle \Psi_1(v, T) | \hat{V}^{-1}| \Psi_1(v, T) \rangle
\end{equation}
and
\begin{equation}
\langle\hat{V}^{1/3}\rangle := \langle \Psi_1(v, T) | \hat{V}^{1/3}| \Psi_1(v, T) \rangle.
\end{equation}
\end{subequations}
(Henceforth, we will drop the subscript index for the expectation value of any operator with respect to the perturbed background quantum state.)
Now, by substituting the decomposition (\ref{BO1-Perturbation}) into Eq.~(\ref{Schroedinger3}), we obtain the following equation for the effects of the perturbed geometry state on the evolution equation of the field:
\begin{align}
i\hbar\partial_T\psi_\mathbf{k} &= \tfrac{1}{2}\Big[\Big(\langle\hat{V}^{-1}\rangle_o+\langle\hat{V}^{-1}\rangle_{\delta}\Big)\hat{P}_{\mathbf{k}}^2   + k^2\Big(\langle\hat{V}^{1/3}\rangle_o+\langle\hat{V}^{1/3}\rangle_{\delta}\Big)\hat{Q}_{\mathbf{k}}^2\Big]\psi_\mathbf{k} \, , \quad 
\label{Schroedinger32}
\end{align}
in which, we have defined
\begin{align}
\langle\hat{V}^{-1}\rangle_{\delta} &:= \langle\Psi_o|\hat{V}^{-1}|\delta\Psi_{\mathbf{k}}\rangle +\langle \delta\Psi_{\mathbf{k}}|\hat{V}^{-1}|\Psi_o\rangle  + \langle \delta\Psi_{ \mathbf{k}}|\hat{V}^{-1}|\delta\Psi_{\mathbf{k}}\rangle, \\
%%%%%%%%%%%%%%%%%%%%%%%%%%% 
\langle\hat{V}^{1/3}\rangle_{\delta} &:=\langle\Psi_o|\hat{V}^{1/3}|\delta\Psi_{\mathbf{k}}\rangle +\langle \delta\Psi_{ \mathbf{k}}|\hat{V}^{1/3}|\Psi_o\rangle   + \langle \delta\Psi_{ \mathbf{k}}|\hat{V}^{1/3}|\delta\Psi_{ \mathbf{k}}\rangle.
\end{align} 
Here, {\small $\langle\hat{V}^{-1}\rangle_{\delta}$} and {\small $\langle\hat{V}^{1/3}\rangle_{\delta}$} are modifications to the background dressed metric, being probed in a BO approximation due to backreaction effects. 
The state $\delta\Psi_{\mathbf{k}}$ is expanded in terms of  the eigenfunctions $\delta\xi^\mu_{\mathbf{k}}(v)$ on the right-hand side of Eq.~(\ref{BO1}). A  numerical analysis\footnote{A thorough numerical investigation in the LQG context will be presented in an upcoming paper \cite{ModifiedDR:2022}.} for computing this state was performed in the Appendix \ref{subleadingT}. Therein, for a normalization procedure, we have used the perturbative eigenfunctions (\ref{def:eigenfunction-mod}) of the WDW theory. It turns out that the leading-order terms in these eigenfunctions depend on $k$. Therefore, the correction terms  {\small $\langle\hat{V}^{-1}\rangle_{\delta}$} and {\small $\langle\hat{V}^{1/3}\rangle_{\delta}$} depend  also on $k$: hence,  they are mode dependent.

The effective equation (\ref{Schroedinger3}) corresponds to an evolution equation for the  scalar perturbation's state, $\psi_\mathbf{k}$, on a dressed background metric
\begin{eqnarray}
\tilde{g}_{ab}dx^adx^b = -\tilde{N}(T)dT^2+\tilde{a}^2(T)d\mathbf{x}^2.
\end{eqnarray}  
By comparison, we find the following relations between the components of the emerging dressed metric and the expectation values of quantum operators of the original spacetime metric:
\begin{eqnarray}
\tilde{N} \tilde{a}^{-3} &=& \langle\hat{V}^{-1}\rangle_o\left(1+\delta_1\right) , \label{dress-eq1}\\
\tilde{N} \tilde{a} &=& \langle\hat{V}^{1/3}\rangle_o \left(1+\delta_2\right),
\label{dress-eq2}
\end{eqnarray}
where
\begin{eqnarray}
\delta_1(k, T) =\frac{\langle\hat{V}^{-1}\rangle_{\delta}}{\langle\hat{V}^{-1}\rangle_o}, \quad    \quad  
\delta_2(k, T) = \frac{\langle\hat{V}^{1/3}\rangle_{\delta}}{\langle\hat{V}^{1/3}\rangle_o}.
\quad \qquad  
\end{eqnarray}
By solving Eqs.~(\ref{dress-eq1}) and (\ref{dress-eq2}), we obtain
\begin{align}
\tilde{N}(k, T) &=    \bar{N}(T)\, f(k, T) \, ,
\label{dressed-metric-eq1-BO2}\\
\tilde{a}(k, T) &=    \bar{a}(T)\, q(k, T) \, ,  \quad\quad
\label{dressed-metric-eq2-BO2}
\end{align}
where
\begin{eqnarray}
	f(k, T) &=&    \big(1+\delta_1\big)^{1/4}\big(1+\delta_2\big)^{3/4},
	\label{BR-func1} \\
	q(k, T) &=&   \left(\frac{1+\delta_2}{1+\delta_1}\right)^{1/4} ,  \quad\quad
	\label{BR-func2}
\end{eqnarray}
are mode-dependent functions representing the backreaction effects in the emerged dressed metric $\tilde{g}$. In the absence of backreaction, $f(k, T)$ and $q(k, T)$ tend to unity. Moreover, $\bar{N}_T$ and $\bar{a}$ are  components of the dressed metric given in a test field approximation (where no backreaction is taken into account):
\begin{eqnarray}
\bar{N}_T(T) &=&   \left[\big\langle\hat{V}^{-1}\big\rangle_o ~\big\langle\hat{V}^{1/3}\big\rangle_o^3 \right]^{\frac{1}{4}}, \label{dressed-metric-eq2a} \\
\bar{a}(T) &=& \left[\big\langle\hat{V}^{1/3}\big\rangle_o\, \big\langle\hat{V}^{-1}\big\rangle_o^{-1}\right]^{\frac{1}{4}}. \quad
\label{dressed-metric-eq2}
\end{eqnarray}
Equations.~(\ref{dressed-metric-eq1-BO2}) and (\ref{dressed-metric-eq2-BO2}) present the mode-dependent components of the dressed metric $\tilde{g}$ emerged in the herein quantum gravity regime. This implies that a  ``rainbow'' metric emerges in the interior region of the collapse background spacetime, due to backreaction effects.

%%%%%%%%%%%%%%%%%%%%%%%%%%%%%%%%%%%%%%%%%%%%%%%%%%%%%%
\section{Effective dynamics of the   dressed metric}
\label{QFT2}

From the point of view of a semiclassical observer, we intend  to find a corresponding evolution equation for the dressed metric component $\tilde{a}$ with respect to a new time coordinate $\tau$ with $d\tau=\tilde{N}(T, k)dT$. 
In particular, we compute the Friedmann equation corresponding to the dressed scale factor $\tilde{a}$ as $\tilde{H}\equiv\partial_\tau\tilde{a}/\tilde{a}$:
\begin{eqnarray}
\tilde{H}  = \frac{1}{4}\left(\frac{\partial_{\tau}\langle \hat{v}^{1/3}\rangle}{\langle \hat{v}^{1/3}\rangle}  - \frac{\partial_{\tau}\langle \hat{v}^{-1}\rangle}{\langle \hat{v}^{-1}\rangle}\right) , \quad 
\label{Friedmann-dressed1}
\end{eqnarray}
where
$\langle \hat{v}^{1/3}\rangle=\alpha_o^{-1/3}\langle \hat{a}\rangle$ and 
$\langle\hat{v}^{-1}\rangle=\alpha_o\langle \hat{a}^{-3}\rangle$ and expectation values are computed with respect to the backreacted background states. To the first-order quantum corrections, one can show that the backreacted dressed Hubble rate  \eqref{Friedmann-dressed1} reduces to\footnote{The details of the calculations are presented in Appendix \ref{App-A}.}
\begin{eqnarray}
\frac{\partial_\tau\tilde{a}}{\tilde{a}} \, \approx \, \langle\hat{H}\rangle \, = \, \frac{1}{3} \langle \widehat{:\dot{v}/v:} \rangle . \quad 
\label{Friedmann-dressed1b}
\end{eqnarray}
The right-hand side of the equation above is given by \cite{Husain:2011tm}
\begin{eqnarray}
\langle\hat{H}\rangle &=& \frac{i}{3\hbar} \big\langle\hat{v}^{-1/2} \big[\hat{{ H}}_{\rm grav}, \hat{v}\big]\hat{v}^{-1/2}\big\rangle  \, =\,  \frac{\pi G}{\alpha_o} \langle\hat{h}\rangle, \quad \quad 
\label{Hubble-h-new}
\end{eqnarray}
where the operator $\hat{h}$ is defined in Eq.~\eqref{operator-r-h}. 

By using the central moments $C^{abc}$, defined in Eq.~\eqref{moment-gen}, we get
\begin{equation}
\langle \hat{H} \rangle^2 = \langle \hat{H}^2 \rangle - (\pi G/\alpha_o)^2G^{002}, \label{eq:hubbledis}
\end{equation}
where $G^{002}\equiv \langle (\delta\hat{h})^2 \rangle$. Taking the expectation value of the total Hamiltonian constraint \eqref{Hamiltonian-constraint} with respect to the background quantum state yields
\begin{eqnarray}
  \big\langle\widehat{:{V}^{-1}{H}_{\rm grav}:}\big\rangle  + \big\langle\widehat{:{V}^{-1}{H}_{T}:}\big\rangle + \big\langle\widehat{:{V}^{-1}{H}_{\mathbf{k}}:}\big\rangle = 0. \quad \quad
  \label{eq:hubbledis1}
\end{eqnarray} 
Also, an energy density related to backreaction can be defined as
\begin{equation}
\rho_k :=\big\langle\widehat{:{V}^{-1}{H}_{\mathbf{k}}:}\big\rangle = \big\langle\widehat{:{V}^{-1}\epsilon^n_{\mathbf{k}}:}\big\rangle = N_k \hbar k \big\langle\widehat{V^{-\frac{4}{3}}}\big\rangle.
\label{eq:hubbledis2}
\end{equation}
Then, from Eqs.~(\ref{eq:hubbledis})-(\ref{eq:hubbledis2}) we get
\begin{eqnarray}
  \langle\hat{\rho}_T\rangle + N_k \hbar k \big\langle\widehat{V^{-\frac{4}{3}}}\big\rangle  = \frac{3\pi G}{2\alpha_o^2}\,  \langle\hat{r}\rangle =: \langle\hat{\rho}\rangle, \label{def:rho-total}
\end{eqnarray} 
where $\hat{r}$ is defined in Eq.~\eqref{operator-r-h} and $\langle\hat{\rho}\rangle$ is the expectation value of the total energy density operator, which turns out to be the sum of the energy densities of dust field and  backreaction.

From Eq.~\eqref{Hubble-h-new} we have that $\langle\hat{H}^2\rangle=(\pi G/\alpha_o )^2 \langle\hat{h}^2\rangle$. Now, by setting this into Eq.~\eqref{eq:hubbledis}, we get
\begin{eqnarray}
\langle \hat{H} \rangle^2 &=& \frac{8\pi G}{3} \langle \hat{\rho} \rangle \left(1 - \frac{\langle\hat{\rho}\rangle}{\rho_{\rm cr}}  \right) 
- \frac{2\pi G}{3}\rho_{\rm cr} \big[G^{002} + 4 G^{020} \big], 
\label{Friedmann-moments1-new}
\end{eqnarray}
where $ \rho_{\rm cr} \equiv 3\pi G/(2\alpha_o^2) $ \cite{Ashtekar:2006rx} and $G^{abc}$'s are defined in terms of the expectation values of the central moments given in Eq.~\eqref{moment-gen}. Then, by substituting $\langle\hat{\rho}\rangle$ from Eq.~\eqref{def:rho-total} into Eq.~\eqref{Friedmann-moments1-new}, the dressed Hubble rate Eq.~\eqref{Friedmann-dressed1b} to the leading order terms becomes
\begin{align}
\tilde{H}^2 &\approx \frac{8\pi G}{3} \left(\langle\hat{\rho}_T\rangle + \rho_k \right)  \left[1 - \frac{\langle\hat{\rho}_T\rangle + \rho_k}{\rho_{\rm cr}}  \right].  
\label{Friedmann-Dressed-new}
\end{align}
The modified Friedmann equation (\ref{Friedmann-Dressed-new}) for the dressed metric $(\tilde{N}, \tilde{a})$ will be sufficient for our purpose in the rest of this paper.
Interestingly,  the energy density of the backreaction behaves as a {\em radiation} fluid. Clearly,  a quantum bounce still occurs at the collapse final state. However, this bounce may occur much earlier, because, in the presence of a radiationlike backreaction, the total energy density of the collapse grows faster and the critical energy density will be reached earlier than the case where  the backreaction is absent. It should be noticed that, the modified Friedmann equation above is the evolution equation for the perturbed background which is explored by the $\mathbf{k}$th mode of the scalar perturbation.

%%%%%%%%%%%%%%%%%%%%%%%%%%%%%%%%%%%%%%%%%%%%%%%%%%%%%%%%%%%%
\section{Exterior geometry: emergence of a rainbow black hole}
\label{exterior}

To explore the whole spacetime structure of the herein model of gravitational collapse, we need to find a suitable  exterior geometry to be matched with the interior dressed spacetime at the boundary of dust. If the pressures at the boundary
of the cloud vanish (as for a pure dust model), then it
is always possible to match the interior collapsing spacetime with an empty Schwarzschild exterior. However, in
the present model, an effective nonzero pressure emerges due to the radiationlike behavior, induced by backreaction, and
the LQG effects (included  in terms proportional to $1/\rho_{\rm cr}$). In the following, we will therefore match the interior region with a generic nonstatic exterior presented by a {\em generalized Vaidya} geometry  which establishes a radiating generalization of the Schwarzschild geometry \cite{Vaidya:1951zza,Joshi:1987wg}.

\subsection{Matching conditions}
\label{matching}

Let us rewrite the interior  metric as
\begin{eqnarray}
d\tilde{s}_-^2 = -d\tau^2 + \tilde{a}^2(\tau)dr^2 + r^2\tilde{a}^2(\tau)d\Omega^2.
\label{FLRW-interior}
\end{eqnarray}
Now, we  define the matching surface $\Sigma$ as the (interior) boundary shell, $\partial M^-$, of the  collapsing cloud with the radius $r=r_{\rm b}=\rm const$. The unit normal vector to the matching surface, $\Sigma$,
is $n^a_-=\tilde{a}^{-1}(\partial/\partial r)^a$, and the induced metric on $\Sigma$ is given by 
\begin{eqnarray}
h_{ab}^-dx^adx^b &=& -d\tau^2 + r^2\tilde{a}^2d\Omega^2. \nonumber 
\end{eqnarray}
The extrinsic curvature $K_{ab}=\frac{1}{2}{\cal L}_nh_{ab}$ for $\Sigma$ at the interior boundary is obtained by
\begin{eqnarray}
K_{ab}^- &=& \tfrac{1}{2}\left(n_-^c\partial_ch_{ab}^- + h_{cb}^-\partial_an_-^c 
+ h_{ac}^-\partial_bn_-^c\right).
\nonumber
\end{eqnarray}

Now, let  the exterior geometry have a Vaidya form\footnote{Here the mass $\tilde{M}$, unlike the mass given by the Schwarzschild metric, is not  necessarily a constant. Then, our choice of the metric may constitute the simplest non-static generalization of the non-radiative Schwarzschild solution to an effective Einstein's field equation. Therefore, we consider a generalized Vaidya metric in which  the mass parameter is extended 
from a constant to a function of the corresponding null coordinates $u, X$, as $\tilde{M}(u, X)$ \cite{Vaidya:1999zz,Vaidya:1999zza,Vaidya:1953zza}.}, which
in Eddington-Finkelstein coordinates its line element reads
\begin{equation}
ds^2_+ = -F(u, X)du^2 + 2dudX + X^2d\Omega^2, \quad \label{def:Vaidya}
\end{equation}
where $F(u, X)$ is the boundary function given by
\begin{equation}
F(u, X) = 1-\frac{2G\tilde{M}(u, X)}{X}\, 
\label{boundaryf0}
\end{equation}
and $\tilde{M}(u, X)$ is the generalized Vaidya mass. We take only a region with $X>X(u)$ as an exterior region of the collapsing cloud, to be matched to the interior dressed FLRW geometry. Once the relations for $u(\tau)$ and $X(\tau)$ at $r_{\rm b}$ are derived, the form of matching surface in  the exterior region, $\mathcal{M}^+$, is determined. So, we consider a general matching surface for $\mathcal{M}^+$ being parametrized by $(u(\tau), X(\tau))$ in terms of $\tau$, which has to be identified with the interior proper time later. The metric on this surface reads
\begin{equation}
h_{ab}^+dx^adx^b = -\Big[F(\partial_\tau u)^2 - 2(\partial_\tau u)(\partial_\tau X)\Big]d\tau^2  + X^2d\Omega^2.   
\end{equation}
The unit normal vector components on this surface are given by
\begin{eqnarray}
n_+^u &=& \frac{1}{\sqrt{1-2G\tilde{M}/X -2(\partial_\tau X)/(\partial_\tau u)}}\ ,
\nonumber \\
n_+^X &=& \frac{1-2G\tilde{M}/X -(\partial_\tau X)/(\partial_\tau u)}{\sqrt{1-2G\tilde{M}/X -2(\partial_\tau X)/(\partial_\tau u)}}\ . 
\end{eqnarray}
Similarly to the interior surface, using these vectors, we can find the explicit components of the extrinsic curvature for the exterior surface.

Following the relations above, we are prepared now to
formulate the junction conditions at $\Sigma$:
\begin{enumerate}[label=\roman*)]
	\item From the condition $h_{\theta\theta}^-=h_{\theta\theta}^+$, we obtain 
	\begin{eqnarray}
	X(\tau)|_{\Sigma} = r_{\rm b}\tilde{a}(\tau) \equiv \tilde{R}(\tau). \label{Match-1} 
	\end{eqnarray}
	That is, for a given interior scale factor, $\tilde{a}(\tau)$, we can determine $X$ on the exterior matching surface for the given   shell $r_{\rm b}$.
	\item From the condition $h_{\tau\tau}^-=h_{\tau\tau}^+$, 
	 a differential equation is found for $X(u)$ as
	\begin{eqnarray}
	(dX/du)^2=(\partial_\tau X)^2\big(1-2G\tilde{M}/X-2dX/du\big), \label{Match-2} 
	\end{eqnarray}
	which determines the relations between the exterior coordinates (for the given   shell $r_{\rm b}$).
	\item The condition $K_{\tau\tau}^-=K_{\tau\tau}^+$ leads to a differential equation for $X(u)$, as
	\begin{equation}
	\quad \quad K_{uu}^+ + 2K_{uX}^+(dX/du)+ K_{XX}^+(dX/du)^2=0 ,
	\label{Match-3} 
	\end{equation}
	which turns out to be automatically satisfied when given the other junction conditions.
	\item From matching $K_{\theta\theta}^-=K_{\theta\theta}^+$, by setting $X=r_{\rm b}\tilde{a}$, an equation is obtained as
	\begin{equation}
	\quad \quad \quad \Big(1-\frac{2G\tilde{M}}{X} - \frac{dX}{du}\Big)^2= 1-\frac{2G\tilde{M}}{X} - 2\frac{dX}{du},
	\label{Match-4}  
	\end{equation}
	by expanding  of which, and using  Eq.~(\ref{Match-2}), we obtain
	\begin{eqnarray}
	\frac{2G\tilde{M}}{X}\Big|_{\Sigma}  = (\partial_\tau X)^2\big|_\Sigma = r_{\rm b}^2(\partial_\tau \tilde{a})^2.
	\label{BH-mass0}
	\end{eqnarray} 
	Thus, we obtain a relation between the generalized Vaidya mass function $\tilde{M}(u, X)$ and the $\tau$-time derivative of the interior dressed scale factor $\tilde{a}$ at $\Sigma$. 
\end{enumerate}
The right-hand side of Eq.~(\ref{BH-mass0}) is already determined by the modified (dressed) Hubble parameter in Eq.~(\ref{Friedmann-Dressed-new}). However, it would be more convenient to rewrite the exterior mass function, $\tilde{M}$, in terms of a quantum mass $\hat{M}$ induced by the interior quantum-gravity-inspired operators. We will evaluate such a quantum mass in the what follows.

From Eqs.~(\ref{Match-1}) and (\ref{dressed-metric-eq2-BO2}), the physical radius of the collapse reads
	\begin{equation}
	\tilde{R}\, =\, r_{\rm b} \tilde{a}\, \approx\, r_{\rm b}\langle \hat{V} \rangle^{1/3}.
	\label{PhysicalRadius}
	\end{equation} 
	Note that, the time dependence in $\tilde{a}$ is encoded in the expectation value $\langle \hat{V} \rangle^{1/3}$ with respect to the perturbed state $\Psi_1$.
Using this, we can write the physical volume of the spherical cloud as
\begin{equation}
\tilde{V}=(4\pi/3)\tilde{R}^3 = (4\pi/3)r_{\rm b}^3\, \langle \hat{V} \rangle. 
\end{equation}	 
We note that, in classical theory, we have $\tilde{V}=V=\ell^3a^3$, where $q(k, T)=1$ and $\tilde{a}=a$, so that  $(4\pi/3)r_\text{b}^3=\ell^3$. Recall that we  set $\ell=1$ throughout previous sections, which yields  $r_\text{b}=(4\pi/3)^{-1/3}$.

Let $M$ be the classical mass of the collapsing cloud given by $M=\rho_TV$. A quantum  mass $\hat{M}$ can then be introduced for the collapsing cloud, generated now by the background quantum (dust) matter source plus quantum corrections induced by backreaction. 
Since both the matter and geometry are quantized in quantum gravity, the quantized mass is given by {\small $\hat{M}=\widehat{\rho V}=-\hat{{H}}_{\rm grav}$}. Now, from  Eq.~\eqref{v-hat-alpha-B}, we  write the expectation value of the quantum mass as
\begin{equation}
\langle  \hat{M} \rangle = \sum_{n=0}^\infty   \beta_n \langle \hat{v} \rangle^{-n}  \Big[\langle\hat{\rho}\rangle  
\langle \hat{{ V}} \rangle \, G^{n00}  + \rho_{\rm cr} \langle \hat{V} \rangle  G^{n10} \Big].   
\label{v-hat-alpha-B0}
\end{equation}
It turns out that, up to {\em zeroth order}  (i.e., $n=0$), the expectation value of the total quantum mass, $\langle  \hat{M} \rangle_0$, is obtained from expectation value  of the quantum mass associated to the quantized dust cloud, and that of the quantum backreaction of each scalar mode, $\mathbf{k}$. It reads
\begin{eqnarray}
\langle  \hat{M} \rangle_0\, :=\,  \langle\hat{\rho}\rangle  
  \langle \hat{V} \rangle  = \langle  \hat{M}_T \rangle +  M_k\, ,
  \label{M0}
\end{eqnarray}
where {\small $\langle  \hat{M}_T \rangle = \langle\hat{\rho}_T \rangle  
\langle \hat{{ V}} \rangle$} is the mass of the dust field and the second term on the right-hand side is the mass generated by backreaction of the $\mathbf{k}$th mode, 
\begin{equation}
	M_k = \rho_k\tilde{V} = \frac{r_{\rm b} N_k \hbar k}{\tilde{R}}.\label{BR-mass}
\end{equation}

In leading-order terms, by ignoring the moments $G^{abc}$, the time evolution of the backreacted, dressed scale factor $\tilde{a}$ 
reads {\small $(\partial_\tau\tilde{a})^2 = \tilde{a}^2 \tilde{H}^2\approx \langle \hat{V} \rangle^{2/3}\tilde{H}^2$}, where $\tilde{H}$ is given by Eq.~(\ref{Friedmann-Dressed-new}). Having that {\small $r_{\rm b}^2(\partial_\tau \tilde{a})^2=(\partial_\tau X)^2|_\Sigma$} [cf. Eq.~(\ref{BH-mass0})], we get
\begin{align}
(\partial_\tau X)^2|_\Sigma\,  \approx\, \frac{2G}{\tilde{R}} \left(\langle\hat{M}_T\rangle +  M_k \right)  \left[1 - \frac{3}{4\pi \rho_{\rm cr}} \frac{\langle\hat{M}_T\rangle + M_k}{\tilde{R}^3} \right].  
\end{align}
For convenience, in what follows we will set {\small $\tilde{R} \approx r_{\rm b}\langle \hat{V} \rangle^{1/3} \equiv q(k, T)\bar{R}$} to distinguish the area radius $\tilde{R}$ in the presence of the backreaction from {\small $\bar{R}\equiv r_{\rm b}\langle \hat{V} \rangle_o^{1/3}$}, in which the backreaction is absent (i.e., $q(k, T)=1$). So, "tilde" and "bar" refer to high-energy and low-energy observers, respectively. 
Now, from Eq.~(\ref{BH-mass0}) we obtain an effective (dressed) mass $\tilde{M}_\text{b}$:
\begin{align}
\tilde{M}_\text{b}  &= \langle\hat{M}\rangle_0 \left[1 - \frac{3}{4 \pi \rho_{\rm cr}} \frac{\langle\hat{M}\rangle_0}{\tilde{R}^3}   \right] \nonumber\\
&= \left(\langle\hat{M}_T\rangle +  \frac{r_{\rm b}N_k \hbar k}{\tilde{R}} \right)  \left[1 - \frac{3}{4\pi \rho_{\rm cr}} \left(\frac{\langle\hat{M}_T\rangle}{\tilde{R}^3} +  \frac{r_{\rm b}N_k \hbar k }{\tilde{R}^4} \right)  \right].
\label{Mass-dressed1}
\end{align}
From Eq.~(\ref{Mass-dressed1}) it is clear that, unlike the relativistic collapse of a dust matter,  the effective mass $\tilde{M}_\text{b}$ here is not a constant and depends on the (dressed backreacted) physical radius $\tilde{R}$ (at the boundary $\Sigma$) and the mode $k$ of the scalar perturbation.  
This is a consequence of the fact that the energy density growth of the (semiclassical interior) matter cloud is accompanied by a negative pressure which does not vanish on the boundary surface.

It should be noticed that $\tilde{M}_\text{b}$ given in Eq.~(\ref{Mass-dressed1})  {\em is not} the total mass in the Vaidya region (i.e., $X\geq \tilde{R}$) but the mass on the boundary surface $\Sigma$. That is why we have denoted it by a subscript "$\text{b}$". Therefore, to find the total Vaidya mass $\tilde{M}(u, X)$, one needs to determine the exact form of the energy-momentum tensor, satisfying the (effective) Einstein field equations and the energy conditions in the exterior region.
Indeed, the reason for calling this exterior geometry a generalized Vaidya is that, in a region sufficiently close to the boundary, we expect that quantum gravitational effects could be effectively described by an effective energy-momentum tensor in GR.

%%%%%%%%%%%%%%%%%%%%%%%%%%%%%%%%%%%%%%%%%%%%%%%
\subsection{Rainbow horizons at the boundary surface}

Let us investigate a qualitative behavior of horizons in the vicinity of the matter shells due to the quantum gravity effects.
The exterior region can be a generalized Vaidya spacetime, supported by finite density and pressures, which vanishes rapidly at large distances. Such a spacetime can be consistently matched with a quantum-gravity-corrected interior region using the discussed boundary conditions. These conditions are effectively satisfied in our model for the dust collapse in the quantum regime. The question of effective dynamics of the exterior spacetime will be answered in the subsequent subsection. Here we address the formation of dynamical horizon, when we know only the dynamics of matching surface $\Sigma$.

\begin{figure}
	\includegraphics[width=4in]{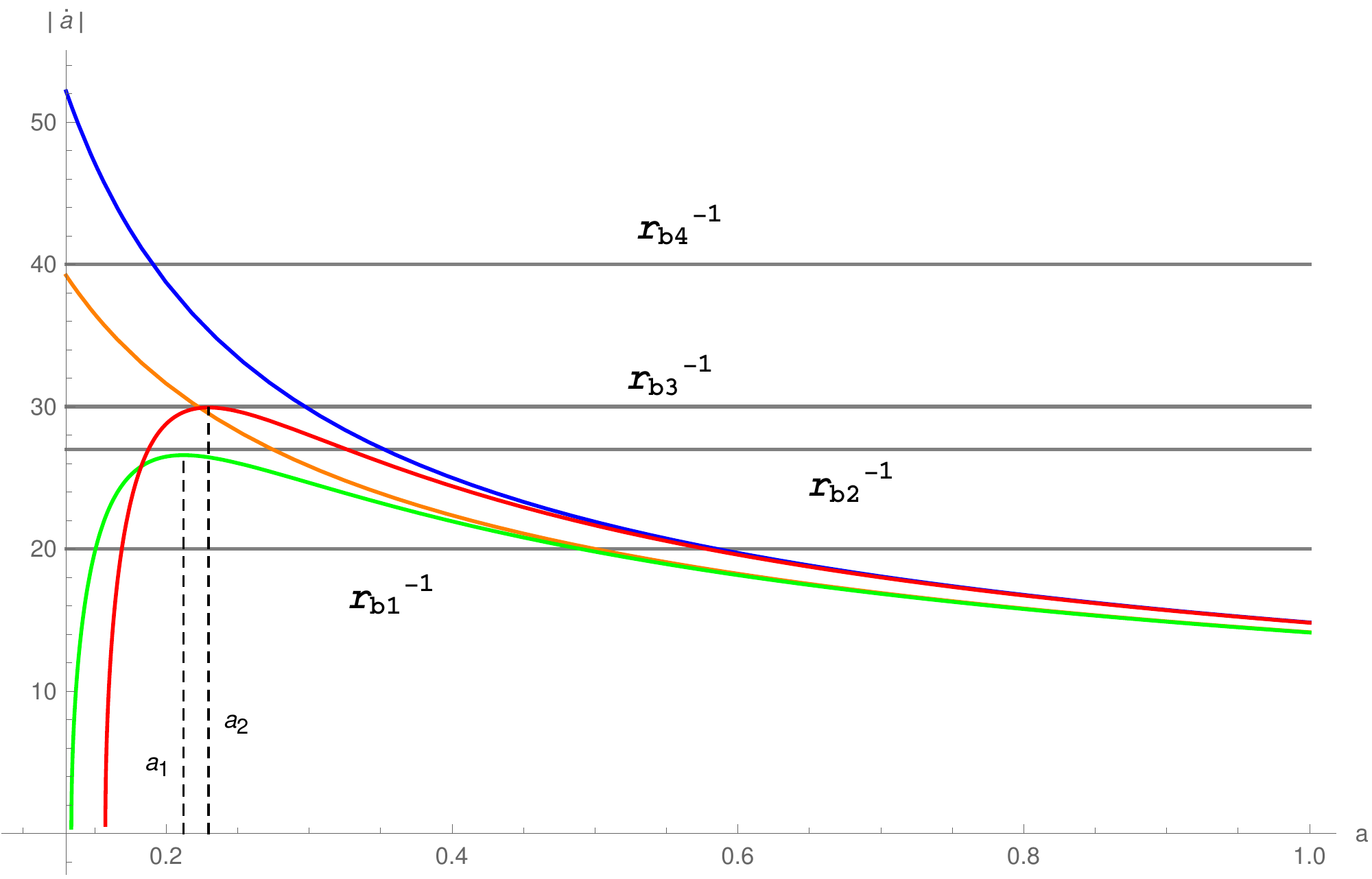}
	\caption{Qualitative behavior of Eq.~\eqref{Friedmann-Dressed-new} is plotted. Horizontal lines correspond to different values of $r_{\rm b}$.
	The colored curves  represent different effective collapse scenarios; a situation where neither backreaction nor loop correction are applied (orange curve); the case where only loop correction is present (green curve); the case where only backreaction effect is considered (blue curve); and the case 
	where both backreaction and loop corrections (i.e., both quantum gravity effects) are applied (red curve).}
	\label{Fig:DH}
\end{figure}

When the condition $F(u, \tilde{R}) = 0$ holds in Eq.~\eqref{boundaryf0}, which is equivalent to $2G \tilde{M}_{\rm b} = X$ for the known mass function (\ref{Mass-dressed1}), a dynamical horizon forms and intersects the boundary surface $\Sigma$ \cite{Hayward:1993mw,Ashtekar:2002ag}. 
From the point of view of the interior spacetime parameters, the
right-hand side of Eq.~\eqref{BH-mass0} is already determined by the
modified (dressed) Hubble parameter in Eq.~\eqref{Friedmann-Dressed-new}. Thus,
when $\partial_{\tau}\tilde{a}$ reaches the value $|\partial_{\tau}\tilde{a}| = r_{\rm b}^{-1}$, a dynamical
horizon forms. In classical GR, $\partial_{\tau}\tilde{a}$ is unbounded
and diverges at the singularity where  $a\to 0$ (blue and orange curve in Fig.~\ref{Fig:DH}). However, in the presence of quantum gravity effects, $\partial_{\tau}\tilde{a}$ changes from a finite initial condition and vanishes at some point where a quantum bounce occurs (red and green curves in Fig.~\ref{Fig:DH}).
In this case, there is a turning point in $|\dot{a}|$. Based on the initial values for $r_{\rm b}$ (i.e. $r_{\rm b1}$, $r_{\rm b2}$, etc.), a horizon may or may not form. In the presence of backreaction (red curve), the  threshold $a_{\rm th}$  for the horizon formation,  given from $|\dot{a}|_{\rm max}(a_{\rm th})=r_{\rm b}^{-1}$, is changed and would happen earlier than the case where the backreaction is ignored (green curve). This is depicted in Fig.~\ref{Fig:DH}, $a_2 > a_1$, where $a_2$ is the threshold in the presence of both backreaction and loop corrections (i.e., terms proportional to $1/\rho_{\rm cr}$), whereas $a_1$ is the threshold for the case of a pure loop correction.  

\subsection{Exterior rainbow geometry}

So far we have found the boundary mass $\tilde{M}_\text{b}(\tilde{R})$ in the Vaidya region due to matching with the interior spacetime [cf. Eq.~(\ref{Mass-dressed1})]. 
However, in order to specify the
exterior geometry, the total mass $\tilde{M}(u, X)$  should be exactly determined in the exterior Vaidya region (i.e., $X\geq \tilde{R}$). This requires the knowledge
about the modified Einstein field equations in the exterior region.

Let us assume that the (effective) energy-momentum tensor in the exterior Vaidya region can be written as \cite{Ziaie:2019klz}
\begin{eqnarray}
T_{\mu\nu} = \sigma N_\mu N_\nu
 + (\varrho+p)(N_\mu L_\nu+N_\nu L_\mu)+pg_{\mu\nu},
\label{typeII}
\end{eqnarray}
with the help of two null vectors,
\begin{eqnarray}
N_\mu=\delta_\mu^{\,0} \quad \text{and} \quad L_\nu=\tfrac{1}{2}F(u, X)\delta_\nu^{\,0}+\delta_\nu^{\,1}, \quad \quad 
\label{ellk}
\end{eqnarray}
such that, $N_\lambda N^\lambda= L_\lambda L^\lambda = 0$ and $N_\lambda L^\lambda=-1$.
Here, $\varrho$ and $p$ are, respectively, the energy density and  pressure associated to $T_{\mu \nu}$, the effective energy-momentum tensor in the exterior region that should satisfy the (effective) Einstein field equations.
Using these parameters, the nonvanishing components of Einstein field equations are
\begin{eqnarray}
\sigma(X,u) &=& \frac{2}{\kappa X^2}\frac{d\tilde{M}}{du},\label{feqs1}\\
\varrho(X,u) &=& \frac{2}{\kappa X^2}\frac{d\tilde{M}}{dX},\label{feqs2}\\
p(X,u) &=& -\frac{1}{\kappa X}\frac{d^2\tilde{M}}{d^2X}. \quad \quad \label{feqs3}
\end{eqnarray}
If we project $T_{\mu \nu}$ to the following orthonormal basis:
\begin{eqnarray}\label{projecttetra}
\textbf{e}_{(a)}^{~~\mu}=\begin{pmatrix} 
\frac{1}{\sqrt{2}}\left(\frac{3}{2}-\frac{\tilde{M}}{X}\right) & -\frac{1}{\sqrt{2}} & 0 & 0\\
\frac{1}{\sqrt{2}}\left(\frac{3}{2}+\frac{\tilde{M}}{X}\right) & \frac{1}{\sqrt{2}} & 0 &0\\
0 & 0& \frac{1}{X} &0\\ 
0 & 0& 0 &\frac{1}{X\sin\theta}
\end{pmatrix}, \quad \quad 
\end{eqnarray}
where $T_{(a)(b)} = T_{\mu \nu} \; \textbf{e}^{~~\mu}_{(a)} \; \textbf{e}^{~~\nu}_{(b)}$, we find,
\begin{eqnarray}
T^{(a)(b)}=\begin{pmatrix} 
\frac{\sigma}{2}+\varrho & \frac{\sigma}{2} & 0 & 0\\
\frac{\sigma}{2} & \frac{\sigma}{2}-\varrho & 0 & 0\\
0 & 0& p &0\\ 
0 & 0& 0 & p
\end{pmatrix}. \quad \quad
\label{projectedemt}
\end{eqnarray}

Assuming an equation of state (EOS) $p=w\varrho$ with $w=\rm const.$, for the fluid and replacing it into Eqs.~\eqref{feqs2} and \eqref{feqs3}, we find the following solution for the mass function:
\begin{eqnarray}
\tilde{M} \left(u, X \right) = 
\alpha F_1(u)X^\beta+F_2(u),
\label{solmassf}
\end{eqnarray}
where $\beta \neq 1/2 $ and $F_1(u),\, F_2(u)$ are two arbitrary functions. Constants $\beta$ and $\alpha$ are related to each other by
\begin{eqnarray}
\beta(w) =1-2w =: \frac{1}{\alpha(w)}\, . \quad \
\label{beta}
\end{eqnarray}
Using the mass function \eqref{solmassf} along with Eqs.~\eqref{feqs1}-\eqref{feqs3}, the effective profiles for the  energy-momentum tensor take the form
\begin{eqnarray}
\sigma(u,X) &=& \frac{2}{\kappa X^2}\left(\frac{{dF}_2(u)}{du}+\alpha X^\beta \frac{d{F}_1(u)}{du}\right),\label{effprofs0}\\
\varrho(u,X) &=& \frac{2}{\kappa}\, F_1(u)X^{\beta-3},  \quad \label{effprofs1}\\
p(u,X) &=& \frac{1-\beta}{\kappa}\, F_1(u)X^{\beta-3}. \quad \quad \quad \quad \label{effprofs}
\end{eqnarray}
Arbitrary functions $F_1(u)$ and $F_2(u)$ can be obtained through matching conditions and should be chosen such that  the energy conditions  are satisfied. Once they are found,  a physical energy-momentum tensor is established for the Vaidya region. 

In the absence of quantum effects (i.e., $\rho_{\rm crit}\to 0$) and backreaction (i.e., $k=0$), the exterior solution should reduce to the standard Schwarzschild geometry. This corresponds to a vacuum region with a vanishing energy-momentum tensor, requiring that $\sigma=\varrho=p=0$, thus,
\begin{subequations}
	\begin{align}
		&\alpha X^\beta\frac{d{F}_1(u)}{du} + \frac{d{F}_2(u)}{du}=0,\\
		& F_1(u)X^{\beta-3} =0,\\
		&F_1(u)(1-\beta)X^{\beta-3}=0.  
	\end{align}
\end{subequations}
The above equations have solution if $F_1(u)$ and $d{F}_2(u)/du$  vanish in the relativistic limit.

On the other hand, as mentioned above, the energy conditions  will establish additional constraints on $F_1$ and $F_2$ \cite{Hawking:1973uf,Ziaie:2019klz}. In particular, the {\em weak} energy condition ($\sigma\geq 0,\, \varrho \geq 0,\, \varrho + p\geq 0$)  requires that
\begin{equation}
F_1(u) \geq 0 \quad \text{and} \quad \beta \leq 2.
\label{wecsec22} 
\end{equation}
The weak and {\em strong} energy conditions together ($\sigma\geq 0,\; \varrho \geq 0,\; p \geq 0$) lead to
\begin{equation}
F_1(u) \geq 0 \quad \text{and} \quad \beta \leq 1.
\label{wecsec23} 
\end{equation}
Finally, the {\em dominant} energy condition ($\sigma \geq 0,\; \varrho \geq 0,\; \varrho \geq |p|$) yields
\begin{equation}
F_1(u) \geq 0 \quad \text{and} \quad  0 \leq \beta \leq 2.
\label{wecsec21} 
\end{equation}
In summary, the  energy conditions acquire the physical ranges  $F_1(u) \geq 0$ and $0 \leq \beta \leq 1$ for the arbitrary parameters $F_1$ and $\beta$ of the model.

On the boundary surface $\Sigma$, where $X = \tilde{R}=r_\text{b}\tilde{a}$, we have
\begin{align}
\tilde{M}(u, X)|_\Sigma &= \tilde{M}_\text{b}(\tilde{R}).
\end{align}
This yields 
\begin{align} 
\alpha F_1(u)\tilde{R}^\beta+F_2(u) &= \langle\hat{M}\rangle_0   \left(1 - \frac{3}{4\pi \rho_{\rm cr}} \frac{\langle\hat{M}\rangle_0}{\tilde{R}^3}\right),
\label{boundary-mass}
\end{align}
where, $\langle\hat{M}\rangle_0$ is given by Eq.~(\ref{M0}):
\begin{equation}
\langle\hat{M}\rangle_0  = \langle\hat{M}_T\rangle +  \frac{r_{\rm b}N_k \hbar k }{\tilde{R}}\, .
\label{M0-1}
\end{equation}
Furthermore, in the absence of backreaction (when $k=0$) and quantum gravity effects (where $\rho_{\rm cr}\to\infty$), the standard Schwarzschild solution should be retrieved. This, together with Eqs.~(\ref{boundary-mass}) and (\ref{M0-1}) gives the the unknown functions as
\begin{align} 
F_1(u) &=  \frac{\beta}{\tilde{R}^\beta}\left(\tilde{M}_{\rm b} -\langle\hat{M}_T\rangle \right)
\nonumber \\
&= \frac{\beta\langle\hat{M}\rangle_0}{\tilde{R}^\beta}   \left(1 - \frac{3}{4\pi \rho_{\rm cr}} \frac{\langle\hat{M}\rangle_0}{\tilde{R}^3}\right) - \frac{\beta\langle\hat{M}_T\rangle}{\tilde{R}^\beta}\, ,  \\ 
F_2(u) &= \langle \hat{M}_T\rangle.
\end{align}
Once the free functions $F_1$ and $F_2$ are found, we can extract the fluid profiles $\sigma, \varrho, p$ and the mass function $\tilde{M}(u, X)$ in the Vaidya region.

For the $\mathbf{k}$th mode of the interior scalar field,  the mass  $\langle\hat{M}\rangle_0$ [cf. Eq.~(\ref{M0-1})]  clearly depends  on $k$. Likewise, the boundary radius $\tilde{R}$ depends implicitly on $k$. Therefore, $F_1$, which is a function of $\langle\hat{M}\rangle_0$ and $\tilde{R}$, will also depend on the mode $k$. 
In other words, for a fixed mode $k$ on the interior region of the collapsing ball, $F_1$ is unique and depends on the value of $k$, so different $k$'s acquire  distinct functions $F_1$.
Thereby, the fluid profiles (\ref{effprofs0})-(\ref{effprofs}), defined as functions of $F_1$ and $F_2$, are uniquely obtained for a fixed value of $k$, as
\begin{eqnarray}
	\sigma_k &=& \frac{2}{\kappa X^2}\frac{d\langle \hat{M}_T\rangle}{du}+\frac{2\alpha\beta}{\kappa X^{2-\beta}}  \frac{d}{du}\left(\frac{\tilde{M}_{\rm b}(k) - \langle\hat{M}_T\rangle}{\tilde{R}^\beta} \right),\label{effprofs0-b}\\
	%%----------------------------------
	\varrho_k &=& \frac{2\beta}{\kappa\tilde{R}^\beta}\, \, \frac{\tilde{M}_{\rm b}(k)-\langle\hat{M}_T\rangle}{X^{3-\beta}},  \quad \label{effprofs1-b}\\
	%%---------------------------------------
	p_k &=&  \frac{\beta(1-\beta)}{\kappa\tilde{R}^\beta}\, \frac{\tilde{M}_{\rm b}(k)-\langle\hat{M}_T\rangle}{X^{3-\beta}}. \quad \quad \quad \quad \label{effprofs0-b}
\end{eqnarray}
These induced profiles establish a unique energy-momentum tensor (\ref{typeII}) in the exterior Vaidya region as
\begin{eqnarray}
	T_{\mu\nu}^{(k)} = \sigma_k N_\mu N_\nu
 + (\varrho_k+p_k)(N_\mu L_\nu+N_\nu L_\mu)+p_kg_{\mu\nu}^{(k)}.
	\label{typeII-mode}
\end{eqnarray}
This implies  that, for matching between the inner FLRW region and the outer Vaidya region to be unique at the boundary, associated with each interior mode $k$ there should exist a unique energy-momentum tensor {\small $T_{\mu\nu}^{(k)}$} in the exterior region. Interestingly, for the specified $k$, the exterior metric {\small $g_{\mu\nu}^{(k)}$},  which appeared in the last term of (\ref{typeII-mode}), is a unique solution to the Einstein field equation outside the dust ball. We will find this unique exterior metric in what follows.

Using the derived  $F_1(u)$ and $F_2(u)$ in Eq.~(\ref{solmassf}),  the mass function $\tilde{M}(u, X)$ in the Vaidya region is achieved.\footnote{There can be other choices for the fluid, such as one with EoS $p=w \varrho^\gamma$, which carry the interior quantum gravity corrections to the exterior Vaidya region.  In that case mass function $\tilde{M} \left(u, X\right)$ will be a polynomial of $X$. In the present model, we are interested in considering gravitational collapse of a {\em radiating fluid} for the exterior spacetime, so the choice $\gamma = 1$ suffices for our purpose.}
It is clear that the loop correction, encoded in terms proportional to $1/\rho_{\rm cr}$, appears in the first term of   $F_1(u)$, whereas the backreaction effect is included in $\langle\hat{M}\rangle_0$ and $\tilde{R}$.
Having found $\tilde{M}(u, X)$,  the exterior function (\ref{boundaryf0}) takes the form
\begin{align}
F_k(u, X) =  1-\frac{2G}{X}\langle \hat{M}_T\rangle -\frac{2G}{X^{1-\beta}}  \tilde{R}^{-\beta} \left[\langle\hat{M}\rangle_0\left(1 - \frac{3}{4\pi \rho_{\rm cr}} \frac{\langle\hat{M}\rangle_0}{\tilde{R}^3}\right) - \langle\hat{M}_T\rangle\right].
\label{vaidyametr2}
\end{align}
As expected, the boundary function $F_k(u, X)$, induced by the fluid profiles in the Vaidya region, is uniquely defined for a given value of $k$ and the fluid EOS, $\beta$. Note that, in order to show this $k$ dependency clearly,  we have set the subscript $k$ in $F_k(u, X)$.

By changing the time coordinate through $du=dt+dX/F_k(u,X)$ in Eq.~(\ref{def:Vaidya}), the exterior generalized Vaidya metric becomes
\begin{eqnarray}
ds^2_+ = -F_k(t,X)dt^2 + F_k^{-1}(t,X)dX^2 + X^2d\Omega^2. \quad
\label{exterior-BH}
\end{eqnarray}
Here, the  physical radius $X$ belongs to the interval $r_\text{b}\left(\alpha_o v_{\rm m}(k)\right)^{1/3}\leq X < +\infty$, where $v_{\rm m}(k)$ is the minimum volume of the collapse at the quantum bounce. 
A  physical interpretation of the emergent  exterior geometry (\ref{exterior-BH}) is as follows.  
The interior quantum gravity effects (provided by the loop correction and the backreaction of field modes) induce a fluid with the energy-momentum  $T_{\mu\nu}$, in the exterior region of the collapsing ball. 
In fact, each mode $\mathbf{k}$ of the field in the inner region induces an energy-momentum {\small $T_{\mu\nu}^{(k)}$} associated with the external fluid, so that 
there exists a one-to-one correspondence between the discrete interior energy-momentum tensor (of each mode $\mathbf{k}$) and the  discretized induced exterior one, {\small $T_{\mu\nu}^{(k)}$}.
The full $T_{\mu\nu}$ of the exterior fluid would consist of the full set of the  interior modes $\mathbf{k}$ on the same lattice $\mathcal{L}$.
The $\mathbf{k}$th mode of the exterior fluid constructs the $k$-dependent spacetime geometry (\ref{exterior-BH}) in the exterior region so that,
different modes of the external fluid experience different geometries; a  {\em rainbow} exterior spacetime background emerges.

By setting $F_k(t, X)=0$ in Eq.~(\ref{exterior-BH}), one finds a set of solutions for  horizons. It turns out that, if  such solutions exist for a given value of $k$ associated with the exterior fluid profile, with a specified EOS parameter $\beta$, a particular set of horizons emerges. Then,  different modes of the fluid would experience different horizons, so that a refraction of black hole horizons can occur in the exterior region. Therefore, a {\em rainbow black hole} can emerge in the framework of the distant observer.

From Eq.~(\ref{vaidyametr2}) it is clear that, even at large distances far from the Planck region ($X\gg\ell_{\rm Pl}$), where the loop effects [i.e., terms proportional to $1/\rho_{\rm cr}$ in Eq.~\eqref{vaidyametr2}] are negligible, a quantum gravity effect will still exist due to backreaction of the $\mathbf{k}$th mode:
\begin{align}
F_k(t, X) =  1-\frac{2G}{X}\langle \hat{M}_T\rangle -\frac{2Gr_{\rm b}}{\tilde{R}^{1+\beta}}\,  \frac{N_k \hbar k }{X^{1-\beta}}\, .
\label{vaidyametr2-b}
\end{align}
We are interested in ranges far from the bounce, where the backreaction effects are still significant while loop corrections are negligible. Therefore, in the rest of the paper, we will explore the physical consequences of the solution \eqref{vaidyametr2} with no loop correction included.

%%%%%%%%%%%%%%%%%%%%%%%%%%%%%%%%%%%%%%%
\section{Gravitational lensing effect}
\label{GL}

So far, we have seen that the quantum gravity effects of the interior dust ball can be carried out to the exterior region through suitable  junction conditions. For each mode $\mathbf{k}$ of the interior matter field,  a generalized Vaidya geometry then emerges in the exterior region through an induced fluid with energy-momentum tensor (\ref{typeII-mode}).  We assume that, the external matter is a photonic fluid with discrete energy-momentum    {\footnotesize $T_{\mu\nu}^{(k)}$} and an EoS $w = 1/3$ (i.e., radiation). 
Consequently,  the exterior spacetime will be modified by the backreaction of the induced photons, thus, photons with different energies experience different spacetime geometries\footnote{The backreaction of each mode of  photons on the exterior background changes the spacetime metric, so different modes will induce different geometries which in turn, can be probed by the same modes. This is similar to the refraction of light wavelengths in a medium  with refractive indices which depend on those wavelengths. Indeed, this is a consequence of the backreaction of light modes on the medium  so that different modes of light feel different refractive indices and hence pass through different trajectories. This leads to a rainbow feature, as emerging from a prism.}. In this section, we will investigate the phenomenological implications of the  propagation of such photons on  the exterior induced spacetime, by studying their  gravitational lensing and calculate, perturbatively, the  effects of backreaction on the Einstein angle.

We are interested only in effects of backreaction on the exterior spacetime. Therefore,  we consider large distances from the Planck scale where loop effects  are negligible, whereas backreaction effects through semiclassical gravity are still present. In this case, the exterior  boundary function is given from Eq.~(\ref{vaidyametr2-b}) as 
\begin{equation}
F(t, X) = 1-\frac{X_{\rm S}}{X} -\frac{R_k^{2/3}(t)}{X^{2/3}}, 
\label{vaidyametr3}
\end{equation}
where, we have defined a Schwarzschild radius as {\small $X_{\rm S} \equiv 2G \langle \hat{M}_T \rangle$} and {\small $R_k^{2/3}(t) \equiv  {2 r_\text{b} N_k k\ell_{\rm Pl}^2/\tilde{R}^{4/3}(t)}$}.
Location of the apparent horizon, $X_{\rm AH}$, can be obtained by solving $F_k(t, X_{\rm AH})=0$:
\begin{eqnarray}
(X - X_{\rm S})^3 - R_k^2 X = 0, 
\label{Horizon}
\end{eqnarray}
which has the following solution:
\begin{equation}
X_{\rm AH}(k) = X_{\rm S} + \frac{(\frac{2}{3})^{1/3} R_k^2}{\mathcal{W}(X_{\rm S};R_k)} + \frac{\mathcal{W}(X_{\rm S};R_k)}{18^{1/3}} .
\end{equation}
where {\small $\mathcal{W}(X_{\rm S};R_k) \equiv \left(9 X_{\rm S} R_k^2 + \sqrt{81 X_{\rm S}^2 R_k^4 - 12 R_k^6}\right)^{1/3}$}.

Now, we intend to compute the deflection angle with respect to the exterior metric \eqref{exterior-BH} with the boundary function (\ref{vaidyametr3}). The generic geodesic equation can be written as
\begin{equation}
\frac{dv^i}{d\vartheta} + \Gamma^i_{jk} v^j v^k = 0,
\label{geodesic1}
\end{equation}
where $v^i \equiv dx^i/d\vartheta$ is the tangent vector to the  geodesic curve and $\vartheta$ is an affine parameter.
For  null curves, with $g_{ij} v^i v^j = 0$, on the exterior background  (\ref{exterior-BH}) with the boundary function (\ref{geodesic1}), the full set of geodesic equations  becomes
\begin{subequations}
	\begin{align}
	 &t^{\prime\prime} + \frac{1}{F_k}\frac{dF_k}{dX}\,  t^\prime X^\prime - \frac{1}{2F_k^3}\frac{dF_k}{dt}\, {X^\prime}^2 + \frac{1}{2F_k}\frac{dF_k}{dt}\,  {t^\prime}^2= 0,
	\quad  \\
	&\varphi^{\prime\prime} + \frac{2}{R}\,  R^\prime\, \varphi^\prime + 2 \cot\theta\, \varphi^\prime\, \theta^\prime = 0,
	\\
	&\theta^{\prime\prime} + \frac{2}{R}\, \theta^\prime\, R^\prime - \sin\theta \cot\theta \left(\varphi^\prime\right)^2 = 0, 
	\\
	&X^{\prime\prime} - \frac{1}{2 F_k}\frac{dF_k}{dX}  \left(X^\prime\right)^2 + \frac{1}{2 }F_k\frac{dF_k}{dX}\left(t^\prime\right)^2 \nonumber \\
	&X F_k \left[\left(\theta^\prime\right)^2 +\sin^2\theta \left(\varphi^\prime\right)^2\right]+\frac{1}{F_k}\frac{dF_k}{dt}\,  t^\prime X^\prime = 0, 
	\end{align}\label{eq:geodesicEq}
\end{subequations}
where a prime stands for a derivative with respect to  $\vartheta$. 

To study lensing in the exterior spacetime, we need to make a stationarity assumption. We assume that, at an early stage of the collapse, the crossing time of photons, $t_p$, is much smaller than the timescale of variation of the lens, $t_l$ (i.e., $t_p \ll t_l$). Then we will derive the lensing observables through perturbative methods. In the weak field limit, we will expand quantities in terms of the expansion parameters $\epsilon_m \equiv X_{\rm S} /X$ and $\epsilon_k \equiv (R_k/X)^{2/3}$ so that the time derivative terms in Eqs.~\eqref{eq:geodesicEq} will become of next order; for example, $dF_k/dt \propto \epsilon_k\,d\tilde{R}/dt \propto \epsilon_k\,d\tilde{a}/dt \propto \epsilon_k\,\epsilon_m^{1/2}$. Being interested in first-order effects, i.e., terms proportional to $\epsilon_m$ or $\epsilon_k$, we ignore terms with time derivatives in Eq.~\eqref{eq:geodesicEq}. In other words, we will consider regimes in which stationary features of spacetime are more important than nonstationary ones.
Without loss of generality, we will work on the equatorial plane $\theta=\pi/2$. Then, Eqs.~\eqref{eq:geodesicEq} reduce to \cite{Weinberg:1972kfs},
\begin{subequations}
	\label{equ:def} 
	\begin{align}
	&t^\prime \simeq C/F_k,
	\\
	&X^2\, \varphi^\prime = b,
	\\
	&F^{-1}_k\left(X^\prime\right)^2 + b^2/X^2 - 1/F_k \simeq -\lambda,
	\end{align}
\end{subequations}
where $C$, $b$ and $\lambda$ are constants of integration. For photons, we set $\lambda = 0$, and for simplicity we set  $C = 1$. Putting everything together in Eqs.~\eqref{equ:def}, we get the following geodesic for photons:
\begin{align} 
\varphi(X_{\rm so})-\varphi(X_{\rm ob}) &= \int\, \frac{dX}{X^2}
\left[\frac{1}{b^2}-\frac{1}{X^2}  +\frac{X_{\rm S}}{X^3} +\frac{R_k^{2/3}}{X^{8/3}}\right]^{-1/2}, 
\label{equ:geodesic}
\end{align}
where, $X_{\rm so}$ and $X_{\rm ob}$ are the locations of source and observer, respectively. We consider a collection of photons (as part of the fluid) which start emitting from the source in $X_{\rm so}$, moving toward the turning point on $X_{0}$ (the closest distance to the lens, i.e. the dust ball here), where $dX/d\varphi = 0$, and keep propagating until they reach  the observer on $X_{\rm ob}$. The deflection angle of the trajectory with respect to a straight line can then be obtained as
\begin{equation}
\Delta\varphi(X_0) = | \varphi(X_{\rm so}) - \varphi(X_{\rm ob}) | - \pi.
\end{equation}
In turning point, we have
\begin{equation}
\frac{F_k(X_0)}{X_0^2} = \frac{1}{b^2}, \label{rel:impactp}
\end{equation}
whence  Eq.~\eqref{equ:geodesic}, in terms of the expansion parameters $\epsilon_m$ and $\epsilon_k$ and a new variable $x=X_0/X$, becomes
\begin{align}
\varphi(X_{\rm so})-\varphi(X_{\rm ob}) &= \left(\int_{x = \frac{X_0}{X_{\rm so}}}^{1} dx + \int_{x = \frac{X_0}{X_{\rm ob}}}^{1} dx \right) (1-x^2)^{-\frac{1}{2}} \nonumber \\
& \quad \quad \times \left[1-\frac{\epsilon_m(1-x^3)}{1-x^2}- \frac{\epsilon_k(1-x^{8/3})}{1-x^2} \right]^{-\frac{1}{2}}. \label{eq:def-integration}
\end{align}
Integration in the right-hand side of Eq.~\eqref{eq:def-integration} can be done perturbatively. Then,  up to first order in $X_0/X_{\rm so}$, $X_0/X_{\rm ob}$, $\epsilon_m$ and $\epsilon_k$,  we get
\begin{align}
\Delta\varphi(X_0) &= \varphi(X_{\rm so})-\varphi(X_{\rm ob}) -\pi  \nonumber \\
& \simeq 2\,\epsilon_m + \epsilon_k \frac{\sqrt{\pi}\; \Gamma\left(11/6\right)}{\Gamma\left(4/3\right)}  - (1+\epsilon_m)\left(\frac{X_0}{X_{\rm so}} + \frac{X_0}{X_{\rm ob}} \right)  - \frac{\epsilon_k}{2} \left(\frac{X_0}{X_{\rm so}} + \frac{X_0}{X_{\rm ob}} \right) + \mathcal{O}(x^2). \label{eq:def-angle}
\end{align}
Deflection angle \eqref{eq:def-angle} contains local and nonlocal  terms. Since the metric function \eqref{vaidyametr3} is not asymptotically flat \cite{Husain:1995bf}, we cannot isolate the whole gravitational system, so it would be better to take into account the local terms. To rewrite Eq.~\eqref{eq:def-angle} in terms of the constant of motion $b$, we solve Eq.~\eqref{rel:impactp} for $X_0$. 
Taking  leading-order terms for  $X_{\rm S}/b$ and $R_k/b$, we get  \cite{Keeton:2005jd},
\begin{equation}
X_0 \simeq b \left( 1 - \frac{X_{\rm S}}{2b} - \frac{R_k^{2/3}}{2b^{2/3}} \right).
\end{equation}
Then, the deflection angle  becomes
\begin{align}
\Delta\varphi(k)  &\simeq 2\frac{X_{\rm S}}{b} + \left( \frac{R_k}{b}\right)^{2/3} \frac{\sqrt{\pi}\; \Gamma\left(11/6\right)}{\Gamma\left(4/3\right)}  -b\left(\frac{1}{X_{\rm so}} + \frac{1}{X_{\rm ob}} \right) + \mathcal{O}(x^2). \label{eq:def-angle1}
\end{align}

Since the backreaction effects will be important in the high-energy regimes, using the perturbative approach to solve the lens equation is not suitable. However, we apply a perturbative treatment to get a sense of quantum gravity effects on light rays propagating on the herein mode-dependent emergent spacetime\footnote{For strong gravity and high energy regimes and including time variation of spacetime into consideration, an independent study is required by using the numerical techniques; this will be studied in a separated work \cite{LensingRBH}.}.
%%%%%%%%%%%%%%%%%%
%%% FIGURE : lensing configuration
%%%%%%%%%%%%%%%%%%
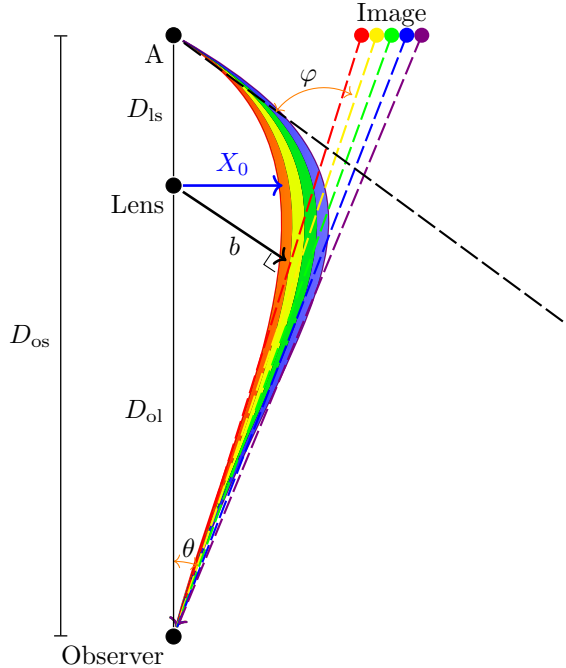
\begin{figure}
	
	\begin{tikzpicture}
		%-------------------------------------
		% defining the nodes 
		%-------------------------------------
		\node (v1) at (-8,-2) {};
		\node (v2) at (-8,4) {};
		\node (v3) at (-8,6) {};
		\node (v4) at (-5.2,6) {};
		\node (v5) at (-2.7,2.1) {};
		\node (v6) at (-6,4.5) {};
		%-------------------------------------
		% naming and painting the nodes 
		%-------------------------------------
		\foreach \v in {1,2,3} {
			\draw  (v\v) circle (0.1);
		}
		
		\fill (v1) circle (0.1) node [below left] {$\text{Observer}$};
		\fill (v2) circle (0.1) node [below left] {$\text{Lens}$};
		\fill (v3) circle (0.1) node [below left] {$\text{A}$};
		%\fill [color=violet] (v4) circle (0.1) node [below right] {$Image$}; % Image-violet
		\fill [color=violet] ($(v4)+(0.5,0)$) circle (0.1) node [below right] {}; % Image-violet
		\fill [color=blue] ($(v4)+(0.3,0)$) circle (0.1) node [below right] {}; % Image-blue
		\fill [color=green] ($(v4)+(0.1,0)$) circle (0.1) node [above] {$\textcolor{black}{\rm Image}$}; % Image-green
		\fill [color=yellow] ($(v4)+(-0.1,0)$) circle (0.1) node [below right] {}; % Image-yellow
		\fill [color=red] ($(v4)+(-0.3,0)$) circle (0.1) node [below right] {}; % Image-red
		% \fill (v5) circle (0.1) node [below right] {};
		%\fill (v6) circle (0.1) node [below right] {$$};
		%-------------------------------------
		% drawing the arrows 
		%-------------------------------------
		%%%%%%% Passing Light Ray
		\draw (v3) [->, thick, color=red, name path=r] [out=328, in=75, looseness=1.0] to (v1); % light ray-red 
		\draw (v3) [->, thick, color=yellow, name path=y] [out=329, in=74, looseness=1.08] to (v1); % light ray-yellow 
		\draw (v3) [->, thick, color=green, name path=g] [out=329, in=74, looseness=1.2] to (v1); % light ray-green 
		\draw (v3) [->, thick, color=blue, name path=b] [out=329, in=74, looseness=1.32] to (v1); % light ray-blue 
		\draw (v3) [->, thick, color=violet, name path=v] [out=330, in=73, looseness=1.4] to (v1); % light ray-violet
		\tikzfillbetween[of=r and y]{left color=red, right color=red!10!yellow}; % filling
		\tikzfillbetween[of=y and g]{left color=yellow, right color=yellow!80!green}; % filling
		\tikzfillbetween[of=g and b]{left color=green, right color=green!80!blue}; % filling
		\tikzfillbetween[of=b and v]{left color=blue, right color=blue!20!white}; % filling
		%%%%%%% Image arrows
		\draw ($(v4)+(-0.3,0)$) [-, thick,color=red, dash pattern={on 7pt off 2pt}] to (v1); % Image-Observer-red
		\draw ($(v4)+(-0.1,0)$) [-, thick,color=yellow, dash pattern={on 7pt off 2pt}] to (v1); % Image-Observer-yellw
		\draw ($(v4)+(0.1,0)$) [-, thick, color=green, dash pattern={on 7pt off 2pt}] to (v1); % Image-Observer-green
		\draw ($(v4)+(0.3,0)$) [-, thick,color=blue, dash pattern={on 7pt off 2pt}] to (v1); % Image-Observer-blue
		\draw ($(v4)+(0.5,0)$) [-, thick, color=violet, dash pattern={on 7pt off 2pt}] to (v1); % Image-Observer-violet
		\draw (v5) [-, thick, dash pattern={on 8pt off 2pt}] to (v3); % v5-Source
		\draw pic["$\varphi$", draw=orange, <->, angle eccentricity=1.2, angle radius=0.8cm]
		{angle=v4--v6--v3}; % alpha angle
		\draw pic["$\theta$", draw=orange, <->, angle eccentricity=1.2, angle radius=1cm]
		{angle=v4--v1--v3}; % theta angle
		%-------------------------------------
		% drawing the distances 
		%-------------------------------------
		%\draw (v2) [->, thick, color=blue] [out=0, in=180] to (-6.8,4) node[midway,above]{$\tau_c$};
		\draw [->, line width = 1pt, color=blue]  (v2) -- (-6.55,4) node[midway,above]{$X_0$};
		\draw [->, line width = 1pt]  (v2) -- (-6.5,3) node[midway,below]{$b$};
		\draw (-6.7, 3.1) -- (-6.8,2.95) -- (-6.65,2.85);
		\draw [-, line width = 0.5pt]  (v1) -- (v2) node[midway,left]{$D_{\rm ol}$};
		\draw [-, line width = 0.5pt]  (v2) -- (v3) node[midway,left]{$D_{\rm ls}$};
		\draw [|-|, line width = 0.5pt]  ($(v1)+(-1.5,0)$) -- ($(v3)+(-1.5,0)$) node[midway,left]{$D_{\rm os}$};

	\end{tikzpicture}
	\caption{\footnotesize{Lensing configuration: deflection of different modes by a point mass, where lens, observer and source are highly aligned. Different modes probe different curvatures and create images at different angular positions, this chromatic gravitational aberration can be called `` Quantum Gravitational Prism". }} \label{fig:lensing}
\end{figure}
%%%%%%%%%%%%%%%%%%
%%% FIGURE
%%%%%%%%%%%%%%%%%%
It is well known that when the observer, source and lens (gravitational field) are aligned, it gives rise to the formation of an Einstein ring. The third term in deflection angle \eqref{eq:def-angle1} can be canceled according to initial values for $\varphi(X_{\rm so})$ and $\varphi(X_{\rm ob})$ at the flat spacetime, when there are no curvature terms. In this case, we make use of lens equation $\theta_E D_{\rm os} = \Delta\phi D_{\rm ls}$, in which $\theta_E$, $D_{\rm os}$ and $D_{\rm ls}$ are the Einstein angle of the image, observer-source and lens-source distances, respectively. In the configuration depicted in Fig.~\ref{fig:lensing}, we  look for the first-order effects, i.e., sub-leading corrections in weak field approximation. We consider terms up to first orders in $\epsilon_m$ and $\epsilon_k$ in the deflection angle \eqref{eq:def-angle1} and then apply a perturbative approach to find a solution for the lens equation. The aim of this investigation is to get a sense of the order of corrections arising from backreactions on the Einstein angle. 

In our case, the black hole spacetime can be described by the metric function \eqref{vaidyametr3}. Thus, angular diameter distances are different from Schwarzschild at spatial infinity and are different from radial coordinates. To calculate the Einstein angle, we use asymptotically flat spacetime relations $b \simeq D_{\rm ol}\, \theta + \mathcal{O}(\epsilon_k^{n})$, $D_{\rm ls} \simeq X_{\rm so} + \mathcal{O}(\epsilon_k^{n})$ and $D_{\rm ol} \simeq X_{\rm ob} + \mathcal{O}(\epsilon_k^{n})$ which bring errors of the order $\mathcal{O}(\epsilon_k^{n+1})$ in the Einstein angle. 
As stated before, we are not looking for exact values, but instead, we are interested in finding the order of corrections using a solution like $\theta_E \approx \theta_0 + \lambda_{\rm pert} \theta_1$ for the lens equation \cite{Sereno:2003nd};  $\lambda_{\rm pert}$ is the perturbation parameter that can be read from Eq.~\eqref{eq:def-angle1} as, 
\begin{align}
\Delta\varphi(k) &\simeq \frac{1}{Z} + \frac{\lambda_{\rm pert}}{Z^{2/3}} \, ,   \\
\lambda_{\rm pert} &= \left( \frac{R_k}{2X_{\rm S}}\right)^{2/3} \frac{\sqrt{\pi}\; \Gamma\left(11/6\right)}{\Gamma\left(4/3\right)}, \label{eq:def-angle2}
\end{align}
where we introduced $Z = b/(2X_{\rm S})$. Using the lens equation, up to the first order in perturbation parameter $\lambda_{\rm pert}$ we find
\begin{eqnarray}
\theta_0 &=& \left(2 X_{\rm S} \frac{D_{\rm ls}}{D_{\rm os} D_{\rm ol}} \right)^{1/2}, \label{eq:theta}\\
\theta_1 &=& \frac{1}{2} \left(\frac{D_{\rm ls}}{D_{\rm os}}\right)^{2/3} \left(\frac{2X_{\rm S}}{D_{\rm ol}}\right)^{1/3},\label{eq:theta1}
\end{eqnarray}
where $D_{\rm ol}$ is the observer-lens distance. Equation~\eqref{eq:theta1} indicates that different modes create images at different angular positions which leads to chromatic aberration for the gravitational lens.
Considering more terms in expansion of $b$ and $\theta$ adds next-order corrections to $\theta_0$ and $\theta_1$. For better precision, it is more convenient to evaluate $\Delta\varphi(k)$ numerically and use the lens equation to find the Einstein angle; these investigations will be presented in a consequent paper \cite{LensingRBH}.

%%%%%%%%%%%%%%%%%%%%%%%%%%%%%%%%%%%%%%%%%%
\subsection{An astronomical example}

Here, we give an example for gravitational chromatic aberration in GRBs  propagating from near a rainbow black hole spacetime. This example will show how the free parameters in the backreaction term of metric function \eqref{vaidyametr3} can be practically interpreted. Each degree of freedom of the electromagnetic field can be treated as a massless scalar field and has an identical backreaction term \cite{Lewandowski:2017cvz}.
In particular, here we consider astronomical data of Cyg~X-1
%\footnote{This is the famous subject of the bet between Hawking and Thorne.} 
for a stellar black hole candidate in our Milky Way galaxy, to study the gravitational chromatic aberration effects within our herein model.

GRBs can be created by gravitational collapse of a star releasing energy (typically $\sim 10^{44-47}$J) as electromagnetic waves \cite{Piron:2015wtz}. Table~\ref{tab:beta0-pert} depicts two sets of data, ``optimistic'' and ``realistic'' sets of expected values for aberration effects of GRBs when they are passing by 
Cyg~X-1. 
In the \textit{optimistic} part, we have presented numerical values made from observations at short distances to Cyg~X-1, a few billions of kilometers, whereas in the \textit{realistic} part, we considered observations from long distances to Cyg~X-1, a few kiloparsecs. 
Although we made many simplifying assumptions in our perturbative analysis, we can still trust the order of corrections for the weak field regime.

%%%%%%%%%%%  Table  %%%%%%%%%%%%%%%% 
\begin{table}[b]
	Optimistic set, $D_{\rm ol} \simeq 10^{-3} $pc, $\mathcal{D} := D_{\rm ls}/D_{\rm os} = 0.5$ \vspace{1mm} \\
	\begin{tabular}{lccc}	
		\Xhline{3\arrayrulewidth}\\ [-2.2ex]  
		$E_k $~~~~~~~~~&$\lambda_{\rm pert}$ &~~~~~~~ $\theta_E$\, (arcsec)~~~~~\\ [.5ex] 
		\hline \\ [-2.ex] 
		0            & 0 & 9.02453456 \vspace{1mm} \\
		$\rm 1\, keV$    & $ 3.29064131 \times 10^{-10}$ & 9.02453460 \vspace{1mm} \\
		$\rm 1\, MeV$          & $ 3.29064131 \times 10^{-7}$ & 9.02456801 \vspace{1mm} \\
		\Xhline{3\arrayrulewidth}  
	\end{tabular} 
	\vspace{4mm}\\
	Realistic set, $D_{\rm ol} \simeq $2 kpc, $\mathcal{D} := D_{\rm ls}/D_{\rm os} = 0.0005$  	\vspace{1mm}\\
	\begin{tabular}{lccc}
		\Xhline{3\arrayrulewidth}  \\ [-2.2ex]  
		$E_k $~~~~~~~~~&$\lambda_{\rm pert}$ &~~~~~~~ $\theta_E$\, ($\mu$ arcsec)~~~~~\\  [.5ex]
		\hline \\ [-2.ex] 
		0            & 0 & 201.79472758 \vspace{1mm}\\  
		$\rm 1\, keV$    & $ 3.29064131 \times 10^{-10}$ & 201.79473023 \vspace{1mm} \\
		$\rm 1\, MeV$          & $ 3.29064131 \times 10^{-7}$ & 201.79738211 \\	
		\Xhline{3\arrayrulewidth}   
	\end{tabular}
	\caption{Optimistic and realistic values for the image positions of the Einstein ring with the source position $\beta = 0$, due to lensing by a stellar black hole with quantum backreaction effects (a rainbow black hole with chromatic aberration effects). The lens is the Milky Way black hole candidate Cyg~X-1 with mass $M \simeq 20 \times M_{\odot}$ \cite{Miller-Jones:2021plh}. We take $r_\text{b}/R_{\rm sch} = 1$ with stationarity assumption $\tilde{R}(t) \sim \tilde{R} \sim r_{\rm b}$ at an early stage of collapse. Here, $E_k$ represents the energy of the massless particle in electron volts with particle number $N_k \sim 10^{55}$; $E = 0$ shows results for the classical case where $R_k = 0$.}
	\label{tab:beta0-pert}
\end{table}
%%%%%%%%%%%  Table  %%%%%%%%%%%%%%%% 

The modification term $R_k^{2/3} \propto N_k \ell_{\rm Pl}^2$ suffers from a huge suppression of the order of $\sim \ell_{\rm Pl}^2$, whereas the free parameter $N_k$ can compensate the squared Planck length, and play the role of amplification parameter that enhances the backreaction effects. In the numerical calculations, we have taken the radius of the boundary shell, $r_\text{b}$, to equal the Schwarzschild radius (note that, this is the least value for the shell radius). Moreover, $N_k$ can be interpreted as the number of photons in an  adiabatic regime. Based on the energy released by GRBs, we take the optimistic value $N_k \sim 10^{55}$ as an approximate value for the number of photons in one energy bound. In Table~\ref{tab:beta0-pert}, we have considered two energy bands in keV and MeV for bursts probing a rainbow black hole. In the optimistic case where rainbow effects can be observed from a nearby source, modification to Einstein ring can be of the order of $\delta \theta_E \sim 100 \; \mu$arcsec, while for the realistic case, in which observation is made from Earth at large distances from the source, i.e., distances of the order of $\sim 2$kps, rainbow effects in $\theta_E$ are minuscule and at best can be of the order of $\delta \theta_E \sim 10^{-3}\mu$arcsec.

\section{Conclusion and discussion}
\label{conclusion}

In this paper, we  considered the gravitational collapse of a (homogeneous) spherically symmetric dust cloud plus a (inhomogeneous) massless scalar perturbation. Classically, when discarding the effects of the {\em homogeneous} sector of the scalar field on the background, this model leads to the formation of a Schwarzschild black hole in the exterior region. In quantum theory in the interior region, the dust field $T$ plays the role of internal time which represents the evolution of the physical Hamiltonian of the gravitational system coupled to the scalar field $\phi$. For each mode of the  scalar perturbation  propagating on this quantized background, the evolution equation  corresponds to evolution for the same field's mode on an effective dressed background. The components of this dressed metric depend on fluctuations of the background quantum geometry. When the backreaction of the quantum modes is taken into account, the emergent dressed background turns out to be mode dependent; a {\em rainbow metric} (with components depending on the energy of the field modes) emerges.

The semiclassical behavior of the interior dressed geometry was presented by employing the quantum backreactions and higher-order corrections due to quantum fluctuations of the spacetime geometry.
The nonclassical features of the interior spacetime were carried out to the exterior region due to convenient matching conditions at the boundary of the  dust cloud: An exterior (nonstatic) black hole geometry [cf. Eq.~\eqref{exterior-BH}] could emerge whose components depend on the mode of the induced fluid in the outer region. Properties of the interior and the induced exterior geometries are summarized as follows.
%%%%%%%%%%%%%%%%%%%%%
\begin{enumerate}[label=\roman*)]
\item At the late stage of collapse inside the dust ball, LQG effects in the interior region are large; i.e., the terms $\propto 1/\rho_{\rm cr}$ in Eq.~\eqref{Friedmann-Dressed-new} are dominant. This indicates that the classical singularity is removed and is replaced by a quantum bounce at the final stage of the collapse. There is an additional correction,  $\propto k\langle V\rangle^{-4/3}$, from the backreaction of each mode $\mathbf{k}$ of the scalar perturbation on the interior quantum spacetime. This term represents the energy density of a radiation fluid that appears to be dominant in very short distances, so that, as the collapse proceeds, the total energy density of the interior region grows faster, compared with the case where a pure dust field is considered.  A thorough numerical analysis would confirm our herein results \footnote{These analyzes will be presented in an upcoming paper \cite{ModifiedDR:2022}.} which imply
that a bounce still occurs in our model, but the backreaction effects speed up its occurrence. 
%%%%%%%
\item This radiationlike effect leads to a unique evolution associated with each mode inside the dust ball. Thus, for each scalar field mode in the interior region of the collapse, a unique dressed, classical-like metric emerges. In other words, different  modes explore different backgrounds and thus, a rainbow geometry emerges in the interior region.
%%%%%%%%%%
\item Outside the dust ball, a generalized Vaidya spacetime, with a suitable choice of (external) fluids, can be matched consistently to each interior mode-dependent spacetime. Therefore, corresponding to each mode $\mathbf{k}$ in the interior region, there exists a unique Vaidya geometry, $g_{\mu\nu}^{+}(k)$ (labeled by the number $k$), provided by an external fluid with an energy-momentum tensor {\footnotesize $T^{(k)}_{\mu \nu}$} satisfying the Einstein field equations outside the dust ball.  This leads to a one-to-one correspondence between the interior field mode $\mathbf{k}$ and an exterior fluid with profiles labeled by $k$.
As a consequence,  different modes and labels, $k^\prime$, of the {\em external} fluid explore different Vaidya backgrounds. This is equivalent to saying that the components of the exterior Vaidya metric depend on the mode or label of the external fluid, being a rainbow geometry. Such a geometry may feature  ``\textit{rainbow horizons}" with  optical properties different from the classical black holes.   
\end{enumerate}

At distances much larger than the scale of the bounce, where the loop effect (i.e., terms proportional to $1/\rho_{\rm cr}$) is negligible, the quantum gravity effects are still significant due to the backreaction effects, through a term $\propto 1/X^{2/3}$ in the exterior metric \eqref{vaidyametr2-b}.
Assuming that the external matter is a {\em radiation} fluid, the  spacetime corresponding to  each mode of this  fluid can be probed as a source of gravitational lensing;  each mode can  provide its own particular Einstein's ring, leading to a chromatic aberration in the gravitational lensing process. Therefore, different modes see different rings, so that a rainbowlike collection of rings can be detected from astrophysical observation of such spacetimes (cf. Fig.~\ref{fig:lensing}).

Finally, we should emphasize that the  results we achieved within this paper are not limited  to the LQG approach only. It can be shown that similar conclusions might arise from other approaches to quantum gravity such as the geometrodynamics approach.

%%%%%%%%%%%%%%%%%%%%%%%%%%%%%
\section*{Acknowledgments}

The work of A.P. was supported in part by the Ministry of Science, Research and Technology of Iran. He is grateful for the support and kind hospitality of University of Warsaw and University of Wroc\l{}aw, where part of this work was completed and he wishes to thank  Professor Mohammad Nouri-Zonoz for comments and helpful discussions concerning this project.
T.P. acknowledges the support by the Polish Narodowe Centrum Nauki (NCN) grants 2012/05/E/ST2/03308 and 2020/37/B/ST2/03604.
The work of Y.T. was supported by the Research deputy of University of Guilan. He also acknowledges the financial support from the Polish Narodowe Centrum Nauki (NCN) through the grant 2012/05/E/ST2/03308. 
J.L. was supported by the Polish Narodowe Centrum Nauki, Grant No. 2011/02/A/ST2/00300.
This paper is based upon work from European Cooperation in Science and Technology (COST) action CA18108 -- Quantum gravity phenomenology in the multi-messenger approach, supported by COST.

%%%%%%%%%%%%%%%%%%%%%%%%%%%%%%%%%%%%%%%%%%%%%%%%
\appendix

\section{Derivation of dressed Hubble rate} \label{App-A}

In order to compute $\partial_\tau\tilde{a}/\tilde{a}$ in Eq.~(\ref{Friedmann-dressed1}), 
we should look for $\partial_{\tau}\langle \hat{v}^{\alpha}\rangle$, $\alpha=-1, 1/3$. This is given by
\begin{equation}
\partial_{\tau}\langle \hat{v}^{\alpha}\rangle =  \frac{\big\langle \big[ \hat{H}_{\rm grav} ,  \hat{v}^{\alpha}  \big] \big\rangle}{-i\hbar} \, ,
\label{evol-alpha}
\end{equation}
in which the scalar field Hamiltonian cancels out in the right-hand side of the equation, because it depends only on the volume operator.
The main tool available at the dynamical level is that of effective equations which describe the evolution of expectation values for a dynamical state. Thus, quantum fluctuations and higher moments act on the evolution of expectation values, described in effective equations by coupling classical and quantum degrees of freedom.

In LQC coupled with dust, the Hamiltonian $\hat{H}_{\rm grav}$ is given by
\begin{eqnarray}
\hat{H}_{\rm grav}\ = \ \frac{3\pi G}{8\alpha_o}\sqrt{|\hat{v}|} \left(\hat{\mathsf{N}}^{2} - \hat{\mathsf{N}}^{-2}\right)^2 \sqrt{|\hat{v}|}\ . \quad 
\label{GR-Hamiltonian}
\end{eqnarray}
Volume operator $\hat{v}$ is defined as $\hat{v}| v \rangle=v| v \rangle$ and $\hat{\mathsf{N}}| v \rangle = e^{i\hat{b}/2}| v \rangle=| v+1 \rangle$ with $[\hat{b}, \hat{v}]=2i$.
The operators $\hat{\mathsf{N}}=\exp(i\hat{b}/2)$,  $\hat{v}$ and $\hat{b}$ satisfy the relations
\begin{eqnarray}
\big[\hat{\mathsf{N}}^n, \hat{v}\big]=-n\hat{\mathsf{N}}^n\ , \quad \quad 
\big[\sin^2(\hat{b}), \hat{v}\big] = 2i\sin(2\hat{b}) \, . \quad \quad 
\end{eqnarray}
By substituting 
$\hat{\mathsf{N}}^{2} - \hat{\mathsf{N}}^{-2} = 2i\sin(\hat{b})$ in Eq.~(\ref{GR-Hamiltonian}) we have
\begin{eqnarray}
\hat{\mathcal{H}}_{\rm grav}\ = \ - \frac{3\pi G}{2\alpha_o}\sqrt{|\hat{v}|} \sin^2(\hat{b}) \sqrt{|\hat{v}|}\, . \quad 
\label{GR-Hamiltonian2}
\end{eqnarray}
It is  convenient to define operators
\begin{eqnarray}
\hat{r} \ :=\ \sin^2(\hat{b}) \quad \quad {\rm and } \quad \quad \hat{h}\ :=\ \sin(2\hat{b})\ ,
\label{operator-r-h}
\end{eqnarray}
with the following commutation relations:
\begin{eqnarray}
[\hat{r}, \hat{v}] =  2i\hbar\hat{h} \quad \quad  {\rm and} \quad \quad [\hat{h}, \hat{v}]=4i(1-2\hat{r})\ .
\label{r-h-v}
\end{eqnarray}
%%
%In order to calculate the commutation relation in Eq.~(\ref{evol-alpha}), we can use of
%\begin{eqnarray}
%\big\langle  [\hat{\mathcal{H}}_{\rm grav}, \widehat{v^{\alpha}}]  \big\rangle &=&
%i\hbar \big\{\langle\hat{\mathcal{H}}_{\rm grav}\rangle, \langle\widehat{v^{\alpha}}\rangle\big\}_{\rm sym} .
%\label{symmetry-Poisson}
%\end{eqnarray}
To determine the evolution equation for observables $\langle\widehat{v^{\alpha}}\rangle$ under Hamiltonian $\langle\hat{H}_{\rm grav}\rangle$, we make use of a background-dependent expansion method. A combination of operators $\hat{D}\big( \hat{v} , \hat{r}, \hat{h} \big)$ can be expanded as (by considering a convenient choice of symmetric ordering)
\begin{eqnarray}
\hat{D} &=& 
D\big(\langle \hat{v} \rangle, \langle\hat{r} \rangle, \langle\hat{h}\rangle\big)
+ \sum_{a,b,c=0}^\infty   \frac{1}{a! b! c!} \frac{\partial^{a+b+c}~D^{klm}}{\partial\langle\hat{v}\rangle^a ~\partial\langle\hat{r}\rangle^b~ \partial \langle\hat{h}\rangle^c} C^{abc} . \quad \quad \quad  
%\nonumber \\
\label{moment-absolute}
\end{eqnarray}
%Note that, $C^{k0}=\tilde{C}^{k0}$.
Note $\hat{D}\big( \hat{v} , \hat{r}, \hat{h} \big)$  is different than the multiplication of expectation values of operators $\hat{v}, \hat{r}$ and $\hat{h}$; to describe the system completely, infinite central fluctuation operators $C^{abc}$ are needed, where  $C^{abc}$  are  the (symmetric ordered) central moments defined by
\begin{eqnarray}
C^{abc} &:=&  (\delta \hat{v})^a(\delta \hat{r})^b (\delta\hat{h})^c
\label{moment-gen}
\end{eqnarray}
and $\delta \hat{v}=\hat{v}-\langle\hat{v}\rangle \mathbb{I}$, $\delta \hat{r}=\hat{r}-\langle\hat{r}\rangle \mathbb{I}$, and $\delta \hat{h}=\hat{h}-\langle\hat{h}\rangle \mathbb{I}$  are fluctuations of the operators $\hat{v}$, $\hat{r}$ and $\hat{h}$ around the background state with expectation values $\langle\hat{v}\rangle$, $\langle\hat{r}\rangle$ and $\langle\hat{h}\rangle$, respectively.

Now, following the definitions above, we can find the expansion of $\langle \hat{v}^\alpha \rangle$ and $\langle\hat{H}_{\rm grav}\rangle$ in terms of operators (\ref{moment-gen}). To do so, we can expand any operator $\hat{v}^\alpha$ as
\begin{eqnarray}
\hat{v}^\alpha &=& \big(\langle \hat{v} \rangle \mathbb{I} + \delta\hat{v}\big)^\alpha   = 
\sum_{n=0}^\infty  \left({\begin{array}{c}
	\alpha \\
	n  \\
	\end{array} } \right) \langle \hat{v} \rangle^{\alpha-n}C^{n00} , \quad \quad 
\label{v-hat-alpha}
\end{eqnarray}
then, by taking its expectation value, we find 
\begin{eqnarray}
\langle \hat{v}^\alpha \rangle &=&   \sum_{k=0}^\infty    \left({\begin{array}{c}
	\alpha \\
	k  \\
	\end{array} } \right) \langle \hat{v} \rangle^{\alpha-k} G^{k00}\ , 
\label{v-alpha-exp}
\end{eqnarray}
where we have defined the moments $G^{k00}$ as 
\begin{eqnarray}
G^{k00} &=& \langle C^{k00} \rangle, \quad  C^{k00} = (\delta \hat{v})^k.
\label{moment-k00}
\end{eqnarray} 
Moreover, in terms of moments (\ref{moment-k00}), by using Eq.~(\ref{v-alpha-exp}),  we can now expand the dressed scale factor as
\begin{eqnarray}
\tilde{a} &=& \alpha_o^{-1/3}\left(\frac{\langle \hat{v}^{1/3} \rangle}{\langle \hat{v}^{-1} \rangle}\right)^{1/4} \nonumber \\
&=& \alpha_o^{-1/3} \langle \hat{v} \rangle^{1/3} \left( \sum_{k=0}^\infty    \left({\begin{array}{c}
	-1 \\
	k  \\
	\end{array} } \right) \langle \hat{v} \rangle^{-k} G^{k00}\right)^{-1/4}    
\left(\sum_{m=0}^\infty    \left({\begin{array}{c}
	\frac{1}{3} \\
	m  \\
	\end{array} } \right) \langle \hat{v} \rangle^{-m} G^{m00}\right)^{1/4}. \quad
\end{eqnarray}
To the zeroth order in quantum fluctuations, $k, m=0$, the dressed scale factor $\tilde{a}$ reduces to $\alpha_o^{-1/3} \langle \hat{v} \rangle^{1/3}$ (it is worth noting that expectation values are taken with respect to the backreacted states, solutions to the full quantum Hamiltonian constraint \eqref{Hamiltonian-constraint} ).

In order to compute the multimoment expansion for the gravitational Hamiltonian:
\begin{eqnarray}
\hat{H}_{\rm grav} &=& - \frac{3\pi G}{2\alpha_o} \big(  \hat{v}^{1/2}\hat{r}\hat{v}^{1/2}  \big) \ ,
\label{grav-Ham-vrv}
\end{eqnarray}
one needs to expand the term $\hat{v}^{1/2}\hat{r}\hat{v}^{1/2}$. 
It should be noted that the status of the symmetry here is similar to that given by the symmetric operator $(\hat{v}\hat{r}+\hat{r}\hat{v})/2$ with integer powers of $\hat{v}$ and $\hat{r}$. Therefore, we expect  that the expansion of the herein Hamiltonian operator (of LQC coupled with dust) will be symmetric automatically and no reordering procedure  is required. 
Despite this analogy, in the latter case,  the binomial  expansion of the symmetric operator $(\hat{v}\hat{r}+\hat{r}\hat{v})/2$ will lead to the finite terms around the background state with expectation values $\langle \hat{v} \rangle$ and $\langle \hat{r} \rangle$, and will be truncated  to a certain order of quantum corrections  provided by   
$\langle \delta\hat{v}\delta\hat{r} + \delta\hat{r}\delta\hat{v} \rangle/2$ once their expectation values are taken. 
However,  in our case, because of one-half power of  the volume operator, expansion of $\hat{v}^{1/2}\hat{r}\hat{v}^{1/2}$ involves infinitely many terms of binomial series. So, the Hamiltonian operator can be expanded as 
\begin{align}
\hat{H}_{\rm grav}  &= - 
\frac{3\pi G}{2\alpha_o}  \sum_{a=0}^\infty \sum_{b=0}^\infty  \left({\begin{array}{c}
	\frac{1}{2} \\
	a  \\
	\end{array} } \right) 
\left({\begin{array}{c}
	\frac{1}{2} \\
	b  \\
	\end{array} } \right) \langle \hat{v} \rangle^{1-a-b}   \Big[\langle\hat{r}\rangle   (\delta\hat{v})^{a+b}   + (\delta\hat{v})^{a}\delta\hat{r}(\delta\hat{v})^{b}   \Big] .
\label{Ham-exp-a}
\end{align}  
%%%%%%%%%%%%%%%%%%%%%%%%%%%%%%%%%%%%%%%%%%%%%%%%%%%%%%%%%%%%%
%%
The  rhs of Eq.~(\ref{Ham-exp-a}) indicates that, the   Hamiltonian operator $\hat{H}_{\rm grav}$ on the full Hilbert space is totally symmetric; that is,
it constitutes all possible reorderings of the operator $\delta\hat{v}$ 
on both sides of the operator $\delta\hat{r}$.
The expectation value of $\hat{H}_{\rm grav}$ can be written now as
\begin{eqnarray}
\langle  \hat{H}_{\rm grav} \rangle &=& - 
\frac{3\pi G}{2\alpha_o}  \sum_{n=0}^\infty \sum_{m=0}^n   \left({\begin{array}{c}
	\frac{1}{2} \\
	m  \\
	\end{array} } \right) 
\left({\begin{array}{c}
	\frac{1}{2} \\
	n-m  \\
	\end{array} } \right) \langle \hat{v} \rangle^{1-n} \Big[\langle\hat{r}\rangle   \big\langle(\delta\hat{v})^{n}\big\rangle   +  \big\langle(\delta\hat{v})^{m}\delta\hat{r}(\delta\hat{v})^{n-m}\big\rangle   \Big] . \quad \quad 
\label{Ham-expn1-a}
\end{eqnarray}
By defining  the central moments  $G^{n 10}$ as
\begin{eqnarray}
G^{n10} &:=&   \frac{1}{\beta_n} \sum_{m=0}^n \left({\begin{array}{c}
	\frac{1}{2} \\
	m  \\
	\end{array} } \right) 
\left({\begin{array}{c}
	\frac{1}{2} \\
	n-m  \\
	\end{array} } \right)   \big\langle  C^{n10} \big\rangle ,\\
C^{n10} &:=& (\delta\hat{v})^{m}\delta\hat{r}(\delta\hat{v})^{n-m},
\label{moment-n10}
\end{eqnarray}
together with
$G^{n00}$ [defined in Eq.~(\ref{moment-k00})],
we can describe the expectation value of the Hamiltonian of the quantum system completely  by
\begin{eqnarray}
\langle  \hat{H}_{\rm grav} \rangle =  -
\frac{3\pi G}{2\alpha_o}  \sum_{n=0}^\infty   \beta_n \langle \hat{v} \rangle^{1-n}  \Big[\langle\hat{r}\rangle G^{n00}  + G^{n10} \Big], \quad  \quad
\label{v-hat-alpha-B}
\end{eqnarray}
where  $\beta_n$ is a normalization constant defined by
\begin{eqnarray}
\beta_n &:=&   \sum_{m=0}^{n} \tilde{\beta}_{nm}\, ; \   \quad  \tilde{\beta}_{nm} = \left({\begin{array}{c}
	\frac{1}{2} \\
	m  \\
	\end{array} } \right) 
\left({\begin{array}{c}
	\frac{1}{2} \\
	n-m  \\
	\end{array} } \right) .   \quad \quad 
\end{eqnarray}

Now, following Eq.~(\ref{evol-alpha}), in order to obtain  the time evolution of $\langle\hat{v}^{\alpha}\rangle$ we should compute commutators
between central fluctuation operators $C^{j00}$ and $C^{n10}$. In particular, we have
\begin{eqnarray}
\langle \big[ \hat{H}_{\rm grav} \; , \hat{v}^{\alpha}\big] \rangle \, =\, - \frac{3\pi G}{2\alpha_o}
\sum_{n=0}^\infty \sum_{j=1}^\infty \sum_{m=0}^{n} \left({\begin{array}{c}
	\alpha \\
	j  \\
	\end{array} } \right) \langle \hat{v} \rangle^{\alpha+1-n-j}  \tilde{\beta}_{nm}  \Big\langle  \langle\hat{r}\rangle \big[ C^{n 0 0}, C^{j00} \big]+ \big[ C^{n10}, C^{j00} \big] \Big\rangle. \quad \quad  \quad 
\label{symmetry-Poisson-1}
\end{eqnarray}
The first bracket on the rhs of the equation above is zero, so our task will be computing only the second bracket. We get
%%%%%%%%%%%%%%%%%%%%%%%%%%%%%%%%%%%%%%%%
\begin{widetext}
	\begin{eqnarray}
	\sum_{m=0}^{n} \tilde{\beta}_{nm} \big\langle \big[ C^{n10},\, C^{j00}\big] \big\rangle  
	&=&    \sum_{m=0}^{n} \tilde{\beta}_{nm}
	\big\langle \big[ (\delta\hat{v})^{m}\delta\hat{r}(\delta\hat{v})^{n-m}  ,\,  (\delta\hat{v})^{j}  \big]\big\rangle \nonumber \\
	%%%%%%%%%%%%%%%%%%%%%%%%%%%%%%%%%%%%%%%%%%%%%%%%%%%%%%%%%%%%%%%%%%%%%%%%%%%%%%%%
	&=& \frac{1}{2}\sum_{m=0}^{n} \sum_{a=0}^m\sum_{b=0}^{n-m} \sum_{c=1}^j\tilde{\beta}_{nm}
	\left({\begin{array}{c}
		m \\
		a  \\
		\end{array} } \right) 
	\left({\begin{array}{c}
		n-m \\
		b  \\
		\end{array} } \right)
	\left({\begin{array}{c}
		j \\
		c  \\
		\end{array} } \right)
	 \nonumber \\
	&& \quad \quad \times (-1)^{n+j-a-b-c} \langle\hat{v}\rangle^{n+j-a-b-c}
	 \big\langle \big[ \hat{v}^{a}\hat{r}\hat{v}^{b},\,  \hat{v}^{c}  \big] \big\rangle 
	\nonumber \\
	%%%%%%%%%%%
	&&  + \frac{1}{2}\sum_{m=0}^{n} \sum_{a=0}^{n-m}\sum_{b=0}^{m} \sum_{c=1}^j\tilde{\beta}_{nm}
	\left({\begin{array}{c}
		n-m \\
		a  \\
		\end{array} } \right) 
	\left({\begin{array}{c}
		m \\
		b  \\
		\end{array} } \right)
	\left({\begin{array}{c}
		j \\
		c  \\
		\end{array} } \right)
	\nonumber \\
	&& \quad \quad \times (-1)^{n+j-a-b-c} 
	\langle\hat{v}\rangle^{n+j-a-b-c}
	 \big\langle \big[\hat{v}^{a}\hat{r}\hat{v}^{b},\, \hat{v}^{c} \big] \big\rangle .\nonumber
	\end{eqnarray}
In deriving the equation above, we have replaced the expectation value of an (nonsymmetric) operator
\begin{eqnarray}
\hat{A}_{n,m} 
&:=& 
\sum_{a=0}^m \sum_{b=0}^{n-m} 
\left({\begin{array}{c}
	m \\
	a  \\
	\end{array} } \right) 
%%%%%%%%%%%%%%%%%%%%%%%%%%%
\left({\begin{array}{c}
	n-m \\
	b  \\
	\end{array} } \right) 
(-1)^{n-a-b} \langle \hat{v} \rangle^{n-a-b} \,
\hat{v}^{a}\hat{r} \hat{v}^{b} , \quad \quad 
%%%%%%%%%%%%%%%%%%%%%%%%%%%%%%%%%%%%%%%%%%%%%%%
\label{abs-moment-sym1}
\end{eqnarray}
by its symmetric counterpart as 
\begin{eqnarray}
\langle \hat{B}_{n, m} \rangle &=& \tfrac{1}{2}\langle \hat{A}_{n,m} + \hat{A}_{n,n-m}\rangle \,. 
\end{eqnarray}
Using linearity and the Leibniz rule, we get
%%
%
%\begin{widetext}
	\begin{align}
	\sum_{m=0}^{n} \tilde{\beta}_{nm} \big\langle \big[ C^{n10},\, C^{j00}\big] \big\rangle   
	&= \frac{1}{2}\sum_{m=0}^{n} \sum_{a=0}^m\sum_{b=0}^{n-m} \sum_{c=1}^j\tilde{\beta}_{nm}
	\left({\begin{array}{c}
		m \\
		a  \\
		\end{array} } \right) 
	\left({\begin{array}{c}
		n-m \\
		b  \\
		\end{array} } \right)
	\left({\begin{array}{c}
		j \\
		c  \\
		\end{array} } \right) \nonumber \\
&	\quad \quad \times (-1)^{n+j-a-b-c} \langle\hat{v}\rangle^{n+j-a-b-c}
\big\langle \hat{v}^{a}\big[\hat{r},\,  \hat{v}^{c} \big]\hat{v}^{b} \big\rangle 
	\nonumber \\
	%%%%%%%%%%%
	& \quad  + \frac{1}{2}\sum_{m=0}^{n} \sum_{a=0}^{n-m}\sum_{b=0}^{m} \sum_{c=1}^j\tilde{\beta}_{nm}
	\left({\begin{array}{c}
		n-m \\
		a  \\
		\end{array} } \right) 
	\left({\begin{array}{c}
		m \\
		b  \\
		\end{array} } \right)
	\left({\begin{array}{c}
		j \\
		c  \\
		\end{array} } \right) \nonumber \\
	&\quad \quad \times
	(-1)^{n+j-a-b-c} \langle\hat{v}\rangle^{n+j-a-b-c}
	\big\langle \hat{v}^{a}\big[\hat{r},\,  \hat{v}^{c} \big]\hat{v}^{b} \big\rangle 
	\nonumber \\
	%%%%%%%%%%%%%%%%%%%%%%%%%%%%%%%%%%%%%%%%%%%%%%%%%%%%%%%%%%%%%%%%%%%%%%%%%%%%%%%
	&= \sum_{m=0}^{n}  \sum_{c=1}^j\tilde{\beta}_{nm}
	\left({\begin{array}{c}
		j \\
		c  \\
		\end{array} } \right)
	(-1)^{j-c} \langle\hat{v}\rangle^{j-c}
	\Big(\big\langle (\delta\hat{v})^{m}\big[\hat{r},\,  \hat{v}^{c} \big](\delta\hat{v})^{n-m} \big\rangle  \Big)  \quad \quad \nonumber \\
	%%%%%%%%%%%%%%%%%%%%%%%%%%%%%%%%%%%%%%%%%%%%%%%%%%%%%%%%%%%%%%%%%%%%%%%%%%%%%%%
	&= \sum_{m=0}^{n}  \sum_{c=1}^j\tilde{\beta}_{nm}
	\left({\begin{array}{c}
		j \\
		c  \\
		\end{array} } \right)
	(-1)^{j-c} \langle\hat{v}\rangle^{j-c}
	\Big( \sum_{l = 0}^{c-1} 
	\left({\begin{array}{c}
		c-1 \\
		l  \\
		\end{array} } \right)
	\big\langle (\delta\hat{v})^{m} \hat{v}^l \big[\hat{r},\,  \hat{v} \big] \hat{v}^{c-1-l} (\delta\hat{v})^{n-m} \big\rangle \Big) 
	\nonumber \\
	%%%%%%%%%%%%%%%%%%%%%%%%%%%%%%%%%%%%%%%%%%%%%%%%%%%%%%%%%%%%%%%%%%%%%%%%%%%%%%%%
	&= 2 i \hbar \sum_{m=0}^{n}  \sum_{c=1}^j\tilde{\beta}_{nm}
	\left({\begin{array}{c}
		j \\
		c  \\
		\end{array} } \right)
	(-1)^{j-c} \langle\hat{v}\rangle^{j-1}
	 \Big( \langle\hat{h}\rangle \; \sum_{l=0}^{c-1}\left({\begin{array}{c}
		c-1 \\
		l  \\
		\end{array} } \right) 
	\sum_{k = 0}^{c-1} 
	\left({\begin{array}{c}
		c-1 \\
		k  \\
		\end{array} } \right)  \langle\hat{v}\rangle^{-k} G^{(n+k)00} 
		\nonumber \\
	%%%%%%%%%%%%%%%%%%%%%%%%%%%%%%%%%%%%%%%%%%%%%%%%%%%%%%%%%%%%%%%%%%%%%%%%%%%%%%%%
	& \quad +\sum_{l = 0}^{c-1} \sum_{k = 0}^{l} \sum_{d = 0}^{c-1-l}
	\left({\begin{array}{c}
		c-1 \\
		l  \\
		\end{array} } \right)
	\left({\begin{array}{c}
		l \\
		k  \\
		\end{array} } \right)
	\left({\begin{array}{c}
		c-1-l \\
		d  \\
		\end{array} } \right) \langle\hat{v}\rangle^{-k-d}
	\big\langle (\delta\hat{v})^{m+k} \delta\hat{h} (\delta\hat{v})^{n-m+d} \big\rangle \Big). \quad 
	\label{moments01}
	\end{align}
	%\end{widetext}
%\end{widetext}

Now we can rewrite Eq.~(\ref{evol-alpha}) as 
\begin{eqnarray}
\frac{\big\langle \big[ \hat{\mathcal{H}}_{\rm grav} ,  \hat{v}^{\alpha}  \big] \big\rangle}{-i\hbar} 
&=:&  \frac{3\pi G}{\alpha_o}\left( {\cal A}(\alpha) \langle \hat{h} \rangle + {\cal B}(\alpha) \right), \quad 
\label{symmetry-Poisson-1-d}
\end{eqnarray}
where  ${\cal A}(\alpha)$ and ${\cal B}(\alpha)$ are given, respectively, by
\begin{eqnarray}
{\cal A}(\alpha) &=&   \sum_{n=0}^\infty \sum_{j=1}^\infty \sum_{c=1}^j  \sum_{k = 0}^{c-1}    \sum_{l=0}^{c-1}\left({\begin{array}{c}
	c-1 \\
	l  \\
	\end{array} } \right)
\left({\begin{array}{c}
	\alpha \\
	j  \\
	\end{array} } \right)
\left({\begin{array}{c}
	j \\
	c  \\
	\end{array} } \right)
\left({\begin{array}{c}
	c-1 \\
	k  \\
	\end{array}} \right) \beta_n (-1)^{j-c} \langle \hat{v} \rangle^{\alpha-n-k} G^{(n+k)00}, \quad \quad 
\\
%%%%%%%%%%%%%%%%%%%%%%%%%%%%%%%%%%%%%%%%%%%%%%%%%%%%%%%%%%%%%%%%%%%%%%%%%%%%
{\cal B} (\alpha) &=&  \sum_{n=0}^\infty \sum_{j=1}^\infty \sum_{m=0}^n  \sum_{c= 1}^{j} \sum_{l= 0}^{c-1} \sum_{k= 0}^{l} \sum_{d= 0}^{c-1-l}   \tilde{\beta}_{nm} \left({\begin{array}{c}
	\alpha \\
	j  \\
	\end{array} } \right)
\left({\begin{array}{c}
	j \\
	c  \\
	\end{array} } \right)
\left({\begin{array}{c}
	c-1 \\
	l  \\
	\end{array} } \right)
\left({\begin{array}{c}
	l \\
	k  \\
	\end{array} } \right) 
\left({\begin{array}{c}
	c-1-l \\
	d  \\
	\end{array} } \right)\nonumber \\
&& \times (-1)^{j-c} \langle \hat{v} \rangle^{\alpha-n-k-d}   \big\langle (\delta\hat{v})^{m+k} \delta\hat{h} (\delta\hat{v})^{n-m+d} \big\rangle.
\end{eqnarray}
By dividing Eq.~\eqref{symmetry-Poisson-1-d} to $\langle \hat{v}^\alpha \rangle$ [given in Eq.~(\ref{v-alpha-exp})] we obtain
\begin{eqnarray}
\frac{1}{3}\frac{\partial_{\tau}\langle \hat{v}^\alpha \rangle}{\langle \hat{v}^\alpha \rangle}
&=&
\frac{\pi G}{\alpha_o}\Big({\cal F}(\alpha)  \langle \hat{h} \rangle +  {\cal G}(\alpha) \Big)
\, , \quad 
\label{Hubble-h0-1}
\end{eqnarray}
where
\begin{eqnarray}
{\cal F}(\alpha) &:=& \left(\sum_{k=0}^\infty    \left({\begin{array}{c}
	\alpha \\
	k  \\
	\end{array} } \right) \langle \hat{v} \rangle^{\alpha-k} G^{k00}\right)^{-1} {\cal A}(\alpha) \, , \\
{\cal G}(\alpha) &:=& \left(\sum_{k=0}^\infty    \left({\begin{array}{c}
	\alpha \\
	k  \\
	\end{array} } \right) \langle \hat{v} \rangle^{\alpha-k} G^{k00}\right)^{-1} {\cal B}(\alpha) \, . \quad \quad \quad 
\end{eqnarray}
Inserting the definition (\ref{Hubble-h-new}) into Eq.~(\ref{Hubble-h0-1}) we obtain
\begin{eqnarray}
\frac{1}{3}\frac{\partial_{\tau}\langle \hat{v}^\alpha \rangle}{\langle \hat{v}^\alpha \rangle}
&=&
{\cal F}(\alpha) \langle \hat{H} \rangle + \frac{\pi G}{\alpha_o} {\cal G}(\alpha) 
\, . \quad 
\label{Hubble-h0}
\end{eqnarray}
%%%
By computing (\ref{Hubble-h0}) for values $\alpha=-1$ and $\alpha=1/3$ and replacing them in Eq.~(\ref{Friedmann-dressed1}) we obtain the dressed Friedmann equation $\tilde{H}$. Let us define 
\begin{eqnarray}
{\cal F}_1 &:=& {\cal F}({\alpha=-1})\, ,  \quad \quad {\cal G}_1 \ :=\ {\cal G}({\alpha=-1})\, ,  \quad \quad \nonumber \\
{\cal F}_2 &:=& {\cal F}({\alpha=1/3})\, ,  \quad \quad {\cal G}_2 \ :=\ {\cal G}({\alpha=1/3})\, ,  \quad \quad 
\end{eqnarray}
and
\begin{eqnarray}
\delta {\cal F} &=& {\cal F}_2 - {\cal F}_1 \, , \quad \quad \delta {\cal G} \ =\ {\cal G}_2 - {\cal G}_1 \, .
\end{eqnarray}
Using these definitions we derive  the dressed Friedmann equation (\ref{Friedmann-dressed1}) as
\begin{eqnarray}
\tilde{H} &=& \frac{3}{4} \left(\delta{\cal F} \cdot \langle \hat{H} \rangle + \frac{\pi G}{\alpha_o} \delta{\cal G}\right).
\label{Hubble-dressed-2}
\end{eqnarray}
To the leading order, we have
\begin{eqnarray}
{\cal F}_1 &\approx& -1 + \mathcal{O}(G^{100}/\langle \hat{v} \rangle),  \quad {\cal G}_1 \approx 0 + \mathcal{O}(G^{101}/\langle \hat{v} \rangle) \nonumber \\
{\cal F}_2 &\approx& \frac{1}{3} + \mathcal{O}(G^{100}/\langle \hat{v} \rangle),  \quad {\cal G}_2 \approx 0 + \mathcal{O}(G^{101}/\langle \hat{v} \rangle). \nonumber
\end{eqnarray}
In Eq.~(\ref{Hubble-dressed-2}), the quantum fluctuations  included in the function ${\cal G}$ are given by moments of the order of $G^{(n+k+d)01}$ which are very small for large volumes. Therefore, the second term is negligible, and the squared dressed Hubble rate can be approximated as $\tilde{H}^2\approx (9/16)(\delta{\cal F})^2 \langle \hat{H} \rangle^2$. Considering only the  leading-order  correction terms, where $\delta \mathcal{F} \approx 4/3$, the dressed Hubble rate reduces to $\tilde{H}^2 \approx \langle \hat{H} \rangle^2$. In this approximation, the dressed Hubble rate has a similar form as the one provided by the effective dynamics of LQC \cite{Husain:2011tk}; however, in the present case, the backreaction should be included. This leading-order term of the modified Friedmann equation for the dressed metric $(\tilde{N}, \tilde{a})$ will be sufficient for our purpose in this paper.

%%%%%%%%%%%%%%%%%%%%%%%%%%%%%%%%%%%%%5
\section{Sub-leading terms in eigenfunctions
 ${\protect\underline{e}}^{\pm}_{\mu,\mathbf{k}}(v)$ } \label{subleadingT}

The rate of convergence for eigenfunctions is a few orders weaker than what is needed for numerical calculations. We found the following corrections improving the convergence rate of functions \eqref{def:eigenfunction-mod}. $b_n$ and $a_n$ are corrections to the phase and amplitude of the eigenfunctions, respectively:
\begin{align}
b_1 &= \frac{ \alpha_o l}{\pi  G \mu^2}, \quad \quad\quad
 b_2 = \frac{\alpha_o^2 l^2}{6 \pi ^2 G^2 \mu^4}, \quad\quad \quad b_3 = \frac{ -36 \pi ^3 G^3 \mu^8+81 \pi ^3 G^3 \mu^4-4 \alpha_o^3 l^3 }{216 \pi ^3 G^3 \mu^6}, \nonumber \\
b_4 &= \frac{-108 \pi ^3 G^3 \mu^8 \alpha_o l-21 \pi ^3 G^3 \mu^4 \alpha_o l+5 \alpha_o^4 l^4}{1080 \pi ^4 G^4 \mu^8}, \quad \quad b_5 = \frac{-108 \pi ^3 G^3 \mu^8 \alpha_o^2 l^2-237 \pi ^3 G^3 \mu^4 \alpha_o^2 l^2-14 \alpha_o^5 l^5}{9072 \pi ^5 G^5 \mu^{10}}, \nonumber \\
b_6 &= \frac{-11664 \pi ^6 G^6 \mu^{16}+29160 \pi ^6 G^6 \mu^{12}-76545 \pi ^6 G^6 \mu^8+480 \pi^3 G^3 \mu^8 \alpha_o^3 l^3+14600 \pi ^3 G^3 \mu^4 \alpha_o^3 l^3+280 \alpha_o^6 l^6}{466560 \pi ^6 G^6 \mu^{12}}, \nonumber \\
b_7 &= \frac{-11664 \pi ^6 G^6 \mu^{16} \alpha_o l+12456 \pi ^6 G^6 \mu^{12} \alpha_o l+53967 \pi ^6 G^6 \mu^8 \alpha_o l-72 \pi ^3 G^3 \mu^8 \alpha_o^4 l^4-9310 \pi ^3 G^3 \mu^4 \alpha_o^4 l^4-88 \alpha_o^7 l^7}{342144 \pi ^7 G^7 \mu^{14}},
\end{align}
and for $a_n$ we find
\begin{align}
 a_1 &= -\frac{\alpha_o l}{6 \pi  G \mu^2},  \quad\quad \quad 
 a_2 = \frac{5 \alpha_o^2 l^2}{72 \pi ^2 G^2 \mu^4},  \quad\quad \quad 
  a_3 = \frac{36 \pi ^3 G^3 \mu^8+27 \pi ^3 G^3 \mu^4-5 \alpha_o^3 l^3}{144 \pi ^3 G^3 \mu^6}, \nonumber \\
 a_4 &= \frac{432 \pi ^3 G^3 \mu^8 \alpha_o l-380 \pi ^3 G^3 \mu^4 \alpha_o l + 65 \alpha_o^4 l^4}{3456 \pi ^4 G^4 \mu^8}, \nonumber\\
 a_5 &= \frac{-216 \pi ^3 G^3 \mu^8 \alpha_o^2 l^2+670 \pi ^3 G^3 \mu^4 \alpha_o^2 l^2-221 \alpha_o^5 l^5}{20736 \pi ^5 G^5 \mu^{10}}, 
\end{align}
where $l = \alpha_o^{-1/3} N_k \hbar k$.
\end{widetext}

\bibliography{References}

%apsrev4-2.bst 2019-01-14 (MD) hand-edited version of apsrev4-1.bst
%Control: key (0)
%Control: author (8) initials jnrlst
%Control: editor formatted (1) identically to author
%Control: production of article title (0) allowed
%Control: page (0) single
%Control: year (1) truncated
%Control: production of eprint (0) enabled
\begin{thebibliography}{48}%
\makeatletter
\providecommand \@ifxundefined [1]{%
 \@ifx{#1\undefined}
}%
\providecommand \@ifnum [1]{%
 \ifnum #1\expandafter \@firstoftwo
 \else \expandafter \@secondoftwo
 \fi
}%
\providecommand \@ifx [1]{%
 \ifx #1\expandafter \@firstoftwo
 \else \expandafter \@secondoftwo
 \fi
}%
\providecommand \natexlab [1]{#1}%
\providecommand \enquote  [1]{``#1''}%
\providecommand \bibnamefont  [1]{#1}%
\providecommand \bibfnamefont [1]{#1}%
\providecommand \citenamefont [1]{#1}%
\providecommand \href@noop [0]{\@secondoftwo}%
\providecommand \href [0]{\begingroup \@sanitize@url \@href}%
\providecommand \@href[1]{\@@startlink{#1}\@@href}%
\providecommand \@@href[1]{\endgroup#1\@@endlink}%
\providecommand \@sanitize@url [0]{\catcode `\\12\catcode `\$12\catcode
  `\&12\catcode `\#12\catcode `\^12\catcode `\_12\catcode `\%12\relax}%
\providecommand \@@startlink[1]{}%
\providecommand \@@endlink[0]{}%
\providecommand \url  [0]{\begingroup\@sanitize@url \@url }%
\providecommand \@url [1]{\endgroup\@href {#1}{\urlprefix }}%
\providecommand \urlprefix  [0]{URL }%
\providecommand \Eprint [0]{\href }%
\providecommand \doibase [0]{https://doi.org/}%
\providecommand \selectlanguage [0]{\@gobble}%
\providecommand \bibinfo  [0]{\@secondoftwo}%
\providecommand \bibfield  [0]{\@secondoftwo}%
\providecommand \translation [1]{[#1]}%
\providecommand \BibitemOpen [0]{}%
\providecommand \bibitemStop [0]{}%
\providecommand \bibitemNoStop [0]{.\EOS\space}%
\providecommand \EOS [0]{\spacefactor3000\relax}%
\providecommand \BibitemShut  [1]{\csname bibitem#1\endcsname}%
\let\auto@bib@innerbib\@empty
%</preamble>
\bibitem [{\citenamefont {Gambini}\ and\ \citenamefont
  {Pullin}(1999)}]{Gambini:1998it}%
  \BibitemOpen
  \bibfield  {author} {\bibinfo {author} {\bibfnamefont {R.}~\bibnamefont
  {Gambini}}\ and\ \bibinfo {author} {\bibfnamefont {J.}~\bibnamefont
  {Pullin}},\ }\bibfield  {title} {\bibinfo {title} {{Nonstandard optics from
  quantum space-time}},\ }\href {https://doi.org/10.1103/PhysRevD.59.124021}
  {\bibfield  {journal} {\bibinfo  {journal} {Phys. Rev.}\ }\textbf {\bibinfo
  {volume} {D59}},\ \bibinfo {pages} {124021} (\bibinfo {year} {1999})},\
  \Eprint {https://arxiv.org/abs/gr-qc/9809038} {arXiv:gr-qc/9809038 [gr-qc]}
  \BibitemShut {NoStop}%
%%CITATION = GR-QC/9809038;%%
\bibitem [{\citenamefont {Alfaro}\ \emph {et~al.}(2000)\citenamefont {Alfaro},
  \citenamefont {Morales-Tecotl},\ and\ \citenamefont
  {Urrutia}}]{Alfaro:1999wd}%
  \BibitemOpen
  \bibfield  {author} {\bibinfo {author} {\bibfnamefont {J.}~\bibnamefont
  {Alfaro}}, \bibinfo {author} {\bibfnamefont {H.~A.}\ \bibnamefont
  {Morales-Tecotl}},\ and\ \bibinfo {author} {\bibfnamefont {L.~F.}\
  \bibnamefont {Urrutia}},\ }\bibfield  {title} {\bibinfo {title} {{Quantum
  gravity corrections to neutrino propagation}},\ }\href
  {https://doi.org/10.1103/PhysRevLett.84.2318} {\bibfield  {journal} {\bibinfo
   {journal} {Phys. Rev. Lett.}\ }\textbf {\bibinfo {volume} {84}},\ \bibinfo
  {pages} {2318} (\bibinfo {year} {2000})},\ \Eprint
  {https://arxiv.org/abs/gr-qc/9909079} {arXiv:gr-qc/9909079 [gr-qc]}
  \BibitemShut {NoStop}%
%%CITATION = GR-QC/9909079;%%
\bibitem [{\citenamefont {Magueijo}\ and\ \citenamefont
  {Smolin}(2004)}]{Magueijo:2002xx}%
  \BibitemOpen
  \bibfield  {author} {\bibinfo {author} {\bibfnamefont {J.}~\bibnamefont
  {Magueijo}}\ and\ \bibinfo {author} {\bibfnamefont {L.}~\bibnamefont
  {Smolin}},\ }\bibfield  {title} {\bibinfo {title} {{Gravity's rainbow}},\
  }\href {https://doi.org/10.1088/0264-9381/21/7/001} {\bibfield  {journal}
  {\bibinfo  {journal} {Class. Quant. Grav.}\ }\textbf {\bibinfo {volume}
  {21}},\ \bibinfo {pages} {1725} (\bibinfo {year} {2004})},\ \Eprint
  {https://arxiv.org/abs/gr-qc/0305055} {arXiv:gr-qc/0305055 [gr-qc]}
  \BibitemShut {NoStop}%
%%CITATION = GR-QC/0305055;%%
\bibitem [{\citenamefont {Mattingly}(2005)}]{Mattingly:2005re}%
  \BibitemOpen
  \bibfield  {author} {\bibinfo {author} {\bibfnamefont {D.}~\bibnamefont
  {Mattingly}},\ }\bibfield  {title} {\bibinfo {title} {{Modern tests of
  Lorentz invariance}},\ }\href {https://doi.org/10.12942/lrr-2005-5}
  {\bibfield  {journal} {\bibinfo  {journal} {Living Rev. Rel.}\ }\textbf
  {\bibinfo {volume} {8}},\ \bibinfo {pages} {5} (\bibinfo {year} {2005})},\
  \Eprint {https://arxiv.org/abs/gr-qc/0502097} {arXiv:gr-qc/0502097 [gr-qc]}
  \BibitemShut {NoStop}%
%%CITATION = GR-QC/0502097;%%
\bibitem [{\citenamefont {Lafrance}\ and\ \citenamefont
  {Myers}(1995)}]{Lafrance:1994in}%
  \BibitemOpen
  \bibfield  {author} {\bibinfo {author} {\bibfnamefont {R.}~\bibnamefont
  {Lafrance}}\ and\ \bibinfo {author} {\bibfnamefont {R.~C.}\ \bibnamefont
  {Myers}},\ }\bibfield  {title} {\bibinfo {title} {{Gravity's rainbow}},\
  }\href {https://doi.org/10.1103/PhysRevD.51.2584} {\bibfield  {journal}
  {\bibinfo  {journal} {Phys. Rev.}\ }\textbf {\bibinfo {volume} {D51}},\
  \bibinfo {pages} {2584} (\bibinfo {year} {1995})},\ \Eprint
  {https://arxiv.org/abs/hep-th/9411018} {arXiv:hep-th/9411018 [hep-th]}
  \BibitemShut {NoStop}%
%%CITATION = HEP-TH/9411018;%%
\bibitem [{\citenamefont {Amelino-Camelia}\ \emph {et~al.}(1998)\citenamefont
  {Amelino-Camelia}, \citenamefont {Ellis}, \citenamefont {Mavromatos},
  \citenamefont {Nanopoulos},\ and\ \citenamefont
  {Sarkar}}]{AmelinoCamelia:1997gz}%
  \BibitemOpen
  \bibfield  {author} {\bibinfo {author} {\bibfnamefont {G.}~\bibnamefont
  {Amelino-Camelia}}, \bibinfo {author} {\bibfnamefont {J.~R.}\ \bibnamefont
  {Ellis}}, \bibinfo {author} {\bibfnamefont {N.~E.}\ \bibnamefont
  {Mavromatos}}, \bibinfo {author} {\bibfnamefont {D.~V.}\ \bibnamefont
  {Nanopoulos}},\ and\ \bibinfo {author} {\bibfnamefont {S.}~\bibnamefont
  {Sarkar}},\ }\bibfield  {title} {\bibinfo {title} {{Tests of quantum gravity
  from observations of gamma-ray bursts}},\ }\href
  {https://doi.org/10.1038/31647} {\bibfield  {journal} {\bibinfo  {journal}
  {Nature}\ }\textbf {\bibinfo {volume} {393}},\ \bibinfo {pages} {763}
  (\bibinfo {year} {1998})},\ \Eprint {https://arxiv.org/abs/astro-ph/9712103}
  {arXiv:astro-ph/9712103 [astro-ph]} \BibitemShut {NoStop}%
%%CITATION = ASTRO-PH/9712103;%%
\bibitem [{\citenamefont {Addazi}\ \emph {et~al.}(2021)\citenamefont {Addazi}
  \emph {et~al.}}]{Addazi:2021xuf}%
  \BibitemOpen
  \bibfield  {author} {\bibinfo {author} {\bibfnamefont {A.}~\bibnamefont
  {Addazi}} \emph {et~al.},\ }\bibfield  {title} {\bibinfo {title} {{Quantum
  gravity phenomenology at the dawn of the multi-messenger era -- A review}},\
  }\href@noop {} {\  (\bibinfo {year} {2021})},\ \Eprint
  {https://arxiv.org/abs/2111.05659} {arXiv:2111.05659 [hep-ph]} \BibitemShut
  {NoStop}%
\bibitem [{\citenamefont {Ashtekar}\ \emph {et~al.}(2009)\citenamefont
  {Ashtekar}, \citenamefont {Kaminski},\ and\ \citenamefont
  {Lewandowski}}]{Ashtekar:2009mb}%
  \BibitemOpen
  \bibfield  {author} {\bibinfo {author} {\bibfnamefont {A.}~\bibnamefont
  {Ashtekar}}, \bibinfo {author} {\bibfnamefont {W.}~\bibnamefont {Kaminski}},\
  and\ \bibinfo {author} {\bibfnamefont {J.}~\bibnamefont {Lewandowski}},\
  }\bibfield  {title} {\bibinfo {title} {{Quantum field theory on a
  cosmological, quantum space-time}},\ }\href
  {https://doi.org/10.1103/PhysRevD.79.064030} {\bibfield  {journal} {\bibinfo
  {journal} {Phys. Rev.}\ }\textbf {\bibinfo {volume} {D79}},\ \bibinfo {pages}
  {064030} (\bibinfo {year} {2009})},\ \Eprint
  {https://arxiv.org/abs/0901.0933} {arXiv:0901.0933 [gr-qc]} \BibitemShut
  {NoStop}%
%%CITATION = ARXIV:0901.0933;%%
\bibitem [{\citenamefont {Dapor}\ \emph {et~al.}(2013)\citenamefont {Dapor},
  \citenamefont {Lewandowski},\ and\ \citenamefont {Puchta}}]{Dapor:2013pka}%
  \BibitemOpen
  \bibfield  {author} {\bibinfo {author} {\bibfnamefont {A.}~\bibnamefont
  {Dapor}}, \bibinfo {author} {\bibfnamefont {J.}~\bibnamefont {Lewandowski}},\
  and\ \bibinfo {author} {\bibfnamefont {J.}~\bibnamefont {Puchta}},\
  }\bibfield  {title} {\bibinfo {title} {{QFT on quantum spacetime: a
  compatible classical framework}},\ }\href
  {https://doi.org/10.1103/PhysRevD.87.104038, 10.1103/PhysRevD.87.129904}
  {\bibfield  {journal} {\bibinfo  {journal} {Phys. Rev.}\ }\textbf {\bibinfo
  {volume} {D87}},\ \bibinfo {pages} {104038} (\bibinfo {year} {2013})},\
  \bibinfo {note} {[Erratum: Phys. Rev.D87,no.12,129904(2013)]},\ \Eprint
  {https://arxiv.org/abs/1302.3038} {arXiv:1302.3038 [gr-qc]} \BibitemShut
  {NoStop}%
%%CITATION = ARXIV:1302.3038;%%
\bibitem [{\citenamefont {Lewandowski}\ \emph {et~al.}(2017)\citenamefont
  {Lewandowski}, \citenamefont {Nouri-Zonoz}, \citenamefont {Parvizi},\ and\
  \citenamefont {Tavakoli}}]{Lewandowski:2017cvz}%
  \BibitemOpen
  \bibfield  {author} {\bibinfo {author} {\bibfnamefont {J.}~\bibnamefont
  {Lewandowski}}, \bibinfo {author} {\bibfnamefont {M.}~\bibnamefont
  {Nouri-Zonoz}}, \bibinfo {author} {\bibfnamefont {A.}~\bibnamefont
  {Parvizi}},\ and\ \bibinfo {author} {\bibfnamefont {Y.}~\bibnamefont
  {Tavakoli}},\ }\bibfield  {title} {\bibinfo {title} {{Quantum theory of
  electromagnetic fields in a cosmological quantum spacetime}},\ }\href
  {https://doi.org/10.1103/PhysRevD.96.106007} {\bibfield  {journal} {\bibinfo
  {journal} {Phys. Rev.}\ }\textbf {\bibinfo {volume} {D96}},\ \bibinfo {pages}
  {106007} (\bibinfo {year} {2017})},\ \Eprint
  {https://arxiv.org/abs/1709.04730} {arXiv:1709.04730 [gr-qc]} \BibitemShut
  {NoStop}%
%%CITATION = ARXIV:1709.04730;%%
\bibitem [{\citenamefont {Dapor}\ \emph {et~al.}(2012)\citenamefont {Dapor},
  \citenamefont {Lewandowski},\ and\ \citenamefont {Tavakoli}}]{Dapor:2012jg}%
  \BibitemOpen
  \bibfield  {author} {\bibinfo {author} {\bibfnamefont {A.}~\bibnamefont
  {Dapor}}, \bibinfo {author} {\bibfnamefont {J.}~\bibnamefont {Lewandowski}},\
  and\ \bibinfo {author} {\bibfnamefont {Y.}~\bibnamefont {Tavakoli}},\
  }\bibfield  {title} {\bibinfo {title} {{Lorentz Symmetry in QFT on Quantum
  Bianchi I Space-Time}},\ }\href {https://doi.org/10.1103/PhysRevD.86.064013}
  {\bibfield  {journal} {\bibinfo  {journal} {Phys. Rev.}\ }\textbf {\bibinfo
  {volume} {D86}},\ \bibinfo {pages} {064013} (\bibinfo {year} {2012})},\
  \Eprint {https://arxiv.org/abs/1207.0671} {arXiv:1207.0671 [gr-qc]}
  \BibitemShut {NoStop}%
%%CITATION = ARXIV:1207.0671;%%
\bibitem [{\citenamefont {Assanioussi}\ \emph {et~al.}(2015)\citenamefont
  {Assanioussi}, \citenamefont {Dapor},\ and\ \citenamefont
  {Lewandowski}}]{Assaniousssi:2014ota}%
  \BibitemOpen
  \bibfield  {author} {\bibinfo {author} {\bibfnamefont {M.}~\bibnamefont
  {Assanioussi}}, \bibinfo {author} {\bibfnamefont {A.}~\bibnamefont {Dapor}},\
  and\ \bibinfo {author} {\bibfnamefont {J.}~\bibnamefont {Lewandowski}},\
  }\bibfield  {title} {\bibinfo {title} {{Rainbow metric from quantum
  gravity}},\ }\href {https://doi.org/10.1016/j.physletb.2015.10.043}
  {\bibfield  {journal} {\bibinfo  {journal} {Phys. Lett.}\ }\textbf {\bibinfo
  {volume} {B751}},\ \bibinfo {pages} {302} (\bibinfo {year} {2015})},\ \Eprint
  {https://arxiv.org/abs/1412.6000} {arXiv:1412.6000 [gr-qc]} \BibitemShut
  {NoStop}%
%%CITATION = ARXIV:1412.6000;%%
\bibitem [{\citenamefont {Garcia-Chung}\ \emph {et~al.}(2021)\citenamefont
  {Garcia-Chung}, \citenamefont {Mertens}, \citenamefont {Rastgoo},
  \citenamefont {Tavakoli},\ and\ \citenamefont
  {Vargas~Moniz}}]{Garcia-Chung:2020zyq}%
  \BibitemOpen
  \bibfield  {author} {\bibinfo {author} {\bibfnamefont {A.}~\bibnamefont
  {Garcia-Chung}}, \bibinfo {author} {\bibfnamefont {J.~B.}\ \bibnamefont
  {Mertens}}, \bibinfo {author} {\bibfnamefont {S.}~\bibnamefont {Rastgoo}},
  \bibinfo {author} {\bibfnamefont {Y.}~\bibnamefont {Tavakoli}},\ and\
  \bibinfo {author} {\bibfnamefont {P.}~\bibnamefont {Vargas~Moniz}},\
  }\bibfield  {title} {\bibinfo {title} {{Propagation of quantum
  gravity-modified gravitational waves on a classical FLRW spacetime}},\ }\href
  {https://doi.org/10.1103/PhysRevD.103.084053} {\bibfield  {journal} {\bibinfo
   {journal} {Phys. Rev. D}\ }\textbf {\bibinfo {volume} {103}},\ \bibinfo
  {pages} {084053} (\bibinfo {year} {2021})},\ \Eprint
  {https://arxiv.org/abs/2012.09366} {arXiv:2012.09366 [gr-qc]} \BibitemShut
  {NoStop}%
\bibitem [{\citenamefont {Bernardini}\ \emph {et~al.}(2015)\citenamefont
  {Bernardini} \emph {et~al.}}]{Bernardini:2014oya}%
  \BibitemOpen
  \bibfield  {author} {\bibinfo {author} {\bibfnamefont {M.~G.}\ \bibnamefont
  {Bernardini}} \emph {et~al.},\ }\bibfield  {title} {\bibinfo {title}
  {{Comparing the spectral lag of short and long gamma-ray bursts and its
  relation with the luminosity}},\ }\href
  {https://doi.org/10.1093/mnras/stu2153} {\bibfield  {journal} {\bibinfo
  {journal} {Mon. Not. Roy. Astron. Soc.}\ }\textbf {\bibinfo {volume} {446}},\
  \bibinfo {pages} {1129} (\bibinfo {year} {2015})},\ \Eprint
  {https://arxiv.org/abs/1410.5216} {arXiv:1410.5216 [astro-ph.HE]}
  \BibitemShut {NoStop}%
\bibitem [{\citenamefont {Yi}\ \emph {et~al.}(2006)\citenamefont {Yi},
  \citenamefont {Liang}, \citenamefont {Qin},\ and\ \citenamefont
  {Lu}}]{Yi:2005ht}%
  \BibitemOpen
  \bibfield  {author} {\bibinfo {author} {\bibfnamefont {T.-F.}\ \bibnamefont
  {Yi}}, \bibinfo {author} {\bibfnamefont {E.-W.}\ \bibnamefont {Liang}},
  \bibinfo {author} {\bibfnamefont {Y.-P.}\ \bibnamefont {Qin}},\ and\ \bibinfo
  {author} {\bibfnamefont {R.-J.}\ \bibnamefont {Lu}},\ }\bibfield  {title}
  {\bibinfo {title} {{On the spectral lags of the short gamma-ray bursts}},\
  }\href {https://doi.org/10.1111/j.1365-2966.2006.10083.x} {\bibfield
  {journal} {\bibinfo  {journal} {Mon. Not. Roy. Astron. Soc.}\ }\textbf
  {\bibinfo {volume} {367}},\ \bibinfo {pages} {1751} (\bibinfo {year}
  {2006})},\ \Eprint {https://arxiv.org/abs/astro-ph/0512270}
  {arXiv:astro-ph/0512270} \BibitemShut {NoStop}%
\bibitem [{\citenamefont {Meszaros}(2006)}]{Meszaros:2006rc}%
  \BibitemOpen
  \bibfield  {author} {\bibinfo {author} {\bibfnamefont {P.}~\bibnamefont
  {Meszaros}},\ }\bibfield  {title} {\bibinfo {title} {{Gamma-Ray Bursts}},\
  }\href {https://doi.org/10.1088/0034-4885/69/8/R01} {\bibfield  {journal}
  {\bibinfo  {journal} {Rept. Prog. Phys.}\ }\textbf {\bibinfo {volume} {69}},\
  \bibinfo {pages} {2259} (\bibinfo {year} {2006})},\ \Eprint
  {https://arxiv.org/abs/astro-ph/0605208} {arXiv:astro-ph/0605208}
  \BibitemShut {NoStop}%
\bibitem [{\citenamefont {Tavakoli}\ \emph {et~al.}(2013)\citenamefont
  {Tavakoli}, \citenamefont {Marto}, \citenamefont {Ziaie},\ and\ \citenamefont
  {Vargas~Moniz}}]{Tavakoli:2013tpa}%
  \BibitemOpen
  \bibfield  {author} {\bibinfo {author} {\bibfnamefont {Y.}~\bibnamefont
  {Tavakoli}}, \bibinfo {author} {\bibfnamefont {J.}~\bibnamefont {Marto}},
  \bibinfo {author} {\bibfnamefont {A.~H.}\ \bibnamefont {Ziaie}},\ and\
  \bibinfo {author} {\bibfnamefont {P.}~\bibnamefont {Vargas~Moniz}},\
  }\bibfield  {title} {\bibinfo {title} {{Semiclassical collapse with tachyon
  field and barotropic fluid}},\ }\href
  {https://doi.org/10.1103/PhysRevD.87.024042} {\bibfield  {journal} {\bibinfo
  {journal} {Phys. Rev.}\ }\textbf {\bibinfo {volume} {D87}},\ \bibinfo {pages}
  {024042} (\bibinfo {year} {2013})}\BibitemShut {NoStop}%
%%CITATION = PHRVA,D87,024042;%%
\bibitem [{\citenamefont {Tavakoli}\ \emph {et~al.}(2014)\citenamefont
  {Tavakoli}, \citenamefont {Marto},\ and\ \citenamefont
  {Dapor}}]{Tavakoli:2013rna}%
  \BibitemOpen
  \bibfield  {author} {\bibinfo {author} {\bibfnamefont {Y.}~\bibnamefont
  {Tavakoli}}, \bibinfo {author} {\bibfnamefont {J.}~\bibnamefont {Marto}},\
  and\ \bibinfo {author} {\bibfnamefont {A.}~\bibnamefont {Dapor}},\ }\bibfield
   {title} {\bibinfo {title} {{Semiclassical dynamics of horizons in
  spherically symmetric collapse}},\ }\href
  {https://doi.org/10.1142/S0218271814500618} {\bibfield  {journal} {\bibinfo
  {journal} {Int. J. Mod. Phys.}\ }\textbf {\bibinfo {volume} {D23}},\ \bibinfo
  {pages} {1450061} (\bibinfo {year} {2014})},\ \Eprint
  {https://arxiv.org/abs/1303.6157} {arXiv:1303.6157 [gr-qc]} \BibitemShut
  {NoStop}%
%%CITATION = ARXIV:1303.6157;%%
\bibitem [{\citenamefont {Goswami}\ \emph {et~al.}(2006)\citenamefont
  {Goswami}, \citenamefont {Joshi},\ and\ \citenamefont
  {Singh}}]{Goswami:2005fu}%
  \BibitemOpen
  \bibfield  {author} {\bibinfo {author} {\bibfnamefont {R.}~\bibnamefont
  {Goswami}}, \bibinfo {author} {\bibfnamefont {P.~S.}\ \bibnamefont {Joshi}},\
  and\ \bibinfo {author} {\bibfnamefont {P.}~\bibnamefont {Singh}},\ }\bibfield
   {title} {\bibinfo {title} {{Quantum evaporation of a naked singularity}},\
  }\href {https://doi.org/10.1103/PhysRevLett.96.031302} {\bibfield  {journal}
  {\bibinfo  {journal} {Phys. Rev. Lett.}\ }\textbf {\bibinfo {volume} {96}},\
  \bibinfo {pages} {031302} (\bibinfo {year} {2006})},\ \Eprint
  {https://arxiv.org/abs/gr-qc/0506129} {arXiv:gr-qc/0506129 [gr-qc]}
  \BibitemShut {NoStop}%
%%CITATION = GR-QC/0506129;%%
\bibitem [{\citenamefont {Bojowald}\ \emph {et~al.}(2005)\citenamefont
  {Bojowald}, \citenamefont {Goswami}, \citenamefont {Maartens},\ and\
  \citenamefont {Singh}}]{Bojowald:2005qw}%
  \BibitemOpen
  \bibfield  {author} {\bibinfo {author} {\bibfnamefont {M.}~\bibnamefont
  {Bojowald}}, \bibinfo {author} {\bibfnamefont {R.}~\bibnamefont {Goswami}},
  \bibinfo {author} {\bibfnamefont {R.}~\bibnamefont {Maartens}},\ and\
  \bibinfo {author} {\bibfnamefont {P.}~\bibnamefont {Singh}},\ }\bibfield
  {title} {\bibinfo {title} {{A Black hole mass threshold from non-singular
  quantum gravitational collapse}},\ }\href
  {https://doi.org/10.1103/PhysRevLett.95.091302} {\bibfield  {journal}
  {\bibinfo  {journal} {Phys. Rev. Lett.}\ }\textbf {\bibinfo {volume} {95}},\
  \bibinfo {pages} {091302} (\bibinfo {year} {2005})},\ \Eprint
  {https://arxiv.org/abs/gr-qc/0503041} {arXiv:gr-qc/0503041 [gr-qc]}
  \BibitemShut {NoStop}%
%%CITATION = GR-QC/0503041;%%
\bibitem [{\citenamefont {Husain}\ and\ \citenamefont
  {Pawlowski}(2012)}]{Husain:2011tk}%
  \BibitemOpen
  \bibfield  {author} {\bibinfo {author} {\bibfnamefont {V.}~\bibnamefont
  {Husain}}\ and\ \bibinfo {author} {\bibfnamefont {T.}~\bibnamefont
  {Pawlowski}},\ }\bibfield  {title} {\bibinfo {title} {{Time and a physical
  Hamiltonian for quantum gravity}},\ }\href
  {https://doi.org/10.1103/PhysRevLett.108.141301} {\bibfield  {journal}
  {\bibinfo  {journal} {Phys. Rev. Lett.}\ }\textbf {\bibinfo {volume} {108}},\
  \bibinfo {pages} {141301} (\bibinfo {year} {2012})},\ \Eprint
  {https://arxiv.org/abs/1108.1145} {arXiv:1108.1145 [gr-qc]} \BibitemShut
  {NoStop}%
%%CITATION = ARXIV:1108.1145;%%
\bibitem [{\citenamefont {Brown}\ and\ \citenamefont
  {Kuchar}(1995)}]{Brown:1994py}%
  \BibitemOpen
  \bibfield  {author} {\bibinfo {author} {\bibfnamefont {J.~D.}\ \bibnamefont
  {Brown}}\ and\ \bibinfo {author} {\bibfnamefont {K.~V.}\ \bibnamefont
  {Kuchar}},\ }\bibfield  {title} {\bibinfo {title} {{Dust as a standard of
  space and time in canonical quantum gravity}},\ }\href
  {https://doi.org/10.1103/PhysRevD.51.5600} {\bibfield  {journal} {\bibinfo
  {journal} {Phys. Rev.}\ }\textbf {\bibinfo {volume} {D51}},\ \bibinfo {pages}
  {5600} (\bibinfo {year} {1995})},\ \Eprint
  {https://arxiv.org/abs/gr-qc/9409001} {arXiv:gr-qc/9409001 [gr-qc]}
  \BibitemShut {NoStop}%
%%CITATION = GR-QC/9409001;%%
\bibitem [{\citenamefont {Bojowald}(2010)}]{Bojowald:2010qpa}%
  \BibitemOpen
  \bibfield  {author} {\bibinfo {author} {\bibfnamefont {M.}~\bibnamefont
  {Bojowald}},\ }\href@noop {} {\emph {\bibinfo {title} {{Canonical Gravity and
  Applications: Cosmology, Black Holes, and Quantum Gravity}}}}\ (\bibinfo
  {publisher} {Cambridge University Press},\ \bibinfo {year}
  {2010})\BibitemShut {NoStop}%
%%CITATION = INSPIRE-1384854;%%
\bibitem [{\citenamefont {Ashtekar}\ \emph
  {et~al.}(2006{\natexlab{a}})\citenamefont {Ashtekar}, \citenamefont
  {Pawlowski},\ and\ \citenamefont {Singh}}]{Ashtekar:2006rx}%
  \BibitemOpen
  \bibfield  {author} {\bibinfo {author} {\bibfnamefont {A.}~\bibnamefont
  {Ashtekar}}, \bibinfo {author} {\bibfnamefont {T.}~\bibnamefont
  {Pawlowski}},\ and\ \bibinfo {author} {\bibfnamefont {P.}~\bibnamefont
  {Singh}},\ }\bibfield  {title} {\bibinfo {title} {{Quantum nature of the big
  bang}},\ }\href {https://doi.org/10.1103/PhysRevLett.96.141301} {\bibfield
  {journal} {\bibinfo  {journal} {Phys. Rev. Lett.}\ }\textbf {\bibinfo
  {volume} {96}},\ \bibinfo {pages} {141301} (\bibinfo {year}
  {2006}{\natexlab{a}})},\ \Eprint {https://arxiv.org/abs/gr-qc/0602086}
  {arXiv:gr-qc/0602086 [gr-qc]} \BibitemShut {NoStop}%
%%CITATION = GR-QC/0602086;%%
\bibitem [{\citenamefont {Ashtekar}\ \emph {et~al.}(2003)\citenamefont
  {Ashtekar}, \citenamefont {Bojowald},\ and\ \citenamefont
  {Lewandowski}}]{Ashtekar:2003hd}%
  \BibitemOpen
  \bibfield  {author} {\bibinfo {author} {\bibfnamefont {A.}~\bibnamefont
  {Ashtekar}}, \bibinfo {author} {\bibfnamefont {M.}~\bibnamefont {Bojowald}},\
  and\ \bibinfo {author} {\bibfnamefont {J.}~\bibnamefont {Lewandowski}},\
  }\bibfield  {title} {\bibinfo {title} {{Mathematical structure of loop
  quantum cosmology}},\ }\href {https://doi.org/10.4310/ATMP.2003.v7.n2.a2}
  {\bibfield  {journal} {\bibinfo  {journal} {Adv. Theor. Math. Phys.}\
  }\textbf {\bibinfo {volume} {7}},\ \bibinfo {pages} {233} (\bibinfo {year}
  {2003})},\ \Eprint {https://arxiv.org/abs/gr-qc/0304074} {arXiv:gr-qc/0304074
  [gr-qc]} \BibitemShut {NoStop}%
%%CITATION = GR-QC/0304074;%%
\bibitem [{\citenamefont {Husain}\ and\ \citenamefont
  {Pawlowski}(2011)}]{Husain:2011tm}%
  \BibitemOpen
  \bibfield  {author} {\bibinfo {author} {\bibfnamefont {V.}~\bibnamefont
  {Husain}}\ and\ \bibinfo {author} {\bibfnamefont {T.}~\bibnamefont
  {Pawlowski}},\ }\bibfield  {title} {\bibinfo {title} {{Dust reference frame
  in quantum cosmology}},\ }\href
  {https://doi.org/10.1088/0264-9381/28/22/225014} {\bibfield  {journal}
  {\bibinfo  {journal} {Class. Quant. Grav.}\ }\textbf {\bibinfo {volume}
  {28}},\ \bibinfo {pages} {225014} (\bibinfo {year} {2011})},\ \Eprint
  {https://arxiv.org/abs/1108.1147} {arXiv:1108.1147 [gr-qc]} \BibitemShut
  {NoStop}%
%%CITATION = ARXIV:1108.1147;%%
\bibitem [{\citenamefont {Giesel}\ \emph {et~al.}(2009)\citenamefont {Giesel},
  \citenamefont {Tambornino},\ and\ \citenamefont {Thiemann}}]{Giesel:2009at}%
  \BibitemOpen
  \bibfield  {author} {\bibinfo {author} {\bibfnamefont {K.}~\bibnamefont
  {Giesel}}, \bibinfo {author} {\bibfnamefont {J.}~\bibnamefont {Tambornino}},\
  and\ \bibinfo {author} {\bibfnamefont {T.}~\bibnamefont {Thiemann}},\
  }\bibfield  {title} {\bibinfo {title} {{Born-Oppenheimer decomposition for
  quantum fields on quantum spacetimes}},\ }\href@noop {} {\  (\bibinfo {year}
  {2009})},\ \Eprint {https://arxiv.org/abs/0911.5331} {arXiv:0911.5331
  [gr-qc]} \BibitemShut {NoStop}%
%%CITATION = ARXIV:0911.5331;%%
\bibitem [{\citenamefont {Bojowald}\ and\ \citenamefont
  {Skirzewski}(2006)}]{Bojowald:2005cw}%
  \BibitemOpen
  \bibfield  {author} {\bibinfo {author} {\bibfnamefont {M.}~\bibnamefont
  {Bojowald}}\ and\ \bibinfo {author} {\bibfnamefont {A.}~\bibnamefont
  {Skirzewski}},\ }\bibfield  {title} {\bibinfo {title} {{Effective equations
  of motion for quantum systems}},\ }\href
  {https://doi.org/10.1142/S0129055X06002772} {\bibfield  {journal} {\bibinfo
  {journal} {Rev. Math. Phys.}\ }\textbf {\bibinfo {volume} {18}},\ \bibinfo
  {pages} {713} (\bibinfo {year} {2006})},\ \Eprint
  {https://arxiv.org/abs/math-ph/0511043} {arXiv:math-ph/0511043} \BibitemShut
  {NoStop}%
\bibitem [{\citenamefont {Berger}(1975)}]{Berger:1975ag}%
  \BibitemOpen
  \bibfield  {author} {\bibinfo {author} {\bibfnamefont {B.~K.}\ \bibnamefont
  {Berger}},\ }\bibfield  {title} {\bibinfo {title} {{Scalar Particle Creation
  in an Anisotropic Universe}},\ }\href
  {https://doi.org/10.1103/PhysRevD.12.368} {\bibfield  {journal} {\bibinfo
  {journal} {Phys. Rev.}\ }\textbf {\bibinfo {volume} {D12}},\ \bibinfo {pages}
  {368} (\bibinfo {year} {1975})}\BibitemShut {NoStop}%
%%CITATION = PHRVA,D12,368;%%
\bibitem [{\citenamefont {Ashtekar}\ \emph
  {et~al.}(2006{\natexlab{b}})\citenamefont {Ashtekar}, \citenamefont
  {Pawlowski},\ and\ \citenamefont {Singh}}]{Ashtekar:2006wn}%
  \BibitemOpen
  \bibfield  {author} {\bibinfo {author} {\bibfnamefont {A.}~\bibnamefont
  {Ashtekar}}, \bibinfo {author} {\bibfnamefont {T.}~\bibnamefont
  {Pawlowski}},\ and\ \bibinfo {author} {\bibfnamefont {P.}~\bibnamefont
  {Singh}},\ }\bibfield  {title} {\bibinfo {title} {{Quantum Nature of the Big
  Bang: Improved dynamics}},\ }\href
  {https://doi.org/10.1103/PhysRevD.74.084003} {\bibfield  {journal} {\bibinfo
  {journal} {Phys. Rev. D}\ }\textbf {\bibinfo {volume} {74}},\ \bibinfo
  {pages} {084003} (\bibinfo {year} {2006}{\natexlab{b}})},\ \Eprint
  {https://arxiv.org/abs/gr-qc/0607039} {arXiv:gr-qc/0607039} \BibitemShut
  {NoStop}%
\bibitem [{\citenamefont {Mena~Marugan}\ \emph {et~al.}(2011)\citenamefont
  {Mena~Marugan}, \citenamefont {Olmedo},\ and\ \citenamefont
  {Pawlowski}}]{MenaMarugan:2011me}%
  \BibitemOpen
  \bibfield  {author} {\bibinfo {author} {\bibfnamefont {G.~A.}\ \bibnamefont
  {Mena~Marugan}}, \bibinfo {author} {\bibfnamefont {J.}~\bibnamefont
  {Olmedo}},\ and\ \bibinfo {author} {\bibfnamefont {T.}~\bibnamefont
  {Pawlowski}},\ }\bibfield  {title} {\bibinfo {title} {{Prescriptions in Loop
  Quantum Cosmology: A comparative analysis}},\ }\href
  {https://doi.org/10.1103/PhysRevD.84.064012} {\bibfield  {journal} {\bibinfo
  {journal} {Phys. Rev. D}\ }\textbf {\bibinfo {volume} {84}},\ \bibinfo
  {pages} {064012} (\bibinfo {year} {2011})},\ \Eprint
  {https://arxiv.org/abs/1108.0829} {arXiv:1108.0829 [gr-qc]} \BibitemShut
  {NoStop}%
\bibitem [{\citenamefont {Parvizi}\ \emph {et~al.}(shed)\citenamefont
  {Parvizi}, \citenamefont {Pawlowski},\ and\ \citenamefont
  {Tavakoli}}]{ModifiedDR:2022}%
  \BibitemOpen
  \bibfield  {author} {\bibinfo {author} {\bibfnamefont {A.}~\bibnamefont
  {Parvizi}}, \bibinfo {author} {\bibfnamefont {T.}~\bibnamefont {Pawlowski}},\
  and\ \bibinfo {author} {\bibfnamefont {Y.}~\bibnamefont {Tavakoli}},\
  }\href@noop {} {\  (\bibinfo {year} {To be published})}\BibitemShut {NoStop}%
\bibitem [{\citenamefont {Vaidya}(1951)}]{Vaidya:1951zza}%
  \BibitemOpen
  \bibfield  {author} {\bibinfo {author} {\bibfnamefont {P.~C.}\ \bibnamefont
  {Vaidya}},\ }\bibfield  {title} {\bibinfo {title} {{Nonstatic Solutions of
  Einstein's Field Equations for Spheres of Fluids Radiating Energy}},\ }\href
  {https://doi.org/10.1103/PhysRev.83.10} {\bibfield  {journal} {\bibinfo
  {journal} {Phys. Rev.}\ }\textbf {\bibinfo {volume} {83}},\ \bibinfo {pages}
  {10} (\bibinfo {year} {1951})}\BibitemShut {NoStop}%
\bibitem [{\citenamefont {Joshi}(1993)}]{Joshi:1987wg}%
  \BibitemOpen
  \bibfield  {author} {\bibinfo {author} {\bibfnamefont {P.~S.}\ \bibnamefont
  {Joshi}},\ }\href@noop {} {\emph {\bibinfo {title} {{Global aspects in
  gravitation and cosmology}}}}\ (\bibinfo  {publisher} {Clarendon Press},\
  \bibinfo {address} {Oxford},\ \bibinfo {year} {1993})\BibitemShut {NoStop}%
\bibitem [{\citenamefont {Vaidya}(1999{\natexlab{a}})}]{Vaidya:1999zz}%
  \BibitemOpen
  \bibfield  {author} {\bibinfo {author} {\bibfnamefont {P.~C.}\ \bibnamefont
  {Vaidya}},\ }\bibfield  {title} {\bibinfo {title} {{The External Field of a
  Radiating Star in Relativity}},\ }\href
  {https://doi.org/10.1023/A:1018871522880} {\bibfield  {journal} {\bibinfo
  {journal} {Gen. Rel. Grav.}\ }\textbf {\bibinfo {volume} {31}},\ \bibinfo
  {pages} {119} (\bibinfo {year} {1999}{\natexlab{a}})}\BibitemShut {NoStop}%
%%CITATION = GRGVA,31,119;%%
\bibitem [{\citenamefont {Vaidya}(1999{\natexlab{b}})}]{Vaidya:1999zza}%
  \BibitemOpen
  \bibfield  {author} {\bibinfo {author} {\bibfnamefont {P.~C.}\ \bibnamefont
  {Vaidya}},\ }\bibfield  {title} {\bibinfo {title} {{The Gravitational Field
  of a Radiating Star}},\ }\href {https://doi.org/10.1023/A:1018875606950}
  {\bibfield  {journal} {\bibinfo  {journal} {Gen. Rel. Grav.}\ }\textbf
  {\bibinfo {volume} {31}},\ \bibinfo {pages} {121} (\bibinfo {year}
  {1999}{\natexlab{b}})}\BibitemShut {NoStop}%
%%CITATION = GRGVA,31,121;%%
\bibitem [{\citenamefont {Vaidya}(1953)}]{Vaidya:1953zza}%
  \BibitemOpen
  \bibfield  {author} {\bibinfo {author} {\bibfnamefont {P.~C.}\ \bibnamefont
  {Vaidya}},\ }\bibfield  {title} {\bibinfo {title} {{Newtonian Time in General
  Relativity}},\ }\href {https://doi.org/10.1038/171260a0} {\bibfield
  {journal} {\bibinfo  {journal} {Nature}\ }\textbf {\bibinfo {volume} {171}},\
  \bibinfo {pages} {260} (\bibinfo {year} {1953})}\BibitemShut {NoStop}%
%%CITATION = NATUA,171,260;%%
\bibitem [{\citenamefont {Hayward}(1993)}]{Hayward:1993mw}%
  \BibitemOpen
  \bibfield  {author} {\bibinfo {author} {\bibfnamefont {S.~A.}\ \bibnamefont
  {Hayward}},\ }\bibfield  {title} {\bibinfo {title} {{Marginal surfaces and
  apparent horizons}},\ }\href@noop {} {\  (\bibinfo {year} {1993})},\ \Eprint
  {https://arxiv.org/abs/gr-qc/9303006} {arXiv:gr-qc/9303006} \BibitemShut
  {NoStop}%
\bibitem [{\citenamefont {Ashtekar}\ and\ \citenamefont
  {Krishnan}(2002)}]{Ashtekar:2002ag}%
  \BibitemOpen
  \bibfield  {author} {\bibinfo {author} {\bibfnamefont {A.}~\bibnamefont
  {Ashtekar}}\ and\ \bibinfo {author} {\bibfnamefont {B.}~\bibnamefont
  {Krishnan}},\ }\bibfield  {title} {\bibinfo {title} {{Dynamical horizons:
  Energy, angular momentum, fluxes and balance laws}},\ }\href
  {https://doi.org/10.1103/PhysRevLett.89.261101} {\bibfield  {journal}
  {\bibinfo  {journal} {Phys. Rev. Lett.}\ }\textbf {\bibinfo {volume} {89}},\
  \bibinfo {pages} {261101} (\bibinfo {year} {2002})},\ \Eprint
  {https://arxiv.org/abs/gr-qc/0207080} {arXiv:gr-qc/0207080} \BibitemShut
  {NoStop}%
\bibitem [{\citenamefont {Ziaie}\ and\ \citenamefont
  {Tavakoli}(2020)}]{Ziaie:2019klz}%
  \BibitemOpen
  \bibfield  {author} {\bibinfo {author} {\bibfnamefont {A.~H.}\ \bibnamefont
  {Ziaie}}\ and\ \bibinfo {author} {\bibfnamefont {Y.}~\bibnamefont
  {Tavakoli}},\ }\bibfield  {title} {\bibinfo {title} {{Null Fluid Collapse in
  Rastall Theory of Gravity}},\ }\href {https://doi.org/10.1002/andp.202000064}
  {\bibfield  {journal} {\bibinfo  {journal} {Annalen Phys.}\ }\textbf
  {\bibinfo {volume} {532}},\ \bibinfo {pages} {2000064} (\bibinfo {year}
  {2020})},\ \Eprint {https://arxiv.org/abs/1912.08890} {arXiv:1912.08890
  [gr-qc]} \BibitemShut {NoStop}%
\bibitem [{\citenamefont {Hawking}\ and\ \citenamefont
  {Ellis}(2011)}]{Hawking:1973uf}%
  \BibitemOpen
  \bibfield  {author} {\bibinfo {author} {\bibfnamefont {S.~W.}\ \bibnamefont
  {Hawking}}\ and\ \bibinfo {author} {\bibfnamefont {G.~F.~R.}\ \bibnamefont
  {Ellis}},\ }\href {https://doi.org/10.1017/CBO9780511524646} {\emph {\bibinfo
  {title} {{The Large Scale Structure of Space-Time}}}},\ Cambridge Monographs
  on Mathematical Physics\ (\bibinfo  {publisher} {Cambridge University
  Press},\ \bibinfo {year} {2011})\BibitemShut {NoStop}%
\bibitem [{\citenamefont {Weinberg}(1972)}]{Weinberg:1972kfs}%
  \BibitemOpen
  \bibfield  {author} {\bibinfo {author} {\bibfnamefont {S.}~\bibnamefont
  {Weinberg}},\ }\href@noop {} {\emph {\bibinfo {title} {{Gravitation and
  Cosmology}: {Principles and Applications of the General Theory of
  Relativity}}}}\ (\bibinfo  {publisher} {John Wiley and Sons},\ \bibinfo
  {address} {New York},\ \bibinfo {year} {1972})\BibitemShut {NoStop}%
\bibitem [{\citenamefont {Husain}(1996)}]{Husain:1995bf}%
  \BibitemOpen
  \bibfield  {author} {\bibinfo {author} {\bibfnamefont {V.}~\bibnamefont
  {Husain}},\ }\bibfield  {title} {\bibinfo {title} {{Exact solutions for null
  fluid collapse}},\ }\href {https://doi.org/10.1103/PhysRevD.53.R1759}
  {\bibfield  {journal} {\bibinfo  {journal} {Phys. Rev. D}\ }\textbf {\bibinfo
  {volume} {53}},\ \bibinfo {pages} {R1759} (\bibinfo {year}
  {1996})}\BibitemShut {NoStop}%
\bibitem [{\citenamefont {Keeton}\ and\ \citenamefont
  {Petters}(2005)}]{Keeton:2005jd}%
  \BibitemOpen
  \bibfield  {author} {\bibinfo {author} {\bibfnamefont {C.~R.}\ \bibnamefont
  {Keeton}}\ and\ \bibinfo {author} {\bibfnamefont {A.~O.}\ \bibnamefont
  {Petters}},\ }\bibfield  {title} {\bibinfo {title} {{Formalism for testing
  theories of gravity using lensing by compact objects. I. Static, spherically
  symmetric case}},\ }\href {https://doi.org/10.1103/PhysRevD.72.104006}
  {\bibfield  {journal} {\bibinfo  {journal} {Phys. Rev. D}\ }\textbf {\bibinfo
  {volume} {72}},\ \bibinfo {pages} {104006} (\bibinfo {year} {2005})},\
  \Eprint {https://arxiv.org/abs/gr-qc/0511019} {arXiv:gr-qc/0511019}
  \BibitemShut {NoStop}%
\bibitem [{\citenamefont {Nouri-Zonoz}\ \emph {et~al.}(shed)\citenamefont
  {Nouri-Zonoz}, \citenamefont {Parvizi},\ and\ \citenamefont
  {Tavakoli}}]{LensingRBH}%
  \BibitemOpen
  \bibfield  {author} {\bibinfo {author} {\bibfnamefont {M.}~\bibnamefont
  {Nouri-Zonoz}}, \bibinfo {author} {\bibfnamefont {A.}~\bibnamefont
  {Parvizi}},\ and\ \bibinfo {author} {\bibfnamefont {Y.}~\bibnamefont
  {Tavakoli}},\ }\bibfield  {title} {\bibinfo {title} {{}},\ }\href@noop {} {\
  (\bibinfo {year} {To be published})}\BibitemShut {NoStop}%
\bibitem [{\citenamefont {Sereno}(2004)}]{Sereno:2003nd}%
  \BibitemOpen
  \bibfield  {author} {\bibinfo {author} {\bibfnamefont {M.}~\bibnamefont
  {Sereno}},\ }\bibfield  {title} {\bibinfo {title} {{Weak field limit of
  Reissner-Nordstrom black hole lensing}},\ }\href
  {https://doi.org/10.1103/PhysRevD.69.023002} {\bibfield  {journal} {\bibinfo
  {journal} {Phys. Rev. D}\ }\textbf {\bibinfo {volume} {69}},\ \bibinfo
  {pages} {023002} (\bibinfo {year} {2004})},\ \Eprint
  {https://arxiv.org/abs/gr-qc/0310063} {arXiv:gr-qc/0310063} \BibitemShut
  {NoStop}%
\bibitem [{\citenamefont {Piron}(2016)}]{Piron:2015wtz}%
  \BibitemOpen
  \bibfield  {author} {\bibinfo {author} {\bibfnamefont {F.}~\bibnamefont
  {Piron}},\ }\bibfield  {title} {\bibinfo {title} {{Gamma-Ray Bursts at high
  and very high energies}},\ }\href
  {https://doi.org/10.1016/j.crhy.2016.04.005} {\bibfield  {journal} {\bibinfo
  {journal} {Comptes Rendus Physique}\ }\textbf {\bibinfo {volume} {17}},\
  \bibinfo {pages} {617} (\bibinfo {year} {2016})},\ \Eprint
  {https://arxiv.org/abs/1512.04241} {arXiv:1512.04241 [astro-ph.HE]}
  \BibitemShut {NoStop}%
\bibitem [{\citenamefont {Miller-Jones}\ \emph {et~al.}(2021)\citenamefont
  {Miller-Jones} \emph {et~al.}}]{Miller-Jones:2021plh}%
  \BibitemOpen
  \bibfield  {author} {\bibinfo {author} {\bibfnamefont {J.~C.~A.}\
  \bibnamefont {Miller-Jones}} \emph {et~al.},\ }\bibfield  {title} {\bibinfo
  {title} {{Cygnus X-1 contains a 21\textendash{}solar mass black
  hole\textemdash{}Implications for massive star winds}},\ }\href
  {https://doi.org/10.1126/science.abb3363} {\bibfield  {journal} {\bibinfo
  {journal} {Science}\ }\textbf {\bibinfo {volume} {371}},\ \bibinfo {pages}
  {1046} (\bibinfo {year} {2021})},\ \Eprint {https://arxiv.org/abs/2102.09091}
  {arXiv:2102.09091 [astro-ph.HE]} \BibitemShut {NoStop}%
\end{thebibliography}%

\end{document}